\documentclass[twocolumn,preprintnumbers,prb,aps,amssymb,amsmath,showpacs,floatfix,superscriptaddress,longbibliography]{revtex4-1}

\usepackage{graphicx}
\usepackage{epsfig}
\usepackage{natbib}
\usepackage{dcolumn}
\usepackage{bm}
\usepackage{soul}
\usepackage{sidecap}
\usepackage[usenames,dvipsnames]{color}
\usepackage{setspace}
\usepackage[colorlinks=true,linkcolor=blue,urlcolor=blue,citecolor=blue]{hyperref}
\usepackage[usenames,dvipsnames,svgnames,table]{xcolor}

\newcommand{\lsco}{La$_{2-x}$Sr$_x$CuO$_4$}

\newcommand{\lnsco}{La$_{1.6-x}$Nd$_{0.4}$Sr$_x$CuO$_4$}
\newcommand{\lesco}{La$_{1.8-x}$Eu$_{0.2}$Sr$_x$CuO$_4$}

\newcommand{\ybco}{YBa$_{2}$Cu$_{3}$O$_{y}$}

\newcommand{\TN}{$T_{\rm N}$}
\newcommand{\Tc}{$T_{\rm c}$}
\newcommand{\Tstar}{$T^{\star}$}
\newcommand{\Trho}{$T_{\rho}$}
\newcommand{\Tnu}{$T_{\nu}$}

\newcommand{\Tmin}{$T_{\rm min}$}
\newcommand{\TB}{$T_{\rm B}$}
\newcommand{\Tonset}{$T^{\rm onset}$}

\newcommand{\TCDW}{$T_{\rm CDW}$}
\newcommand{\TSDW}{$T_{\rm SDW}$}

\newcommand{\Tnem}{$T_{\rm nem}$}
\newcommand{\Tmag}{$T_{\rm mag}$}

\newcommand{\Hc}{$H_{\rm c2}$}

\newcommand{\pN}{$p_{\rm N}$}
\newcommand{\pstar}{$p^{\star}$}
\newcommand{\pc}{$p_{\rm c}$}
\newcommand{\pcone}{$p_{\rm c1}$}
\newcommand{\pctwo}{$p_{\rm c2}$}

\newcommand{\pFS}{$p_{\rm FS}$}
\newcommand{\pcdwone}{$p_1^{\rm CDW}$}
\newcommand{\pcdwtwo}{$p_2^{\rm CDW}$}
\newcommand{\psdw}{$p_{\rm SDW}$}

\newcommand{\RH}{$R_{\rm H}$}
\newcommand{\nH}{$n_{\rm H}$}

\newcommand{\ie}{{\it i.e.}}
\newcommand{\eg}{{\it e.g.}}

\begin{document}

\title{Pseudogap temperature \texorpdfstring{$T^\star$}{T*} of cuprate superconductors from the Nernst effect}

\author{O.~ Cyr-Choini\`{e}re}
\altaffiliation{Present address: Yale School of Engineering and Applied Science, Yale University, New Haven, Connecticut 06511}
%\altaffiliation{Present address: Department of Physics, McGill University, Montreal, Qu\'{e}bec H3A 2T8, Canada}
\affiliation{Institut quantique, D\'{e}partement de physique  \&  RQMP, Universit\'{e} de Sherbrooke, Sherbrooke,  Qu\'{e}bec J1K 2R1, Canada}

\author{R.~Daou}
\altaffiliation{Present address: Laboratoire CRISMAT, CNRS, Caen, France.}
\affiliation{Institut quantique, D\'{e}partement de physique  \&  RQMP, Universit\'{e} de Sherbrooke, Sherbrooke,  Qu\'{e}bec J1K 2R1, Canada}
  
\author{F.~Lalibert\'{e}}
\affiliation{Institut quantique, D\'{e}partement de physique  \&  RQMP, Universit\'{e} de Sherbrooke, Sherbrooke,  Qu\'{e}bec J1K 2R1, Canada}
  
\author{C.~Collignon}
\affiliation{Institut quantique, D\'{e}partement de physique  \&  RQMP, Universit\'{e} de Sherbrooke, Sherbrooke,  Qu\'{e}bec J1K 2R1, Canada}
   
\author{S.~Badoux}
\affiliation{Institut quantique, D\'{e}partement de physique  \&  RQMP, Universit\'{e} de Sherbrooke, Sherbrooke,  Qu\'{e}bec J1K 2R1, Canada}
 
\author{D.~LeBoeuf}
\altaffiliation{Present address: Laboratoire National des Champs Magn\'{e}tiques Intenses, UPR 3228, (CNRS-INSA-UJF-UPS), Grenoble 38042, France}
\affiliation{Institut quantique, D\'{e}partement de physique  \&  RQMP, Universit\'{e} de Sherbrooke, Sherbrooke,  Qu\'{e}bec J1K 2R1, Canada}
 
\author{J.~Chang}
\altaffiliation{Present address: Department of Physics, University of Zurich, Winterthurerstrasse 190, 8057 Zurich, Switzerland}
\affiliation{Institut quantique, D\'{e}partement de physique  \&  RQMP, Universit\'{e} de Sherbrooke, Sherbrooke,  Qu\'{e}bec J1K 2R1, Canada}
  
\author{B.~J.~Ramshaw}
\altaffiliation{Present address: Department of Physics, Cornell University, 531 Clark Hall, Ithaca, NY, 14853-2501, USA}
\affiliation{Department of Physics and Astronomy, University of British Columbia, Vancouver, British Columbia V6T 1Z4, Canada}

\author{D.~A.~Bonn}
\affiliation{Department of Physics and Astronomy, University of British Columbia, Vancouver, British Columbia V6T 1Z4, Canada}
\affiliation{Canadian Institute for Advanced Research, Toronto, Ontario M5G 1Z8, Canada}

\author{W.~N.~Hardy}
\affiliation{Department of Physics and Astronomy, University of British Columbia, Vancouver, British Columbia V6T 1Z4, Canada}
\affiliation{Canadian Institute for Advanced Research, Toronto, Ontario M5G 1Z8, Canada}

\author{R.~Liang}
\affiliation{Department of Physics and Astronomy, University of British Columbia, Vancouver, British Columbia V6T 1Z4, Canada}
\affiliation{Canadian Institute for Advanced Research, Toronto, Ontario M5G 1Z8, Canada}

\author{J.-Q.~Yan}
\affiliation{Ames Laboratory, Ames, Iowa 50011, USA}

\author{J.-G.~Cheng}
\affiliation{University of Texas - Austin, Austin, Texas 78712, USA}

\author{J.-S.~Zhou}
\affiliation{University of Texas - Austin, Austin, Texas 78712, USA}
  
\author{J.~B.~Goodenough}
\affiliation{University of Texas - Austin, Austin, Texas 78712, USA}
 
\author{S.~Pyon}
\affiliation{Department of Advanced Materials, University of Tokyo, Kashiwa 277-8561, Japan} 

\author{T.~Takayama}
\affiliation{Department of Advanced Materials, University of Tokyo, Kashiwa 277-8561, Japan}
 
\author{H.~Takagi}
\altaffiliation{Present address: Max Planck Institute for Solid State Research, 70569 Stuttgart, Germany}
\affiliation{Department of Advanced Materials, University of Tokyo, Kashiwa 277-8561, Japan} 
\affiliation{RIKEN (The Institute of Physical and Chemical Research), Wako, 351-0198, Japan} 

\author{N.~Doiron-Leyraud}
\affiliation{Institut quantique, D\'{e}partement de physique  \&  RQMP, Universit\'{e} de Sherbrooke, Sherbrooke,  Qu\'{e}bec J1K 2R1, Canada}

\author{Louis Taillefer}
\email{louis.taillefer@usherbrooke.ca}
\affiliation{Institut quantique, D\'{e}partement de physique  \&  RQMP, Universit\'{e} de Sherbrooke, Sherbrooke,  Qu\'{e}bec J1K 2R1, Canada}
\affiliation{Canadian Institute for Advanced Research, Toronto, Ontario M5G 1Z8, Canada}

\date{\today}

\begin{abstract}

We use the Nernst effect to delineate the boundary of the pseudogap phase in 
the temperature-doping phase diagram of hole-doped cuprate superconductors.
New data for the Nernst coefficient $\nu(T)$ of \ybco{} (YBCO), \lesco{} (Eu-LSCO) and \lnsco{} (Nd-LSCO)
are presented and compared with previously published data on YBCO, Eu-LSCO, Nd-LSCO, and \lsco{} (LSCO). 
The temperature $T_\nu$ at which $\nu$\,/\,$T$ deviates from its high-temperature linear behaviour is found 
to coincide with the temperature at which the resistivity $\rho(T)$ deviates from its linear-$T$ dependence, 
which we take as the definition of the pseudogap temperature $T^\star$
-- in agreement with the temperature at which the antinodal spectral gap detected in angle-resolved photoemission spectroscopy (ARPES) opens.
We track $T^\star$ as a function of doping and find that 
it decreases linearly vs $p$ in all four materials, having the same value in the three LSCO-based cuprates,
irrespective of their different crystal structures.
At low $p$, \Tstar~is higher than the onset temperature of the various orders observed in underdoped cuprates,
suggesting that these orders are secondary instabilities of the pseudogap phase.
A linear
extrapolation of \Tstar($p$) to $p$\,=\,$0$ yields \Tstar($p$\,$\to$\,$0$)\,$\simeq$\,\TN(0), the N\'eel temperature for the onset of antiferromagnetic order at $p$\,=\,$0$, 
suggesting that there is a link between pseudogap and antiferromagnetism.
With increasing $p$, \Tstar($p$) extrapolates linearly to zero at $p$\,$\simeq$\,\pctwo, 
the critical doping below which superconductivity emerges
at high doping, suggesting that the conditions which favour pseudogap formation also favour pairing.
We also use the Nernst effect to investigate how far superconducting fluctuations extend above
the critical temperature \Tc, as a function of doping, 
and find that a narrow fluctuation regime tracks \Tc, and not \Tstar. 
This confirms that the pseudogap phase is not a form of precursor superconductivity, 
and fluctuations in the phase of the superconducting order parameter are not what causes \Tc{} to fall on the underdoped side of the \Tc{} dome.

\end{abstract}


\maketitle

%%%%%%%%%%%%%%%%%%%%%%%%%%%%%%%%%%%%%%%%%%%%%%%%%%%%%%%%%%%%%%%%%%%%%%%
%%%%%% Introduction
%%%%%%%%%%%%%%%%%%%%%%%%%%%%%%%%%%%%%%%%%%%%%%%%%%%%%%%%%%%%%%%%%%%%%%%

\section{Introduction}
\label{sec:Intro}

Understanding the mechanisms responsible for superconductivity in cuprates requires that we elucidate 
the nature of the enigmatic pseudogap phase that coexists with the superconducting phase in
their temperature-doping phase diagram.
The pseudogap is a partial gap in the spectral function that opens at the Fermi energy
in $k$-space locations $(\pm \pi,0)$ and $(0,\pm \pi)$, the so-called anti-nodal regions of the first Brillouin zone, 
as measured by angle-resolved photoemission spectroscopy (ARPES)~[\onlinecite{Damascelli2003}]. 
It is essential to know the boundary of the pseudogap phase, \ie~the location of the pseudogap temperature \Tstar{} as a function
of doping $p$ and of the critical doping \pstar{} where the pseudogap phase ends at $T$\,=\,$0$.

Nd-LSCO is the only cuprate material for which this information is complete.
Here, the critical point has been 
located at 
\pstar\,=\,$0.23$\,$\pm$\,$0.01$, 
from in-plane resistivity~[\onlinecite{Daou2009},\onlinecite{Collignon2017}], 
out-of-plane resistivity~[\onlinecite{Cyr-Choiniere2010}]
and Hall effect~[\onlinecite{Collignon2017}]. 
This location is consistent with ARPES measurements at low temperature that find a large pseudogap
at $p$\,=\,$0.20$ but none at $p$\,=\,$0.24$~[\onlinecite{Matt2015}]. 
Moreover, in Nd-LSCO the temperature \Trho{} below which the resistivity $\rho(T)$ 
deviates from its linear-$T$ dependence at high $T$~[\onlinecite{Daou2009},\onlinecite{Collignon2017}] 
agrees with the onset temperature for the opening of the pseudogap measured by ARPES~[\onlinecite{Matt2015}]. 
This shows that resistivity measurements can be used to track \Tstar\,=\,\Trho{} vs $p$ in Nd-LSCO.

In only two other cuprates is the location of \pstar{} well established.
In YBCO, recent high-field Hall measurements in the $T$\,=\,$0$ limit find 
\pstar\,=\,$0.195$\,$\pm$\,$0.005$~[\onlinecite{Badoux2016}], 
in agreement with earlier analyses that yield \pstar\,=\,$0.19$\,$\pm$\,$0.01$~[\onlinecite{Tallon2001}]. 
However, there are no ARPES measurements of \Tstar{} in YBCO,
so one typically relies on \Trho{} determined from resistivity
without spectroscopic confirmation,
and there is some debate as to where \Trho{} crosses the superconducting temperature \Tc~[\onlinecite{Rullier-Albenque2011}]. 
In LSCO, high-field resistivity measurements in the $T$\,=\,$0$ limit~[\onlinecite{Boebinger1996,Cooper2009,Laliberte2016}] yield
\pstar\,=\,$0.18$\,$\pm$\,$0.01$~[\onlinecite{Laliberte2016}]. 
However, there is no consensus on the location of the 
\Tstar$(p)$ line in the phase diagram of LSCO~[\onlinecite{Ando2004},\onlinecite{Hussey2011}]. 

In 
Bi$_2$Sr$_{2-x}$La$_x$CuO$_{6 + \delta}$ (Bi-2201)~[\onlinecite{Kondo2011}] and 
Bi$_2$Sr$_{2}$CaCu$_2$O$_{8 + \delta}$ (Bi-2212)~[\onlinecite{Vishik2012}], 
ARPES measurements have delineated the \Tstar$(p)$ line quite well, and it is found to agree with \Trho{} from resistivity. 
However, there is no agreement on the location of \pstar.
In Bi-2201, STM measurements suggest that \pstar\,$>$\,\pctwo, 
the critical doping below which superconductivity emerges at high doping~[\onlinecite{He2014}], 
while NMR measurements show that \pstar\,$<$\,\pc~[\onlinecite{Kawasaki2010}]. 
In Bi-2212, STM measurements find that \pstar\,=\,$0.19$ (in the superconducting state)~[\onlinecite{Fujita2014}],
while Raman measurements find \pstar\,=\,$0.22$ (in the normal state)~[\onlinecite{Benhabib2015}]. 

In this Article, we show that the Nernst effect can be used to detect \Tstar,
not only in YBCO and 
HgBa$_2$CuO$_{4 + \delta}$ (Hg-1201), 
as shown previously~[\onlinecite{Daou2010},\onlinecite{Doiron-Leyraud2013}], 
but also in the LSCO-based cuprates (Fig.~\ref{sketch}). 
We present new data on 
YBCO, Nd-LSCO and Eu-LSCO, and combine these with
published data on LSCO, Nd-LSCO and Eu-LSCO to determine the pseudogap boundary in all four materials.
We find that the three LSCO-based cuprates have the same 
\Tstar$(p)$ line up to $p$\,$\simeq$\,$0.17$, 
irrespective of their different crystal structures.
This suggests that the interactions responsible for the pseudogap have the same strength.
From the fact that \pstar{} is quite different in LSCO and Nd-LSCO ($0.18$ vs $0.23$), 
we infer that additional mechanisms must dictate the location of the $T$\,=\,$0$ critical point. 
\Tstar{} lies on a line that connects \TN{} at $p$\,=\,$0$, the N\'eel temperature for antiferromagnetic order
at zero doping, to \pctwo.
In YBCO, we again find that \Tstar{} lies on a line connecting \TN{} and \pctwo,
even if \TN{} is now a factor 1.5 larger.
In other words, \Tstar{} in YBCO is 1.5 times larger than in LSCO.
This suggests a link between antiferromagnetism, pseudogap and superconductivity.

The Article is organized as follows.
In sec.~\ref{sec:Nernst}, we give a brief introduction to the Nernst effect.
In sec.~\ref{sec:Methods}, we provide information on the experimental measurement of the Nernst effect. 
In sec.~\ref{sec:YBCO}, we establish the \Tstar$(p)$~line for YBCO.
In sec.~\ref{sec:LSCO}, we establish the \Tstar$(p)$~line for LSCO, Nd-LSCO and Eu-LSCO.
We show in detail how \Tstar~is independent of crystal structure.
In the discussion (sec.~\ref{sec:Discussion}), we compare YBCO and LSCO, and draw 
general observations about the pseudogap phase.
We also plot the onset temperatures of various orders 
on the phase diagrams of YBCO and LSCO and discuss the implications.
In the Appendix (sec.~\ref{sec:Appendix}), we show how superconducting fluctuations
in YBCO, LSCO, Hg-1201, Bi-2212 and Bi-2201 
are limited to a region close to \Tc, well below \Tstar$(p)$,
and explain why previous interpretations suggested a much wider
regime of fluctuations.

%%%%%%%%%%%%%%%%%%%%%%%%%%%%%%%%%%%%%%%%%%%%%%%%%%%%%%%%%%%%%%%%%%%%%%%
%%%%%% The Nernst effect
%%%%%%%%%%%%%%%%%%%%%%%%%%%%%%%%%%%%%%%%%%%%%%%%%%%%%%%%%%%%%%%%%%%%%%%

\section{The Nernst effect}
\label{sec:Nernst}

The Nernst effect is the development of a transverse electric field $E_y$ 
across the width ($y$ axis) of a metallic sample when a temperature gradient $\partial T$\,/\,$\partial x$ 
is applied along its length ($x$ axis) in the presence of a perpendicular magnetic field $H$ (along the $z$ axis). 
Two mechanisms can give rise to a Nernst signal $N$\,$\equiv$\,$E_y$\,/\,$(-$\,$\partial T$\,/\,$\partial x)$~[\onlinecite{Behnia2009,Behnia2015,Behnia2016}]: 
superconducting fluctuations~[\onlinecite{Huebner1979,Wang2006,Chang2012}], 
which give a positive signal, 
and charge carriers (quasiparticles), which can give a signal of either sign. 
The focus of this Article is on the quasiparticle contribution to the Nernst effect in cuprates.

In the \hyperref[sec:Appendix]{Appendix}, we discuss the contribution of superconducting fluctuations 
to the Nernst signal in cuprates and explain how the traditional assumption that it
is the only significant contribution is mistaken. 
We discriminate between the superconducting signal and the quasiparticle signal by using
the fact that only the former is suppressed by a magnetic field. 
We show that the regime of significant
superconducting fluctuations is a relatively narrow band that tracks \Tc,
completely distinct from \Tstar.
This confirms that the pseudogap phase is not caused by
fluctuations in the phase and\,/\,or the amplitude of the superconducting order parameter.

The Nernst signal is related to the conductivity $\tensor\sigma$ and thermoelectric $\tensor\alpha$ tensors via
\begin{align}
	N &= {\frac{\alpha_{xy}\sigma_{xx} - \alpha_{xx}\sigma_{xy}}{\sigma_{xx}^2 + \sigma_{xy}^2}} 
	&\simeq {\frac{\alpha_{xy}}{\sigma_{xx}}} - S {\frac{\sigma_{xy}}{\sigma_{xx}}},
	\label{eq:NKamran2}
\end{align}
where $S$\,$\equiv$\,$\alpha_{xx}$\,/\,$\sigma_{xx}$ is the Seebeck coefficient. 
In-plane isotropy is assumed ($\sigma_{xx}$\,=\,$\sigma_{yy}$) 
and the approximate expression on the right holds for $\sigma_{xx}^2$\,$\gg$\,$\sigma_{xy}^2$. 

The sign of $N$ will thus depend on the relative magnitude of $\alpha_{xy}\sigma_{xx}$ and $\alpha_{xx}\sigma_{xy}$.
In a single-band metal with an energy-independent Hall angle $\theta_{\rm H} $, 
where $\tan\theta_{\rm H}$\,$\equiv$\,$\sigma_{xy}$\,/\,$\sigma_{xx}$, 
the two terms are equal and thus $N$\,=\,$0$~[\onlinecite{Behnia2009,Behnia2015,Behnia2016}]. 
This is the so-called Sondheimer cancellation.
An energy dependence of $\theta_{\rm H}$ will offset this equality in a direction
that is difficult to predict, 
resulting in a finite $N$ whose sign can be either positive or negative~[\onlinecite{Behnia2009,Behnia2015,Behnia2016}].
In general, the sign of $N$ in metals is not understood. 
Even in single-band metals like overdoped cuprates, it is 
unclear why $N$\,$>$\,$0$ in the electron-doped material Pr$_{2-x}$Ce$_x$CuO$_4$ (PCCO)~[\onlinecite{Li2007}] 
and $N$\,$<$\,$0$ in the hole-doped material Nd-LSCO~[\onlinecite{Cyr-Choiniere2009}], since both have a 
positive Hall coefficient.

At low temperature, the magnitude of the quasiparticle Nernst signal is given approximately by~[\onlinecite{Behnia2009,Behnia2015,Behnia2016}]:
\begin{align}
	\frac{|\nu|}{T} \approx \frac{\pi^2}{3}\frac{k^{2}_{\rm B}}{e}\frac{\mu}{\epsilon_{\rm F}},
	\label{eq:nuKamran}
\end{align}
where $\nu$\,$\equiv$\,$N$\,/\,$H$ is the Nernst coefficient, $H$ is the magnetic field, $T$ is the temperature, 
$k_{\rm B}$ is Boltzmann's constant, $e$ is the electron charge, $\mu$ is the carrier mobility, 
and $\epsilon_{\rm F}$ is the Fermi energy.
Eq.~\ref{eq:nuKamran} works remarkably well 
as a universal expression for the Nernst coefficient of metals at $T$\,$\to$\,$0$, 
accurate within a factor two or so in a wide range of materials~[\onlinecite{Behnia2009}]. 
It explains why a phase transition that reconstructs a large Fermi surface into small pockets 
(with small $\epsilon_{\rm F}$) can cause a major enhancement of $\nu$. The heavy-fermion metal 
URu$_2$Si$_2$ provides a good example of this. As the temperature drops below its transition 
to a metallic state with reconstructed Fermi surface at $17$\,K, the carrier density $n$ (or $\epsilon_{\rm F}$) 
falls and the mobility rises, both by roughly a factor 10, and $\nu$\,/\,$T$ increases by a factor 100 or so~[\onlinecite{Bel2004}]. 
Note that the electrical resistivity $\rho(T)$ is affected only weakly by these dramatic changes~[\onlinecite{Hassinger2008}], 
since mobility and carrier density are modified in ways that compensate in the conductivity 
$\sigma$\,=\,$1$\,/\,$\rho$\,=\,$n e \mu$. 
This is why the Nernst effect can be a more sensitive probe of electronic transformations, such as density-wave transitions, than the resistivity. 
Here we use it to study the pseudogap phase of cuprate superconductors.
%

%%%%%%%%%%%%%%%%%%%%%%%%%%%%%%%%%%%%%%%%%%%%%%%%%%%%%%%%%%%%%%%%%%%%%%%
%%%%%%        METHODS
%%%%%%%%%%%%%%%%%%%%%%%%%%%%%%%%%%%%%%%%%%%%%%%%%%%%%%%%%%%%%%%%%%%%%%%

\section{Methods}
\label{sec:Methods}

The YBCO samples measured here ($p$\,=\, $0.078$ and $p$\,=\,$0.085$) 
were single crystals  prepared at the University of British Columbia by flux growth~[\onlinecite{Liang2012}].
The detwinned samples are uncut, unpolished thin platelets, with gold evaporated contacts (of resistance $< 1\,\Omega$), in a six-contact geometry. 
Typical sample dimensions are 20-50 $\times$ 500-800 $\times$ 500-1000 $\mu$\,m$^3$ (thickness $\times$ width $\times$ length). 
Their hole concentration (doping) $p$ was determined from a relationship between the $c$-axis lattice constant and 
the superconducting transition temperature $T_{\rm c}$~[\onlinecite{Liang2006}], defined as the temperature below which the zero-field resistance is zero.

The Nd-LSCO samples ($x$\,=\,$0.20$ and $0.21$) and the Eu-LSCO samples ($x$\,=\,$0.08$, $0.10$ and $0.21$) measured here 
were grown using a travelling float-zone technique in an image furnace at the University of Texas and the University of Tokyo, respectively. 
%The \lnsco{} (Nd-LSCO) samples ($x$\,=\,$0.20$) were grown at the University of Texas. 
%These samples were grown using a travelling float zone technique. 
$ab$-plane single crystals were cut from boules into small rectangular platelets with typical dimensions of $1$\,mm in length and $0.5$\,mm in width 
(in the basal plane of the tetragonal structure), with a thickness of $0.2$\,mm along the $c$ axis. 
Orientation was checked via Laue diffraction. 
The doping $p$ is taken to equal the Sr content $x$, to within $\pm$\,0.005. 
%The length $L$ is measured between the contacts used to measure the temperature gradient. 
%The width $w$ is the distance between the transverse voltage contacts and %t% is the thickness. 
The $T_{\rm c}$ of our samples was determined via resistivity measurements 
as the temperature where $\rho(T)$ goes to zero. 
Electrical contacts on the Nd/Eu-LSCO samples were made to the crystal surface using Epo-Tek H20E silver epoxy,
cured at $180^\circ$C for 5 min and then annealed at $500^\circ$C in flowing oxygen for 1 hr. 
This resulted in contact resistances of less than $1\,\Omega$ at room temperature. 
The longitudinal contacts were wrapped around all four sides of the sample. The current contacts covered the end faces. 
Nernst (transverse) contacts were placed opposite each other in the middle of the sample, extending along the length of the $c$ axis, on the sides. 
The uncertainty in the length $L$ of the sample (between longitudinal contacts) reflects the width of the voltage\,/\,temperature contacts along the $x$ axis.

%%%%%%%%%%%%%%%%%%%%%%%%%     FIGURE 1    %%%%%%%%%%%%%%%%%%%%%%%%%%%%%%%%%%%%%%

\begin{figure}
\centering
\includegraphics[width=0.45\textwidth]{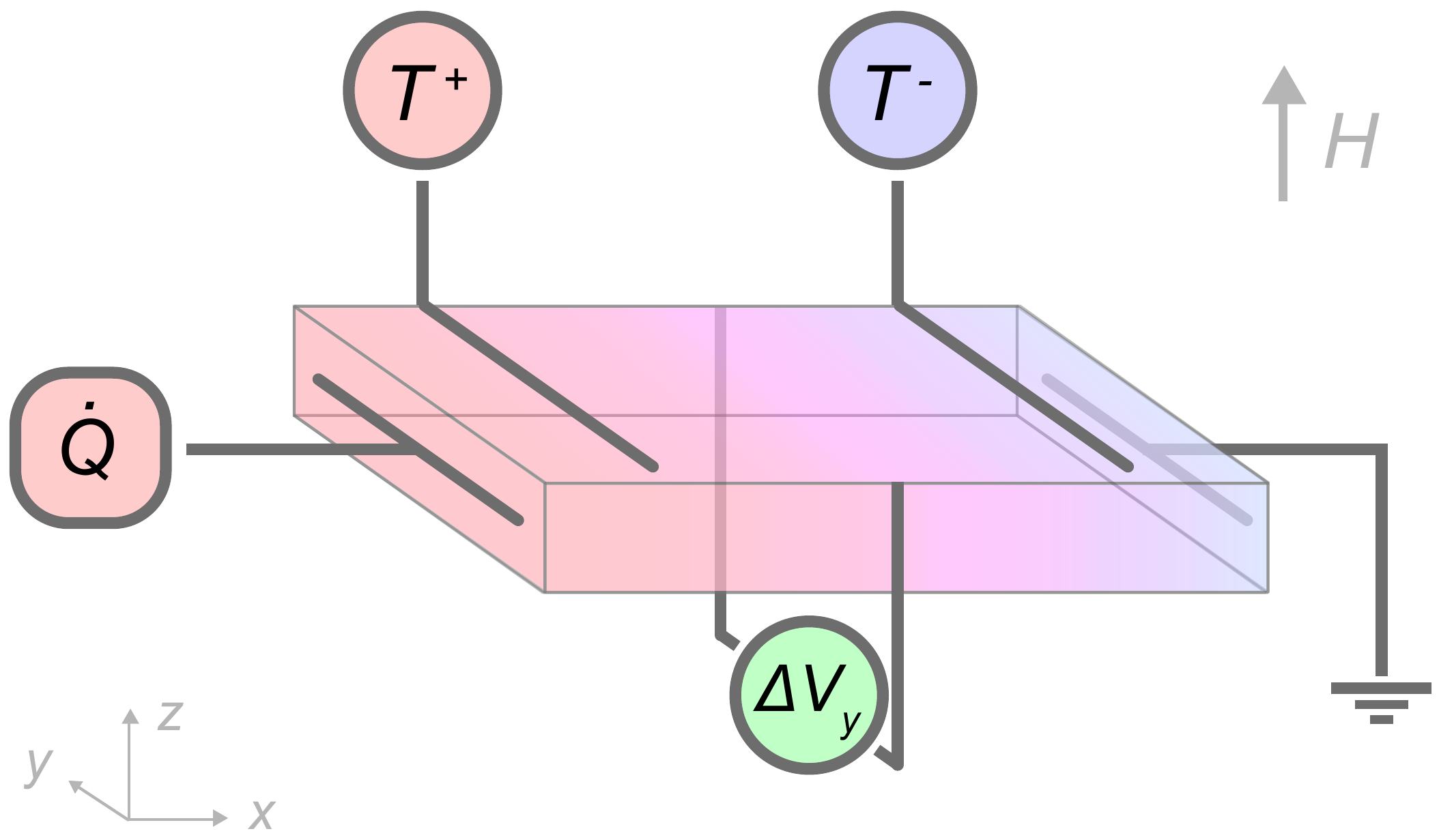}
\caption{(Color online)
Sketch of how the Nernst effect is measured on a sample in the shape of a thin platelet.
A longitudinal temperature gradient along $x$ is generated by applying heat to one end of the sample,
while the other end is kept cold.
A given heat current ($\dot{Q}$) produces a temperature difference ($\Delta T_x = T^+ - T^-$) 
that can be measured either with resistance thermometers or thermocouples.
When a magnetic field ($H$) is applied along $z$, a transverse (Nernst) voltage ($\Delta V_y$) is generated.
The Nernst signal $N$ is the ratio of $\Delta V_y$ over $\Delta T_x$ (Eq.~\ref{eq:Nmeas}).
}
\label{Sketch-Nernst-measurement}
\end{figure}

%%%%%%%%%%%%%%%%%%%%%%%%%%%%%%%%%%%%%%%%%%%%%%%%%%%%%%%%%%%%%%%%%%%%%

Fig.~\ref{Sketch-Nernst-measurement} summarizes how the Nernst signal is measured. 
The Nernst signal was measured by applying a steady heat current through the sample (along the $x$ axis). 
The longitudinal thermal gradient was measured using two uncalibrated Cernox chip thermometers (Lakeshore), referenced to a further calibrated Cernox. 
Alternatively on some samples, the longitudinal thermal gradient was measured using one differential and one absolute type-E thermocouple 
made of chromel and constantan wires known to have a weak magnetic field dependence. 
The temperature of the experiment was stabilized at each point to within $\pm$\,$10$\,mK. 
The temperature and voltage were measured with and without applied thermal gradient ($\Delta$\,$T$) for calibration. 
The magnetic field $H$, applied along the $c$ axis ($H$\,$\parallel$\,$c$), was then swept with the heat on, 
from $-H_{\rm max}$ to $+H_{\rm max}$ (where $H_{\rm max}$\,=\,$10$, $15$ or $16$\,T depending on sample), at $0.4$\,T\,/\,min, continuously taking data. 
The thermal gradient was monitored continuously and remained constant during the course of a sweep. 
The Nernst signal $N$ was extracted from that part of the measured voltage which is anti-symmetric with respect to the magnetic field: 
%$N = E_y / \left( \partial T/\partial x\right) = \left[\Delta V_y(+H) / \Delta T_x - \Delta V_y(-H) / \Delta T_x  \right](L / 2w)$,
%
\begin{align}
	N &= \frac{E_y}{\partial T/\partial x} = \left(\frac{\Delta V_y(+H)}{\Delta T_x} - \frac{\Delta V_y(-H)}{\Delta T_x}\right)\frac{L}{2w},
	\label{eq:Nmeas}
\end{align}
where $\Delta V$ is the difference in the voltage measured with and without thermal gradient. 
$L$ is the length (between contacts along the $x$ axis) and $w$ the width (along the $y$ axis) of the sample. 
This anti-symmetrization procedure removes any longitudinal thermoelectric contribution from the sample and a constant background from the measurement circuit. 
The uncertainty on $N$ comes mostly from the uncertainty in measuring $L$ and $w$, giving a typical error bar of $\pm$\,$10\%$ on $N$.
%\\

%%%%%%%%%%%%%%%%%%%%%%%%%%%%%%%%%%%%%%%%%%%%%%%%%%%%%%%%%%%%%%%%%%%%%%%
%%%%%%        YBCO
%%%%%%%%%%%%%%%%%%%%%%%%%%%%%%%%%%%%%%%%%%%%%%%%%%%%%%%%%%%%%%%%%%%%%%%

\section{YBCO}
\label{sec:YBCO}

%%%%%%%%%%%%%%%%%%%%%%%%%     FIGURE 2    %%%%%%%%%%%%%%%%%%%%%%%%%%%%%%%%%%%%%%
%
\begin{figure}
\centering
\includegraphics[width=0.49\textwidth]{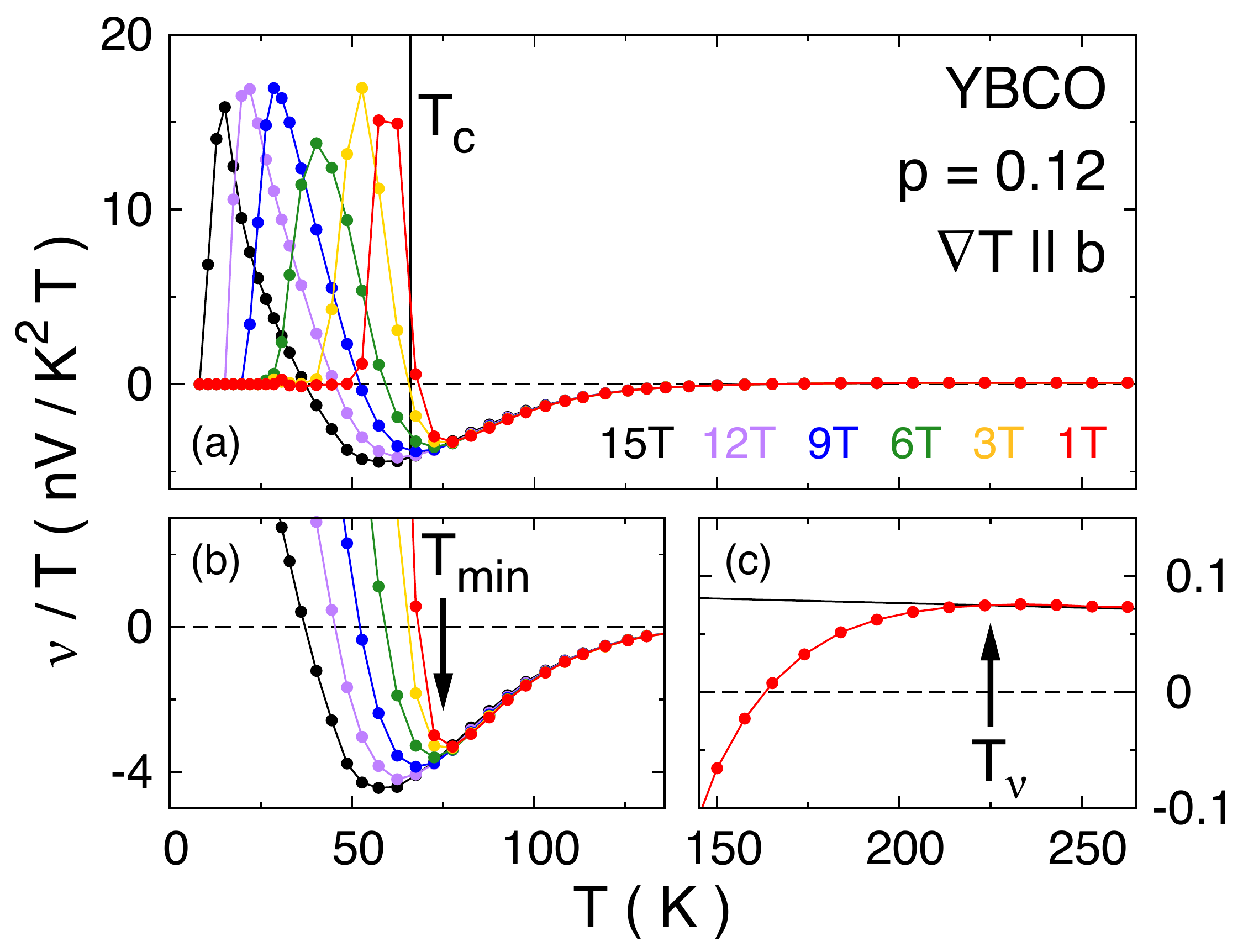}
\caption{(Color online) 
Nernst coefficient $\nu$ of YBCO at a hole doping of $p$\,=\,$0.12$,
plotted as $\nu$\,/\,$T$
versus temperature $T$ for different magnetic fields ($H$\,=\,$1$\,T to $15$\,T), as indicated. 
The thermal gradient is applied in the $b$ direction of the orthorhombic crystal structure. 
Data are reproduced from Ref.~[\onlinecite{Daou2010}]. 
(a)
The vertical line marks the superconducting transition temperature at $H$\,=\,$0$, \Tc\,=\,$66.0$\,K.
(b)
Zoom near \Tc, to show how $T_{\rm min}$ is defined: 
it is the temperature at which the Nernst signal at $H$\,=\,$1$\,T goes through a minimum,
at the foot of the large positive peak due to superconductivity.
(c) 
Zoom at high temperature, where only quasiparticles contribute to the Nernst signal.
$T_\nu$ (arrow) is defined as the temperature below which $\nu(T)$\,/\,$T$ starts to deviate downwards
from its high-temperature linear behaviour.
}
\label{nu-YBCO-one8}
\end{figure}
%
%%%%%%%%%%%%%%%%%%%%%%%%%%%%%%%%%%%%%%%%%%%%%%%%%%%%%%%%%%%%%%%%%%%%%%%

%%%%%%%%%%%%%%%%%%%%%%%%%     FIGURE 3    %%%%%%%%%%%%%%%%%%%%%%%%%%%%%%%%%%%%%%
%
\begin{figure}
\centering
\includegraphics[width=0.47\textwidth]{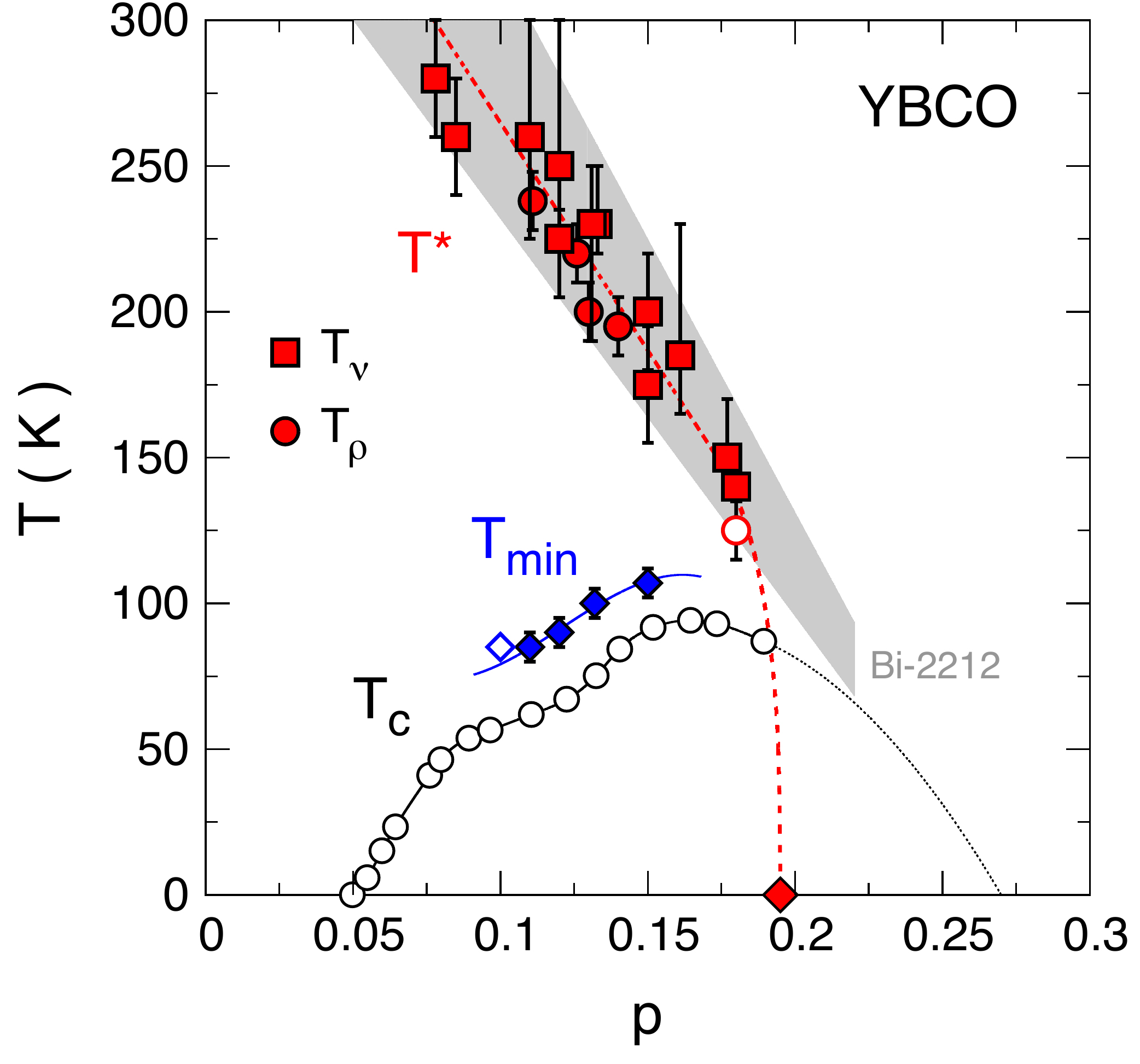}
\caption{(Color online) 
Temperature-doping phase diagram of YBCO, showing three characteristic temperatures. 
The transition temperature \Tc~(open black circles [\onlinecite{Liang2006}]) marks the onset of superconductivity
in zero magnetic field, below which the electrical resistivity is zero.
The solid black line is a guide to the eye through the \Tc{} data points.
The dotted black line is a smooth extension of this line assuming that the superconducting phase 
ends at a critical doping \pc\,=\,$0.27$.
Blue diamonds mark $T_{\rm min}$ (defined  in Fig.~\ref{nu-YBCO-one8}(b)), 
the temperature below which superconducting fluctuations become significant
(from $a$-axis data in Ref.~[\onlinecite{Daou2010}]). 
The open diamond shows $T_{\rm min}$ for a previously measured sample with $p $\,=\,$ 0.1$~[\onlinecite{Rullier-Albenque2006}]. 
The solid blue line is a guide to the eye.
Red circles mark $T_\rho$, the temperature below which the resistivity $\rho(T)$ deviates 
from its high-temperature linear dependence (from data in Ref.~[\onlinecite{Ando2004}]), 
a standard definition of the pseudogap temperature $T^\star$ in YBCO~[\onlinecite{Ito1993}] (see Fig.~\ref{rho-YBCO-LNSCO}(a)).
The open red circle shows $T_\rho$ for a sample with $p $\,=\,$ 0.18$ in which
a high level of disorder scattering was introduced by electron irradiation~[\onlinecite{Rullier-Albenque2008}]. 
In this case, $T_\rho$ marks the onset of an upturn in $\rho(T)$ (see text).
Red squares mark $T_\nu$ (defined in Fig.~\ref{nu-YBCO-one8}(c)), the temperature below which the quasiparticle Nernst signal
departs from its high-temperature behaviour (from present work and Ref.~[\onlinecite{Daou2010}]). 
One can see that within error bars, $T_\nu$\,$\simeq$\,$T_\rho$, both measures of \Tstar.
The red dashed line is a linear fit through the \Tstar{} data points.
Beyond $p$\,=\,$0.18$, it is a guide to the eye extending smoothly to reach $p$\,=\,\pstar{} at $T$\,=\,$0$ (red diamond).
\pstar{} is the critical doping where the pseudogap phase ends at $T $\,=\,$ 0$ in the absence of superconductivity.
In YBCO, \pstar\,=\,$ 0.195$\,$\pm$\,$0.005$~[\onlinecite{Badoux2016}]. 
The grey band marks the range of \Tstar~values measured in Bi-2212 
from spectroscopic probes (ARPES, STS and SIS)~[\onlinecite{Vishik2012}], detected up to $p$\,$\simeq$\,$0.22$.
%
%
%Error bars on \Tnu~and \Trho~are defined in [\onlinecite{Daou2010}]; for \Tnu~error bars at $p = 0.078$ and $p = 0.085$, see Fig.~\ref{nu-YBCO-6p45}.
}
\label{YBCO-phasediag}
\end{figure}
%

%%%%%%%%%%%%%%%%%%%%%%%%%%%%%%%%%%%%%%%%%%%%%%%%%%%%%%%%%%%%%%%%%%%%%%%

%%%%%%%%%%%%%%%%%%%%%%%%%     FIGURE 4   %%%%%%%%%%%%%%%%%%%%%%%%%%%%%%%%%%%%%%
%
\begin{figure}
\centering
\includegraphics[width=0.40\textwidth]{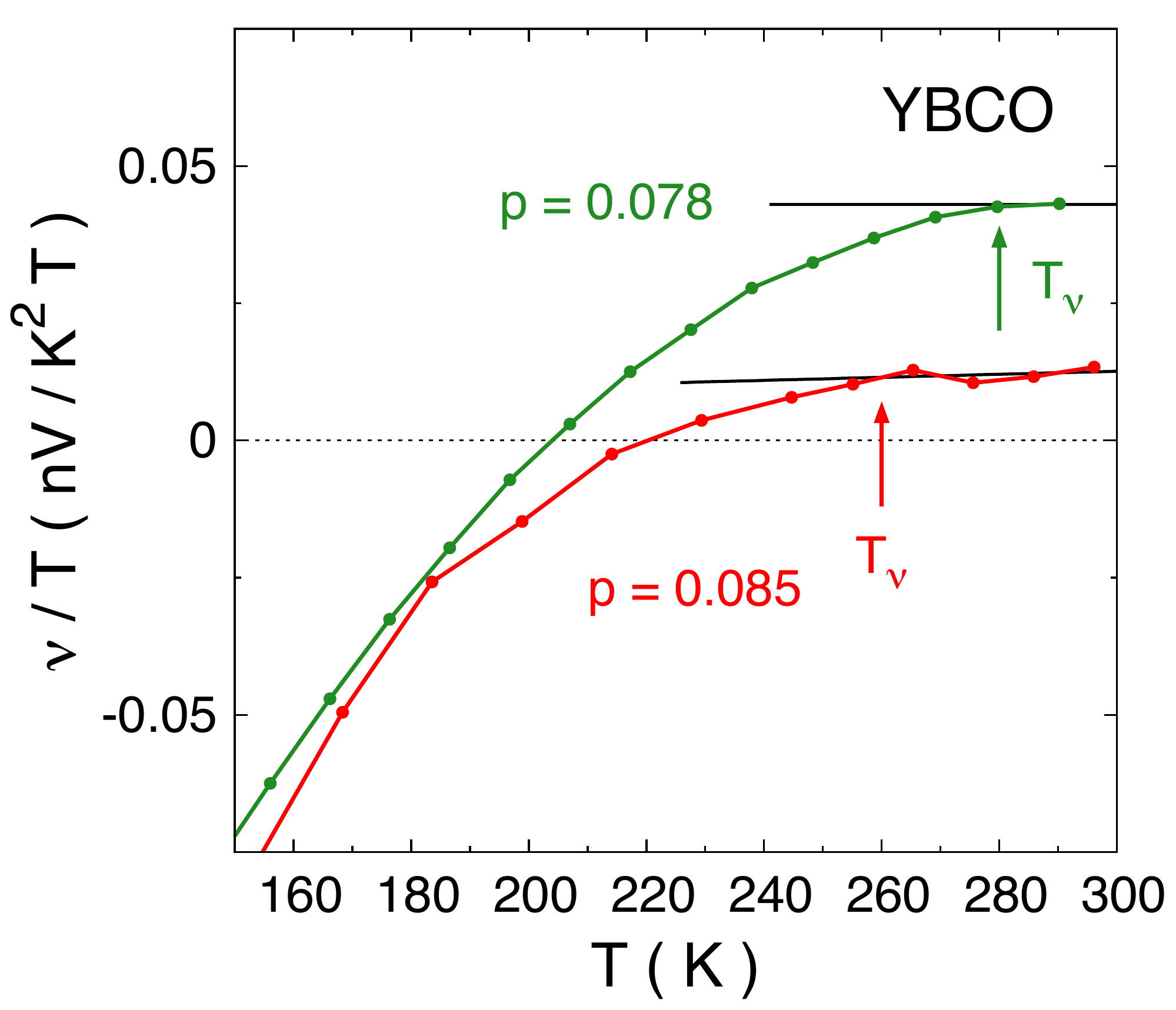}
\caption{(Color online) 
High-temperature Nernst coefficient $\nu$ of YBCO at dopings $p$\,=\,$ 0.078$ (green) and $p$\,=\,$ 0.085$ (red),
plotted as $\nu$\,/\,$T$ versus $T$. 
The thermal gradient was applied in the $a$ direction. 
The color-coded arrows mark \Tnu, the temperature below which $\nu(T)$\,/\,$T$ starts to deviate downwards
from its small, roughly constant value at high temperature: 
$T_\nu$\,=\,$280$\,$\pm$\,$20$\,K and $260$\,$\pm$\,$20$\,K for $p$\,=\,$0.078$ and $0.085$, respectively.
Error bars on \Tnu{} represent the uncertainty in identifying the start of the downturn. 
}
\label{nu-YBCO-6p45}
\end{figure}
%
%%%%%%%%%%%%%%%%%%%%%%%%%%%%%%%%%%%%%%%%%%%%%%%%%%%%%%%%%%%%%%%%%%%%%%%

Nernst data taken on untwinned single crystals of YBCO have been reported for 
a range of dopings, from $p $\,=\,$ 0.11$ to $p $\,=\,$ 0.18$~[\onlinecite{Daou2010}]. 
A typical set of Nernst data is reproduced in Fig.~\ref{nu-YBCO-one8} 
as $\nu$\,/\,$T$ vs $T$, for a sample with $p $\,=\,$ 0.12$. 
Two separate contributions are clearly seen: 
1) a positive and magnetic-field-dependent signal which rises below a temperature $T_{\rm min}$ close to $T_{\rm c}$; 
2) a field-independent signal which goes from small and positive at high temperature to large and negative 
at lower temperature, as it drops below a temperature $T_\nu$. The first is due to superconducting fluctuations, 
the second is due to quasiparticles. In Fig.~\ref{YBCO-phasediag}, the two onset temperatures 
$T_{\rm min}$ and $T_{\nu}$ are plotted on a phase diagram. 
The 10 data points for \Tnu{} (red squares) at $p$\,$>$\,$0.1$ are reproduced from Ref.~[\onlinecite{Daou2010}]; 
they include data taken with $\Delta T$\,$||$\,$a$ and $\Delta T$\,$||$\,$b$ -- both yield the same \Tnu~[\onlinecite{Daou2010}].
In Fig.~\ref{nu-YBCO-6p45}, we report new data for dopings 
$p $\,=\,$ 0.078$ and $p $\,=\,$ 0.085$ which allow us to extend \Tnu{} to low doping. 

\begin{figure}
\centering
\includegraphics[width=0.47\textwidth]{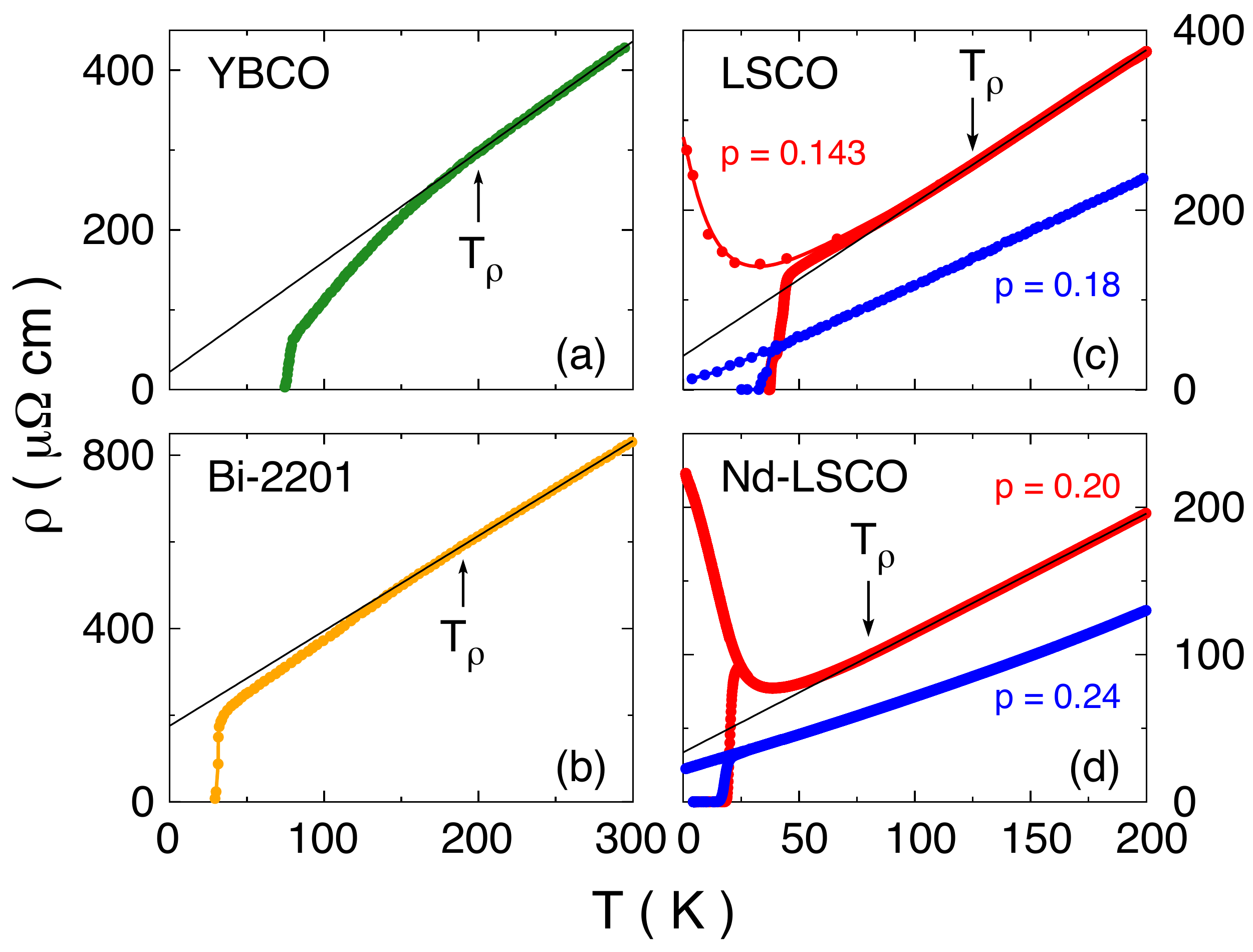}
\caption{(Color online) 
Resistivity $\rho(T)$ as a function of temperature for four cuprate materials:
(a) 
YBCO at $p$\,=\,$0.13$~[\onlinecite{Ando2004}];
(b) 
Bi-2201, underdoped with \Tc\,=\,$27$\,K~[\onlinecite{Kondo2011}];
(c) 
LSCO at $p$\,=\,$0.143$~(red; [\onlinecite{Laliberte2016}]); and $p$\,=\,$0.18$ (blue; [\onlinecite{Cooper2009}])
(d) 
Nd-LSCO at $p$\,=\,$0.20$~(red) and $p$\,=\,$0.24$~(blue)~[\onlinecite{Daou2009}].
The black line is a linear fit of the high-temperature region and 
a zoom enables the extraction of
\Trho~(arrow), the temperature below which $\rho(T)$ deviates from this linear dependence --
a standard criterion for the pseudogap temperature \Tstar.
For LSCO (c) and Nd-LSCO (d), the comparison between two dopings on either side of the pseudogap critical point \pstar{} 
reveals the effect on $\rho(T)$ of the drop in carrier density (from $n$\,=\,$1$\,$+$\,$p$ to $n$\,=\,$p$) 
caused by the pseudogap present at $p$\,$<$\,\pstar~[\onlinecite{Collignon2017},\onlinecite{Laliberte2016}].
}
\label{rho-YBCO-LNSCO}
\end{figure}

In YBCO, a standard criterion for the pseudogap temperature $T^\star$ 
is the temperature $T_\rho$ below which the $a$-axis resistivity $\rho(T)$ 
deviates from its linear temperature dependence at high temperature~[\onlinecite{Ito1993}]. 
An example is shown in Fig.~\ref{rho-YBCO-LNSCO}(a), where we extract $T_\rho $\,=\,$ 200$\,$\pm$\,$10$\,K from published data 
at $p $\,=\,$ 0.13$~[\onlinecite{Ando2004}]. Values for $T_\rho$ at different dopings are plotted on the phase diagram 
of Fig.~\ref{YBCO-phasediag}, where we see that $T_\nu $\,=\,$ T_\rho$, within error bars.

%%%%%%%%%%%%%%%%%%%%%%%%%%     FIGURE 5   %%%%%%%%%%%%%%%%%%%%%%%%%%%%%%%%%%%%%%
%
%
%%%%%%%%%%%%%%%%%%%%%%%%%%%%%%%%%%%%%%%%%%%%%%%%%%%%%%%%%%%%%%%%%%%%%%%

%%%%%%%%%%%%%%%%%%%%%%%%%%     FIGURE 6   %%%%%%%%%%%%%%%%%%%%%%%%%%%%%%%%%%%%%%
%
\begin{figure}
\centering
\includegraphics[width=0.47\textwidth]{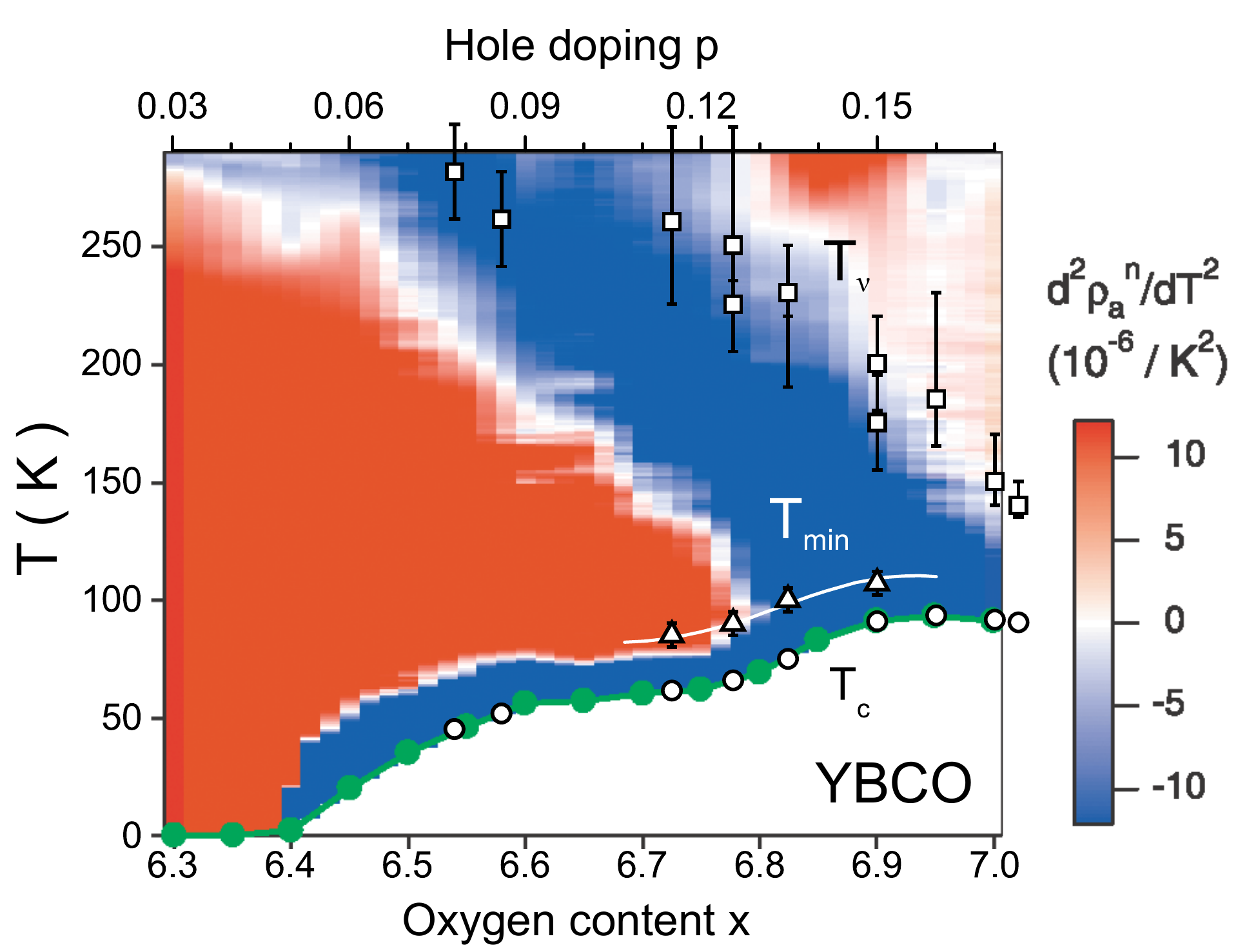}
\caption{(Color online) 
Resistivity curvature map from Ando {\it et al.}~[\onlinecite{Ando2004}]
showing the second temperature derivative of their resistivity data on YBCO,
plotted as a function of temperature (vertical axis) and oxygen doping $x$ (bottom horizontal axis).
The green dots mark \Tc.
The top axis shows the approximate hole doping $p$, estimated from the  \Tc~values [\onlinecite{Liang2006}]. 	
White regions correspond to linear, 
blue ones to sub-linear (downward; $d^2\rho/dT^2$\,$<$\,$0$) 
and red ones to super-linear (upward; $d^2\rho/dT^2$\,$>$\,$0$) 
behavior of resistivity with temperature. 
The boundary of the pseudogap region ($T^\star$) is the lower limit of the white region in the upper right corner.
$T_\nu$, $T_{\rm min}$ and $T_{\rm c}$ from present work and data of Ref.~[\onlinecite{Daou2010}] 
are added respectively as open squares, triangles and circles. 
$T_\nu$ points agree reasonably well with the resistivity criterion (as in Fig.~\ref{YBCO-phasediag}). 
The narrow blue region that tracks $T_{\rm c}$ represents the paraconductivity regime where resistivity drops 
due to superconducting fluctuations just above $T_{\rm c}$. 
$T_{\rm min}$ (triangles), our criterion for the onset of significant superconducting fluctuations in the Nernst effect,
is seen to agree with the onset of paraconductivity, clearly observable at $x < 6.8$ (or $p < 0.13$).
}
\label{Ando-map-YBCO}
\end{figure}
%
%%%%%%%%%%%%%%%%%%%%%%%%%%%%%%%%%%%%%%%%%%%%%%%%%%%%%%%%%%%%%%%%%%%%%%%

As a probe of the pseudogap phase in YBCO, the Nernst effect has an advantage over the resistivity. 
Pseudogap and superconductivity have opposite effects on $\nu(T)$: the former causes it to fall 
to negative values upon cooling, the latter causes it to rise, while for resistivity, 
both phenomena yield a downturn in $\rho(T)$ 
(see Fig.~\ref{rho-YBCO-LNSCO}(a), Fig.~\ref{Ando-map-YBCO}, and paragraph below). 
This makes the separation of the two contributions in the Nernst effect unambiguous, 
and allows us to track their respective onset temperatures.

In Fig.~\ref{Ando-map-YBCO}, we plot $T_\nu$ and $T_{\rm min}$ on the ``curvature map'' 
produced by Ando and Segawa~[\onlinecite{Ando2004}] from the second temperature derivative of their $\rho(T)$ data. 
As already seen in Fig.~\ref{YBCO-phasediag}, the lower bound of the linear-$T$ region 
(white region in the upper right corner of Fig.~\ref{Ando-map-YBCO}) coincides with $T_\nu$ 
and defines the boundary of the pseudogap phase. Below $T^\star$, the initial drop in $\rho(T)$ 
shows up as a blue band, followed by an upturn (in red) (for $p$\,$<$\,$0.13$, in Fig.~\ref{Ando-map-YBCO}). 
Superconducting fluctuations above $T_{\rm c}$ also cause a downturn in $\rho(T)$ (called ``paraconductivity''), 
producing another blue band, which simply tracks $T_{\rm c}$. 
For $p$\,$<$\,$0.13$, the onset of paraconductivity coincides reasonably well with $T_{\rm min}$. 
Therefore $T_{\rm min}$ is the temperature below which superconducting fluctuations (above $T_{\rm c}$) 
start to show up significantly in the Nernst signal. 
For $p$\,$>$\,$0.13$, the two blue bands merge and become indistinguishable - 
the pseudogap downturn flows smoothly into the paraconductivity downturn (see Fig.~\ref{Ando-map-YBCO}). 
This makes it difficult to reliably track $T^\star$ above $p$\,=\,$0.13$, 
and  to say from the resistivity whether there is still a pseudogap phase 
(with $T^\star$\,$>$\,$T_{\rm c}$) beyond optimal doping. %($x $\,=\,$ 6.95$ in Fig.~\ref{Ando-map-YBCO}). 
From the Nernst data, however, the answer is clearly yes, 
with $T^\star$\,$\simeq$\,$140$\,K 
and $T_{\rm c}$\,=\,$90$\,K at $p$\,=\,$0.18$.

\begin{figure}
\centering
\includegraphics[width=0.49\textwidth]{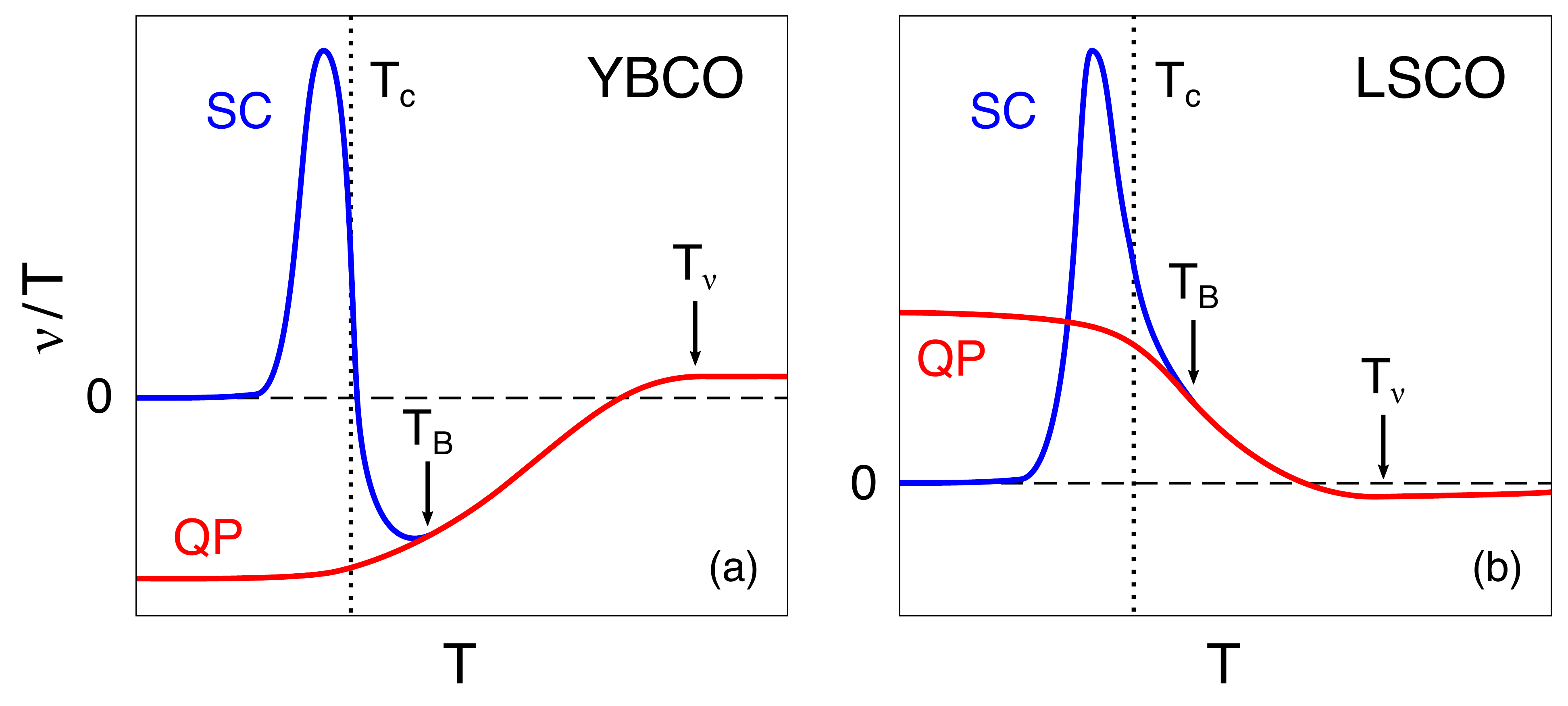}
\caption{(Color online) 
Cartoon illustrating the behaviour of the Nernst coefficient $\nu$ 
in cuprate superconductors, plotted as $\nu$\,/\,$T$ vs $T$.
The quasiparticle signal (QP, red) goes from small at high $T$ to large at low $T$,
with a change of sign.
It is independent of magnetic field.
The change occurs upon entering the pseudogap phase, by crossing below a temperature \Tnu\,=\,\Tstar{} (arrow).
%$\nu/T$ starts to deviate from its high-temperature behaviour.
%
In YBCO (and Hg-1201), $\nu$ is positive at high $T$ (left panel), 
while in LSCO (and Nd/Eu-LSCO), $\nu$ is negative at high $T$.
The superconducting signal (SC, blue) develops below a temperature \TB{} (arrow) 
slightly above the zero-field \Tc{} (vertical dotted line).
It is always positive and it is suppressed by a magnetic field.
}
\label{sketch}
\end{figure}

While in YBCO the signature of \Tstar{} is a downturn in both $\rho(T)$ and $\nu$\,/\,$T$,
we shall see below that the corresponding signature in LSCO is an upturn in those two quantities (see Fig.~\ref{sketch}).
We attribute this difference to a difference in the relative importance of two effects of the pseudogap:
the loss of carrier density and the loss of inelastic scattering.
At $T$\,=\,$0$, there is no inelastic scattering and so only the first effect is relevant.
It has recently become clear that in the normal state at $T$\,=\,$0$ the opening of the pseudogap 
at $p$\,=\,\pstar~causes a rapid drop in the carrier density $n$ from 
$n $\,=\,$1$\,$+$\,$p$ (at $p$\,$>$\,\pstar) 
to $n$\,=\,$p$ (at $p$\,$<$\,\pstar)~[\onlinecite{Collignon2017},\onlinecite{Badoux2016},\onlinecite{Laliberte2016}]. 
%
%Because the associated change in mobility is negligible
The consequence is that $\rho$ at $T$\,$\to$\,$0$ is larger than it would be without the pseudogap 
by a factor $\sim$\,$(1$\,+\,$p$)\,/\,$p$~[\onlinecite{Collignon2017},\onlinecite{Laliberte2016}].
This drop in carrier density is what causes the upturn in $\rho(T)$ seen at $T$\,$\to$\,$0$ 
in LSCO (Fig.~\ref{rho-YBCO-LNSCO}(c))~[\onlinecite{Boebinger1996},\onlinecite{Laliberte2016}], 
Bi-2201~[\onlinecite{Ono2000}], and Nd-LSCO (Fig.~\ref{rho-YBCO-LNSCO}(d))~[\onlinecite{Daou2009},\onlinecite{Collignon2017}], 
when superconductivity is suppressed by a large magnetic field.
In Bi-2201, in addition to a pronounced upturn as $T$\,$\to$\,$0$~[\onlinecite{Ono2000}], 
$\rho(T)$ also exhibits a (slight) downturn below \Tstar{} (Fig.~\ref{rho-YBCO-LNSCO}(b))~[\onlinecite{Ando2004},\onlinecite{Kondo2011}]
showing that the two effects of the pseudogap -- loss of inelastic scattering and loss of carrier density -- do co-exist. 

In order to see an upturn in $\rho(T)$ starting right at \Tstar, 
the loss of inelastic scattering (causing a downturn) must be a small effect compared to the loss of carriers (causing an upturn). 
This is the case in sufficiently disordered samples.
A nice demonstration of this can be seen in YBCO at $p$\,=\,$0.18$. 
In clean samples, \Tnu\,=\,$140$\,$\pm$\,$10$\,K from the Nernst coefficient (Fig.~\ref{YBCO-phasediag}), 
but little is seen in $\rho(T)$ across \Tstar.
However, in a disordered sample at the same doping, a clear {\it upturn} is observed in $\rho(T)$,
beginning at \Trho\,=\,$ 130$\,$\pm$\,$10$\,K (open circle in Fig.~\ref{YBCO-phasediag})~[\onlinecite{Rullier-Albenque2008}].
This upturn is definitely due to the pseudogap since no upturn is observed in $\rho(T)$ when $p$\,$>$\,\pstar,
even for disorder levels large enough to entirely suppress superconductivity~[\onlinecite{Momono1994}].
Calculations without vertex corrections, perhaps appropriate when disorder scattering dominates, 
do get an upturn in the resistivity~[\onlinecite{Bergeron2011}].

In summary, the Nernst effect is a sensitive probe of the pseudogap phase because a key property 
of that phase is a loss of carrier density $n$~[\onlinecite{Badoux2016}], 
and $\nu$\,/\,$T$\,$\sim$\,$1$\,/\,$n$.
Because the pseudogap also causes a drop in inelastic scattering, the two effects 
reinforce each other in the Nernst signal, since $\nu$\,/\,$T$\,$\sim$\,$1$\,/\,$\Gamma$,
while they oppose each other in the resistivity, since $\rho$\,$\sim$\,$\Gamma$\,/\,$n$.
The Nernst effect is also an unambiguous probe of \Tstar{} in YBCO, because here
the quasiparticle and superconducting contributions to the Nernst signal have opposite sign (Fig.~\ref{sketch}).
(Note that an early proposal for the negative Nernst signal in YBCO as being due to the CuO chains in that material~[\onlinecite{Ong2004}] turns out to be incorrect,
as the very same negative signal is observed in the tetragonal material Hg1201~[\onlinecite{Doiron-Leyraud2013}],  
which is free of such chains.)

%%%%%%%%%%%%%%%%%%%%%%%%%     FIGURE 7    %%%%%%%%%%%%%%%%%%%%%%%%%%%%%%%%%%%%%%
%
%
%%%%%%%%%%%%%%%%%%%%%%%%%%%%%%%%%%%%%%%%%%%%%%%%%%%%%%%%%%%%%%%%%%%%%%%

The resulting phase diagram of YBCO is shown in Fig.~\ref{YBCO-phasediag}, where the boundary
of the pseudogap phase is clearly delineated (dashed red line).
It decreases linearly with doping up to $p$\,$\simeq$\,$0.18$ and then drops rapidly to reach its critical point
at \pstar\,=\,$0.195$ (red diamond).
The aprupt drop of \Tstar~at \pstar~could reflect a first-order transition,
as found in some calculations~[\onlinecite{Sordi2012}]. 
It is instructive to compare \Tnu\,=\,\Trho{} in YBCO with the pseudogap temperature \Tstar{} 
measured by spectroscopic means in Bi-2212.
In Fig.~\ref{YBCO-phasediag}, we plot as a grey band the value of \Tstar{}  vs $p$ measured in Bi-2212
by ARPES, SIS tunneling, STS and NMR~[\onlinecite{Vishik2012}].
We see that the \Tstar{} line is essentially the same in YBCO and Bi-2212, two bilayer cuprates with similar \Tc{} domes.
The only difference is in the value of \pstar{} in the normal state, namely
\pstar\,=\,$ 0.195$\,$\pm$\,$0.005$ in YBCO and  
\pstar\,=\,$ 0.22$\,$\pm$\,$0.1$ in Bi-2212.

%%%%%%%%%%%%%%%%%%%%%%%%%%     FIGURE 8   %%%%%%%%%%%%%%%%%%%%%%%%%%%%%%%%%%%%%%
%
\begin{figure}
\centering
\includegraphics[width=0.44\textwidth]{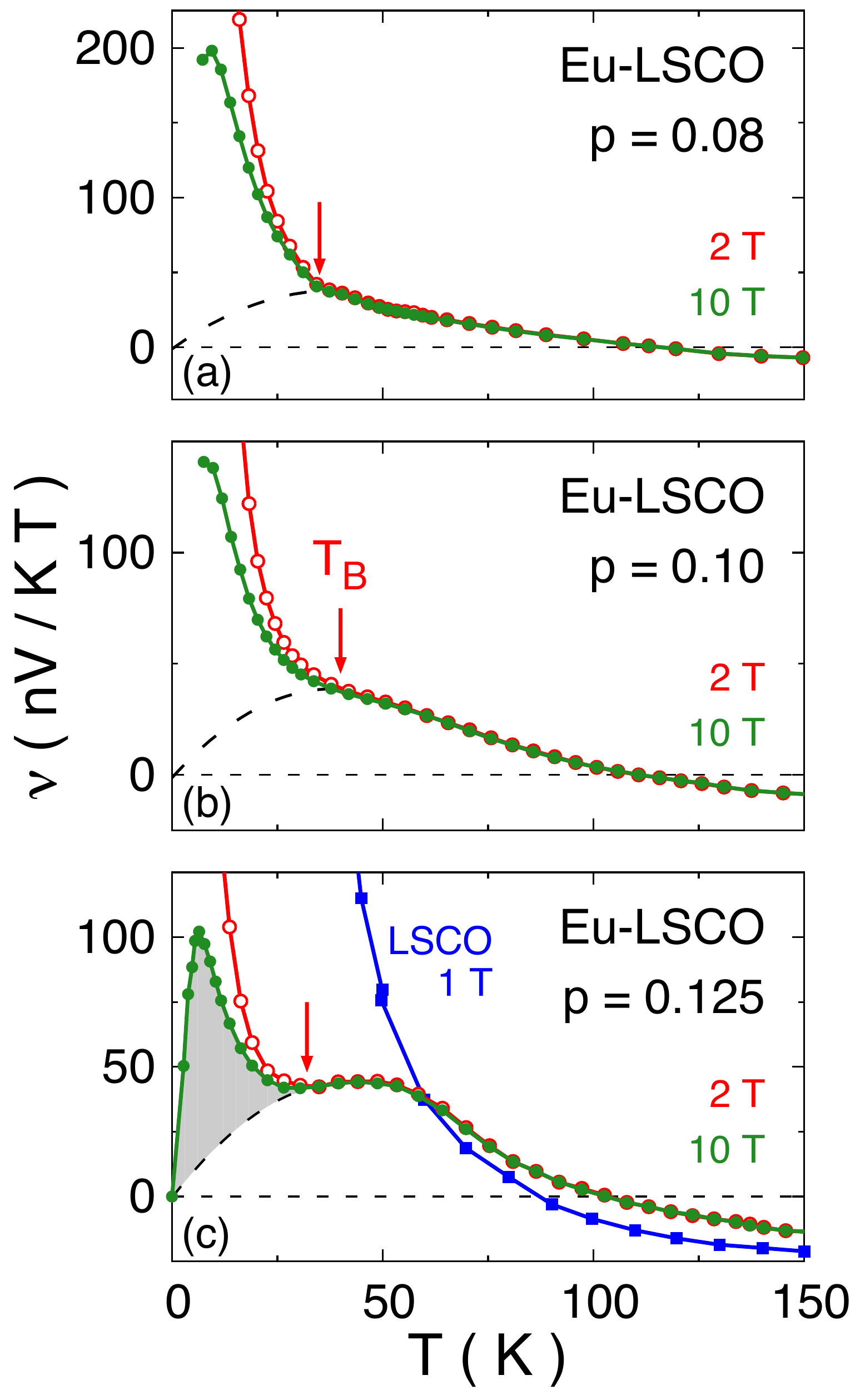}
\caption{
(Color online) Nernst coefficient $\nu$ of Eu-LSCO at dopings $p$\,=\,$0.08$ (a), 
$0.10$ (b) and $0.125$ (c) versus $T$, at $H$\,=\,$2$\,T (open red circles) and $10$\,T (filled green circles). 
The data at $p=0.125$ are taken from Ref.~[\onlinecite{Cyr-Choiniere2009}].
A two-peak structure is seen clearly at $p=0.125$.
At the other two dopings, it shows up as a breaking point in the slope of the data,
at $T$\,$\simeq$\,$35$\,K. 
This two-peak structure reveals the two distinct contributions to the Nernst effect: 
one from superconducting fluctuations, seen as a narrow positive peak at low temperature 
(grey shading in bottom panel), and the other from quasiparticles, seen as a broad positive peak at 
higher temperature. 
The dashed line is a guide to the eye for delimiting the quasiparticle peak.
In panel (c), we also plot LSCO data at $p=0.12$ and  $H$\,=\,$1$\,T (blue; from Ref.~[\onlinecite{Wang2006}]), for comparison.
In LSCO, we see that the two separate contributions flow smoothly one into the other. 
%to show that its higher \Tc~pushes the superconducting peak 
%to higher temperature, so that it lies on top of the quasiparticle peak.
%
The red arrow marks \TB, the temperature above which the field dependence of $\nu$ becomes negligible,
the signature of a negligible superconducting signal.
}
\label{nu-LESCO-low-x}
\end{figure}

%%%%%%%%%%%%%%%%%%%%%%%%%%%%%%%%%%%%%%%%%%%%%%%%%%%%%%%%%%%%%%%%%%%%%%%

%%%%%%%%%%%%%%%%%%%%%%%%%%%%%%%%%%%%%%%%%%%%%%%%%%%%%%%%%%%%%%%%%%%%%%%
%%%%%%%%%%%%%%%%%%%%%%%%%%%%%%%%%%%%%%%%%%%%%%%%%%%%%%%%%%%%%%%%%%%%%%%

%%%%%%%%%%%%%%%%%%%%%%%%%%%%%%%%%%%%%%%%%%%%%%%%%%%%%%%%%%%%%%%%%%%%%%%
%%%%%%        LSCO
%%%%%%%%%%%%%%%%%%%%%%%%%%%%%%%%%%%%%%%%%%%%%%%%%%%%%%%%%%%%%%%%%%%%%%%

%\section{LSCO, Nd-LSCO \& Eu-LSCO}
\section{LSCO, N\texorpdfstring{\lowercase{d}}{d}-LSCO \& E\texorpdfstring{\lowercase{u}}{u}-LSCO}
\label{sec:LSCO}

%%%%%%%%%%%%%%%%%%%%%%%%%%%%%%%%%%%%%%%%%%%%%%%%%%%%%%%%%%%%%%%%%%%%%%%

We now turn to a different family of cuprates, based on La$_2$CuO$_4$.
Three materials will be discussed:
LSCO,
Nd-LSCO and
Eu-LSCO.
In all three materials, the quasiparticle Nernst signal in the pseudogap phase at low temperature is positive, therefore of the same
sign as the superconducting signal.

As illustrated in Fig.~\ref{sketch}, this makes it more difficult than in YBCO to separate the two contributions, 
and this difficulty is what led to early misinterpretations of the positive Nernst signal detected 
in LSCO up to 150 K as being due to vortex-like excitations  in underdoped samples
with \Tc\,$\simeq$\,$0$~[\onlinecite{Xu2000}].
We discuss this issue in more detail in the \hyperref[sec:Appendix]{Appendix}.

%%%%%%%%%%%%%%%%%%%%%%%%     FIGURE 9   %%%%%%%%%%%%%%%%%%%%%%%%%%%%%%%%%%%%
%
\begin{figure}
\centering
\includegraphics[width=0.45\textwidth]{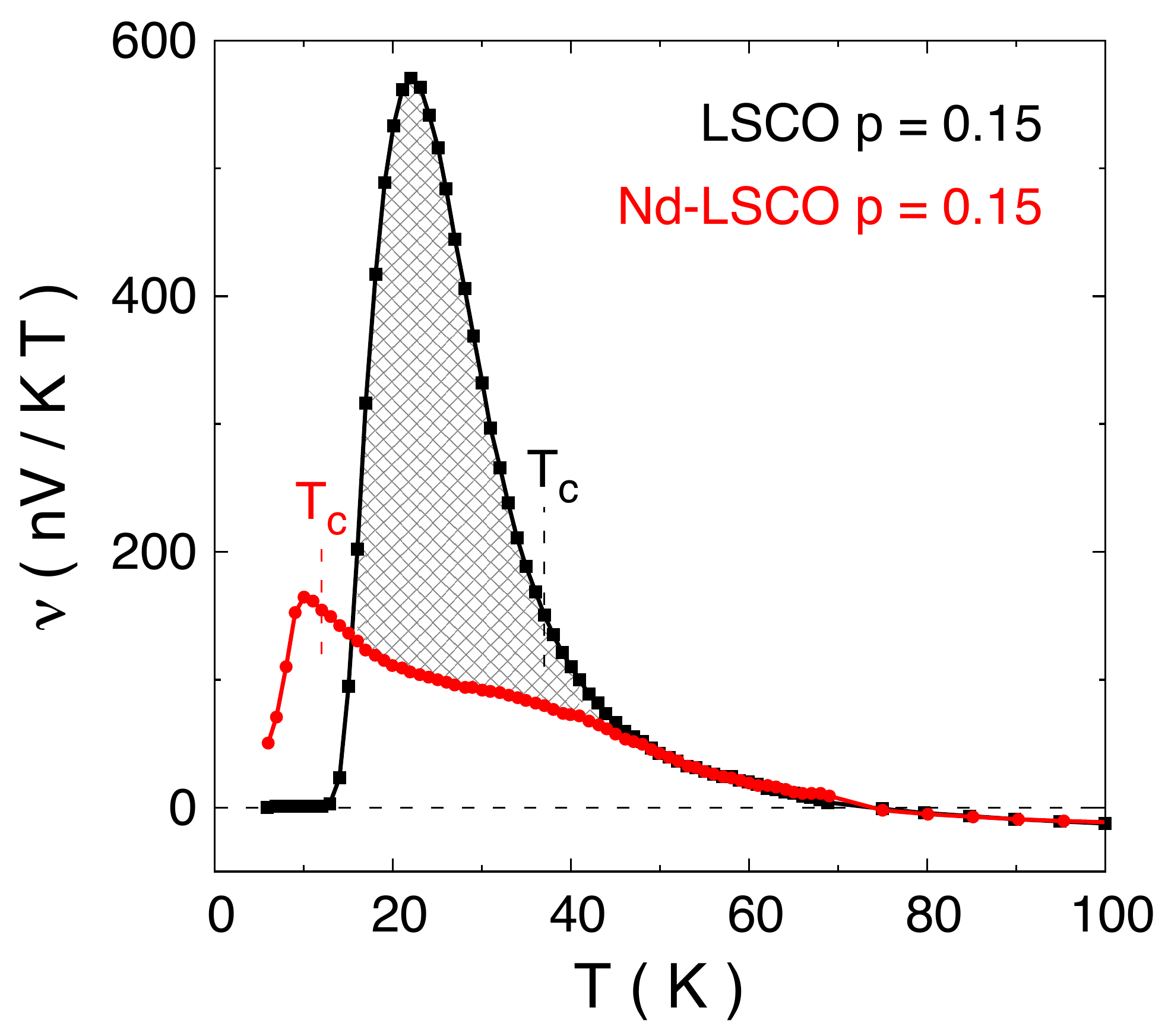}
\caption{(Color online)
Nernst coefficient $\nu$ of Nd-LSCO (red circles) and LSCO (black squares)
at $p$\,=\,$0.15$, as a function of  temperature (data from Ref.~[\onlinecite{Fujii2010}], at $H $\,=\,$ 9$~T).
Down to 50~K or so, the two data sets are virtually identical
(see also Fig.~\ref{Comparison-0p15}).
Note the small anomaly in the Nd-LSCO data at $T$\,$\simeq$\,$70$\,K, due to the LTT transition, a structural transition not present in LSCO.
Below 50\,K, the superconducting signal in LSCO starts to deviate upwards. 
The difference between the two curves (cross-hatched region) is attributable to their different 
$T_{\rm c}$ values ($37$\,K and $12$\,K);
it is the superconducting contribution to the Nernst signal in LSCO. 
}
\label{LSCOp15}
\end{figure}
%
%%%%%%%%%%%%%%%%%%%%%%%%%%%%%%%%%%%%%%%%%%%%%%%%%%%%%%%%%%%%%%%%%%%%%%%

Nernst data taken on single crystals have been reported for 
Nd-LSCO at $p$\,=\,$0.20$ and 0.24 
and for
Eu-LSCO at $p$\,=\,$0.125$ and 0.16~[\onlinecite{Cyr-Choiniere2009}].
The new data reported here were taken on 
Eu-LSCO at $p$\,=\,$0.08$, 0.10 and 0.21, 
and on Nd-LSCO at $p$\,=\,$0.20$ and 0.21.
We start by reviewing published data on Eu-LSCO at $p$\,=\,$0.125$ (from Ref.~[\onlinecite{Cyr-Choiniere2009}]),
displayed in Fig.~\ref{nu-LESCO-low-x}(c), 
as their double-peak structure reveals most clearly 
the presence of two separate contributions to the Nernst signal $\nu(T)$: 
1) a narrow positive peak at low temperature (shaded in grey), attributed to superconducting fluctuations because of its strong field dependence; 
2) a broad positive peak at higher temperature, attributed to quasiparticles because it is independent of field. 
By applying a magnetic field of 28 T, the superconducting peak is entirely suppressed 
and only the quasiparticle peak remains (dashed line)~[\onlinecite{Chang2012}].

A double-peak structure is also observed in Nd-LSCO at $p$\,=\,$0.15$~[\onlinecite{Fujii2010}] (see Fig.~\ref{LSCOp15})
and in the electron-doped cuprate PCCO at $x$\,=\,$0.13$~[\onlinecite{Li2007}].
In all cases, the two peaks in $\nu(T)$ can be resolved because $T_{\rm c}$ is sufficiently low, roughly 10~K. 
By contrast, in LSCO at $p $\,=\,$0.12$ ($p$\,=\,$0.15$), where $T_{\rm c}$\,$\simeq$\,$30$\,K ($37$\,K), the superconducting peak in $\nu$ 
is moved up in temperature so that it lies on top of the quasiparticle peak (Figs.~\ref{nu-LESCO-low-x}(c)~and~\ref{LSCOp15}). 
This unfortunate overlap is what led to the initial misinterpretation of the LSCO data by the Princeton group~[\onlinecite{Xu2000},\onlinecite{Wang2001}]. 

Even when two peaks cannot be resolved, one can still identify a temperature $T_{\rm B}$
above which the Nernst coefficient is independent of magnetic field, a good indication that the superconducting Nernst signal is negligible.
In Fig.~\ref{nu-LESCO-low-x}, we see that the Nernst signal at 2~T splits off from the 10~T data
below $T_{\rm B}$\,$\simeq$\,$30$\,-\,$40$\,K, for all three dopings.
Above $T_{\rm B}$, the Nernst signal is therefore all due to quasiparticles, to a good approximation, 
and this is the signal we will use to pin down the onset temperature \Tstar{} of the pseudogap phase 
in the three LSCO-based cuprates.

%%%%%%%%%%%%%%%%%%%%%%%%%%     FIGURE 10  %%%%%%%%%%%%%%%%%%%%%%%%%%%%%%%%%%%%%%
%
\begin{figure}
\centering
\includegraphics[width=0.48\textwidth]{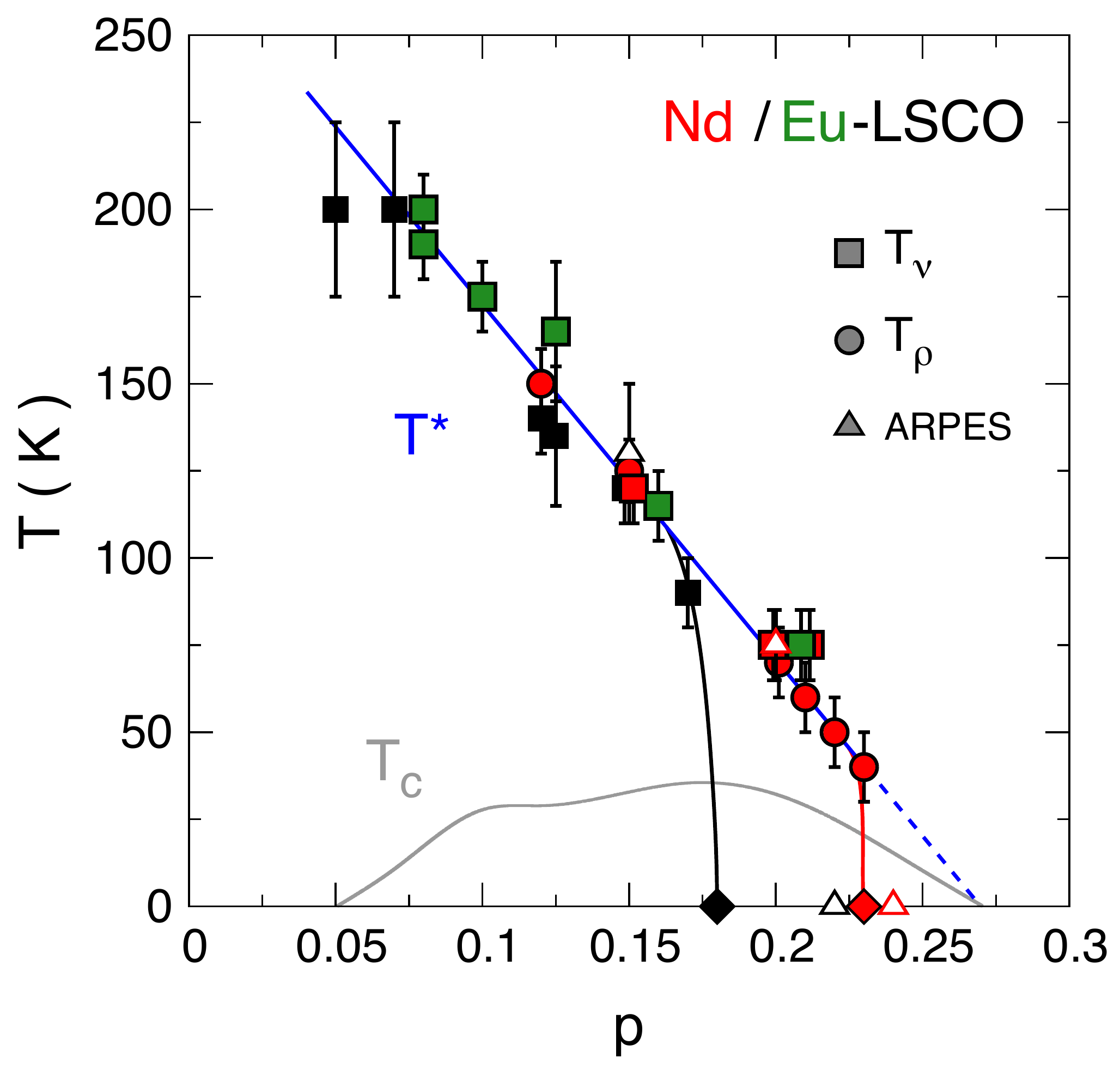}
\caption{(Color online)
Temperature-doping phase diagram of LSCO (black), Nd-LSCO (red) and Eu-LSCO (green),
showing the pseudogap temperature \Tstar~(blue line)
and the superconducting transition temperature \Tc~(of LSCO; grey line).
$T_\nu$ (filled squares, from this work and Refs.~[\onlinecite{Wang2006},\onlinecite{Cyr-Choiniere2009},\onlinecite{Xu2000},\onlinecite{Fujii2010},\onlinecite{Wang2001}]) 
is the temperature below which the quasiparticle Nernst signal starts to increase 
toward large positive values (Fig.~\ref{nu-Nd-Eu-LSCO}). 
$T_\rho$ (filled circles, from Refs.~[\onlinecite{Daou2009},\onlinecite{Collignon2017},\onlinecite{Ichikawa2000}]) 
is the temperature below which the resistivity $\rho(T)$ deviates from linearity (Fig.~\ref{rho-YBCO-LNSCO}). 
The open triangles show \Tstar~detected by ARPES as the temperature below which the anti-nodal pseudogap opens,
in LSCO (black)~[\onlinecite{Yoshida2009}] and Nd-LSCO (red)~[\onlinecite{Matt2015}].
We see that \Tnu\,$\simeq$\,\Trho\,$\simeq$\,\Tstar, within error bars.
Note how the pseudogap phase comes abruptly to an end, at a critical doping 
\pstar\,=\,$0.18$\,$\pm$\,$0.01$ for LSCO (black diamond)~[\onlinecite{Cooper2009},\onlinecite{Laliberte2016}],
and at a much higher doping,
\pstar\,=\,$0.23$\,$\pm$\,$0.01$, for Nd-LSCO (red diamond)~[\onlinecite{Daou2009},\onlinecite{Collignon2017}].
The dashed blue line is a linear extension of the solid blue line. 
%Around $p$\,$\simeq$\,$0.17$, the $T^\star$ dashed line splits in two; 
%black for LSCO where the pseudogap QCP is believed to be at $p^\star $\,=\,$ 0.18$\,$\pm$\,$0.01$~[\onlinecite{Cooper2009}], 
%and red for Nd-LSCO where it lies at $p^\star $\,=\,$ 0.235$\,$\pm$\,$0.05$~[\onlinecite{Daou2009},\onlinecite{Cyr-Choiniere2010}].
%
}
\label{LSCO-phasediag}
\end{figure}
%
%%%%%%%%%%%%%%%%%%%%%%%%%%%%%%%%%%%%%%%%%%%%%%%%%%%%%%%%%%%%%%%%%%%%%%%

%%%%%%%%%%%%%%%%%%%%%%%%%%     FIGURE 11   %%%%%%%%%%%%%%%%%%%%%%%%%%%%%%%%%%%%%%
% Nernst_0p20&0p24_16T
%
\begin{figure}
\centering
\includegraphics[width=0.45\textwidth]{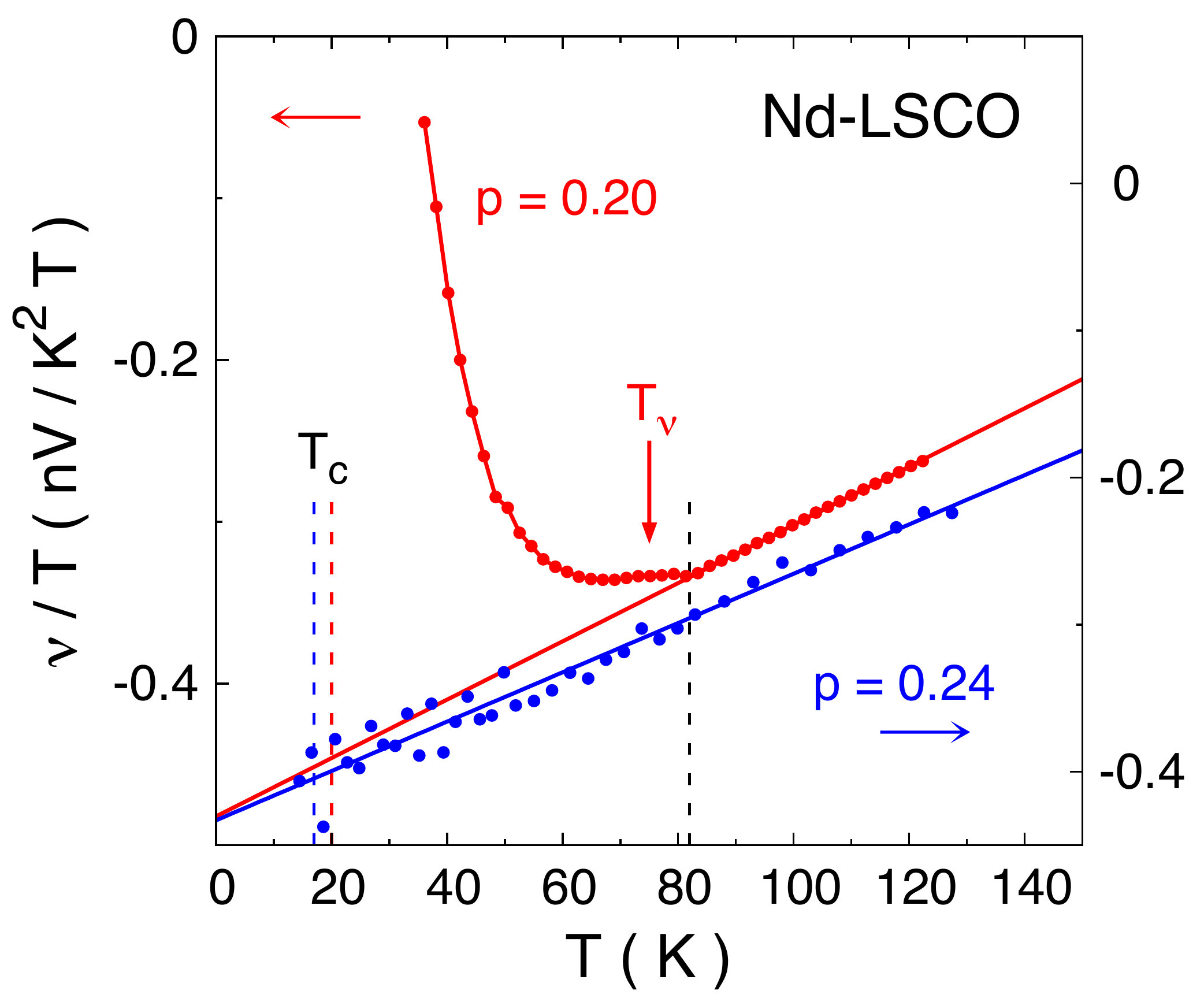}
\caption{(Color online)
Nernst coefficient $\nu$ of Nd-LSCO at $p $\,=\,$ 0.20$ (left axis, red dots, $H $\,=\,$ 16$\,T; this work) 
and $p $\,=\,$ 0.24$ (right axis, blue dots, $H $\,=\,$ 10$\,T~[\onlinecite{Cyr-Choiniere2009}]), 
plotted as $\nu$\,/\,$T$ vs $T$.
The red and blue vertical dashed lines mark \Tc($H $\,=\,$ 0$\,T) at $p $\,=\,$ 0.20$ ($20$\,K) and $0.24$ ($17$\,K), respectively.
The black vertical dashed line marks the transition to the low-temperature tetragonal (LTT) structure, at  
$T_{\rm LTT} $\,=\,$ 82$\,K for $p $\,=\,$ 0.20$;
the transition only causes a small kink in the (red) data (see also Fig.~\ref{LSCOp15}). 
The solid color-coded lines are linear fits to the data above 82~K, extended down to $T$\,=\,$0$. 
This comparison shows the effect of the pseudogap on the Nernst coefficient: 
a large upturn below \Tnu\,=\,\Tstar~(red arrow)) at $p$\,=\,$0.20 <$~\pstar,
in contrast with the continuous linear decrease at $p$\,=\,$0.24 >$~\pstar.
}
\label{Nernst-0p20}
\end{figure}
%
%%%%%%%%%%%%%%%%%%%%%%%%%%%%%%%%%%%%%%%%%%%%%%%%%%%%%%%%%%%%%%%%%%%%%%%

It is convenient to begin with Nd-LSCO, whose temperature-doping phase diagram is shown in Fig.~\ref{LSCO-phasediag}
(red symbols),
because its properties in the vicinity of
the critical doping \pstar{} below which the pseudogap phase appears at $T $\,=\,$ 0$ (red diamond) have been
thoroughly characterized.
In particular, ARPES measurements establish that the anti-nodal pseudogap in Nd-LSCO opens 
below a temperature \Tstar\,=\,$75$\,$\pm$\,$5$\,K at $p$\,=\,$0.20$ (white triangle, Fig.~\ref{LSCO-phasediag}), 
and that there is no pseudogap at $p$\,=\,$0.24$~[\onlinecite{Matt2015}].

The onset of the pseudogap phase has a dramatic impact on the electrical resistivity of Nd-LSCO~[\onlinecite{Daou2009}],
as seen in Fig.~\ref{rho-YBCO-LNSCO}(d).
At $p$\,=\,$0.24$, where there is no pseudogap, the normal-state $\rho(T)$ (measured in high fields)
is linear from $T$\,$\simeq$\,$80$\,K down to $T$\,$\simeq$\,$0$~[\onlinecite{Daou2009},\onlinecite{Collignon2017}]. 
At $p$\,=\,$0.20$, $\rho(T)$ undergoes a huge upturn as $T$\,$\to$\,$0$, increasing its value by a factor $\sim$\,$6$ 
relative to the value $\rho_0$ it would have in the absence of a pseudogap~[\onlinecite{Daou2009},\onlinecite{Collignon2017}].
We define \Trho{} as the temperature where the upturn starts,
relative to the linear-$T$ dependence observed at higher temperature~[\onlinecite{Daou2009},\onlinecite{Collignon2017}].
Using this definition, resistivity data yield the six red circles in Fig.~\ref{LSCO-phasediag}~[\onlinecite{Daou2009},\onlinecite{Collignon2017}]. 
At $p$\,=\,$0.20$, \Trho\,=\,$70$\,$\pm$\,$10$\,K, so that \Trho~$\,=\,$~\Tstar, within error bars,
thereby confirming the interpretation of the low-$T$ upturn in $\rho(T)$ as being due to the pseudogap.

Using measurements of both the in-plane and out-of-plane ($c$-axis) resistivities, the upturn in $\rho(T)$
was tracked vs doping to pinpoint the precise location of the critical point~[\onlinecite{Collignon2017},\onlinecite{Cyr-Choiniere2010}] 
at 
$p^\star$\,$\simeq$\,$0.23$\,$\pm$\,$0.01$~(red diamond in Fig.~\ref{LSCO-phasediag}). 
This type of upturn was first detected in LSCO twenty years ago, as illustrated in Fig.~\ref{rho-YBCO-LNSCO}(c)~[\onlinecite{Boebinger1996}].
Its origin was only recently shown to be a drop in the carrier density from $n$\,=\,$1$\,+\,$p$ above \Tstar{} to $n$\,=\,$p$ 
at $T$\,=\,$0$, combined with a negligible change in carrier mobility $\mu$~[\onlinecite{Laliberte2016}].
In Nd-LSCO, this interpretation is confirmed by Hall effect measurements that indeed find
a drop in the $T$\,=\,$0$ Hall number from $n_{\rm H} $\,=\,$1$\,+\,$p$ above \pstar{} to $n_{\rm H}$\,=\,$p$ 
below \pstar~[\onlinecite{Daou2009},\onlinecite{Collignon2017}], precisely as observed in YBCO~[\onlinecite{Badoux2016}].

The large and abrupt drop in $n$ below \pstar{} should cause 
$\nu$\,/\,$T$ to increase, just as $\rho$ and \RH{} do, since all three quantities go as $1$\,/\,$n$ (at $T$\,=\,$0$).
This is indeed the case.
(A large enhancement of $\nu$, from small and negative to large and positive, is also found in calculations of Fermi-surface reconstruction
by commensurate~[\onlinecite{Hackl2009}] and incommensurate~[\onlinecite{Hackl2010}] antiferrromagnetic order.)
In Fig.~\ref{Nernst-0p20}, we show Nernst data for Nd-LSCO at $p$\,=\,$0.20$ and $p$\,=\,$0.24$,
plotted as $\nu$\,/\,$T$ vs $T$.
The data in this figure are limited to those temperatures where no field dependence is detected, and
are therefore purely a quasiparticle signal.
The difference in behavior is striking.
At $p$\,=\,$0.24$,
$\nu$\,/\,$T$ decreases linearly as $T$\,$\to$\,$0$, down to at least 15\,K,
remaining negative all the way.
This is analogous to the linear-$T$ decrease in $\rho(T)$ at that doping (Fig.~\ref{rho-YBCO-LNSCO}(d)).
The value $\nu$\,/\,$T$ extrapolates to at $T$\,=\,$0$, $- 0.42$\,nV\,/\,K$^2$T, is in reasonable agreement 
with expectation.
Indeed, using the second term in Eq.~\ref{eq:NKamran2}, we estimate
$\nu$\,/\,$T$\,=\,$-\mu S$\,/\,$T$ at $T$\,$\to$\,$0$, with the mobility $\mu$\,=\,$(\rho_{xy}$\,/\,$H)$\,/\,$\rho_{xx}$, 
to yield 
$\nu$\,/\,$T$\,=\,$-0.6$\,nV\,/\,K$^2$T,
given that 
$S$\,/\,$T$\,=\,$+0.3$\,$\mu$V\,/\,K$^2$~[\onlinecite{Daou2009a}] 
and 
$\mu$\,=\,$+0.002$\,T$^{-1}$~[\onlinecite{Collignon2017}] in Nd-LSCO at $p$\,=\,$0.24$. 
The fact that the measured $\nu$\,/\,$T$ is slightly less negative than the calculated one means that 
the first (positive) term in Eq.~\ref{eq:NKamran2} acts to partially reduce its magnitude. 
In the end, 
$\nu$\,/\,$T$\,$\simeq$\, $-(\frac{2}{3})$\,$\mu S$\,/\,$T$, the value given by simple formula in Eq.~\ref{eq:nuKamran},
since
$S$\,/\,$T$\,$\approx$\,$({\pi^2}/{2})$\,$({k^{2}_{\rm B}}$\,/\,${e})$\,$(1$\,/\,$\epsilon_{\rm F})$.
All this means that in Nd-LSCO at $p$\,=\,$0.24$, just as the small (positive) Hall coefficient 
reflects the large hole-like Fermi surface, with a Hall number equal to the carrier density ($n_{\rm H}$\,=\,$1$\,$+$\,$p$)~[\onlinecite{Daou2009}], 
so do the small Seebeck and Nernst coefficients.

%%%%%%%%%%%%%%%%%%%%%%%%%%     FIGURE 12   %%%%%%%%%%%%%%%%%%%%%%%%%%%%%%%%%%%%%%
% Comparison-0p21
\begin{figure}
\centering
\includegraphics[width=0.45\textwidth]{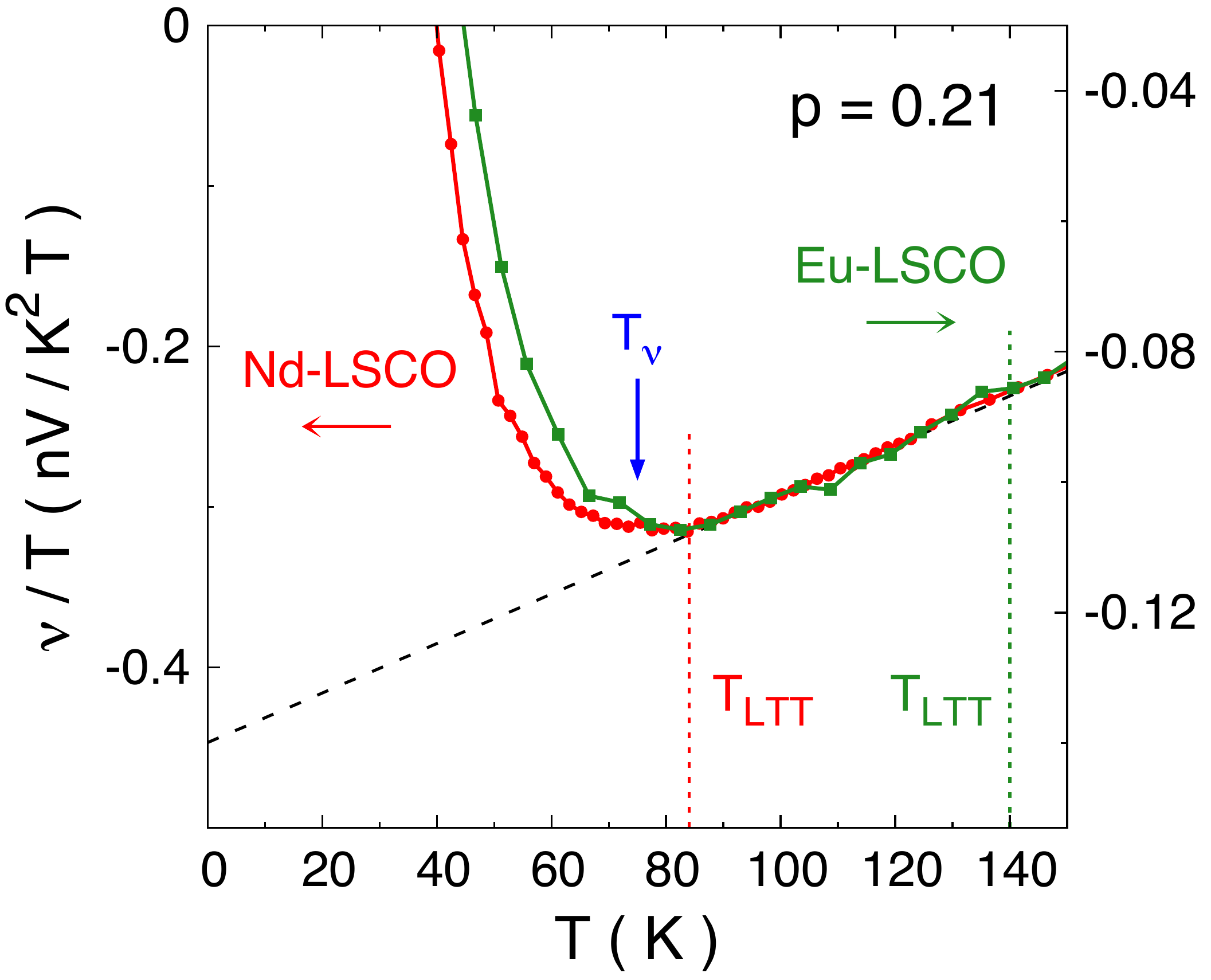}
\caption{(Color online) 
Nernst coefficient $\nu$ of Nd-LSCO (red circles; left axis; $H$\,=\,$16$~T) and Eu-LSCO (green squares; right axis; $H$\,=\,$10$~T),
both at  $p$\,=\,$0.21$, plotted as $\nu$\,/\,$T$ versus $T$. 
Above 40\,K, $\nu$ is independent of magnetic field.
Vertical dotted lines mark the structural transitions to the LTT structure at low $T$.
The black dashed line is a linear fit to the Nd-LSCO data above 85\,K, extended down to $T$\,=\,$0$. 
Eu-LSCO data also show linearity in the same temperature range. 
Data deviate upwards from the linear fit below a temperature $T_\nu $\,=\,$75$\,$\pm$\,$10$\,K for Nd-LSCO (blue arrow) and
$T_\nu $\,=\,$80$\,$\pm$\,$10$\,K for Eu-LSCO.
The very different LTT temperatures of the two materials implies that the upturn in $\nu$\,/\,$T$ observed at roughly
 the same temperature in both is not caused by this structural transition, but instead by the pseudogap opening.
}
\label{Comparison-0p21}
\end{figure}
%
%%%%%%%%%%%%%%%%%%%%%%%%%%%%%%%%%%%%%%%%%%%%%%%%%%%%%%%%%%%%%%%%%%%%%%%

At $p$\,=\,$0.20$\,$<$\,\pstar, 
$\nu$\,/\,$T$ also decreases linearly down to 80\,K, with a similar slope, but below 80\,K, 
it undergoes a dramatic rise to positive values (Fig.~\ref{Nernst-0p20}).
This upturn in $\nu$\,/\,$T$ is analogous to the upturn in $\rho(T)$ at that doping (Fig.~\ref{rho-YBCO-LNSCO}(d)).
It is a second signature of the pseudogap phase.
In other words, just as the parallel drops in $\rho(T)$ and $\nu$\,/\,$T$ observed in YBCO are two signatures of \Tstar,
so the parallel rises in $\rho(T)$ and $\nu$\,/\,$T$ observed in Nd-LSCO are the signature of \Tstar{} in that material -- confirmed
in this case by a direct spectroscopic measurement~[\onlinecite{Matt2015}]. 
Note that in our previous work on the Nernst effect in Nd-LSCO~[\onlinecite{Cyr-Choiniere2009}] 
we attributed the rise in the Nernst coefficient at $p$\,=\,$0.20$ 
to the onset of stripe order (combined charge-density and spin-density waves) at low temperature.
(Note that no charge order has been detected at $p$\,=\,$0.20$, but spin order is seen by neutron diffraction below 20\,K~[\onlinecite{Tranquada1997}], 
with a slowing down of spin fluctuations detected by NQR below 40\,K~[\onlinecite{Hunt2001}].)
The recent ARPES study showing a pseudogap opening at 75\,K~[\onlinecite{Matt2015}], 
precisely where the upturn in $\rho(T)$~[\onlinecite{Collignon2017}] and in $\nu/T$ (Fig.~\ref{Nernst-0p20}) begins, has clarified the cause of the upturns.

%%%%%%%%%%%%%%%%%%%%%%%%%%     FIGURE 13   %%%%%%%%%%%%%%%%%%%%%%%%%%%%%%%%%%%%%%
% Comparison-0p15
\begin{figure}
\centering
\includegraphics[width=0.45\textwidth]{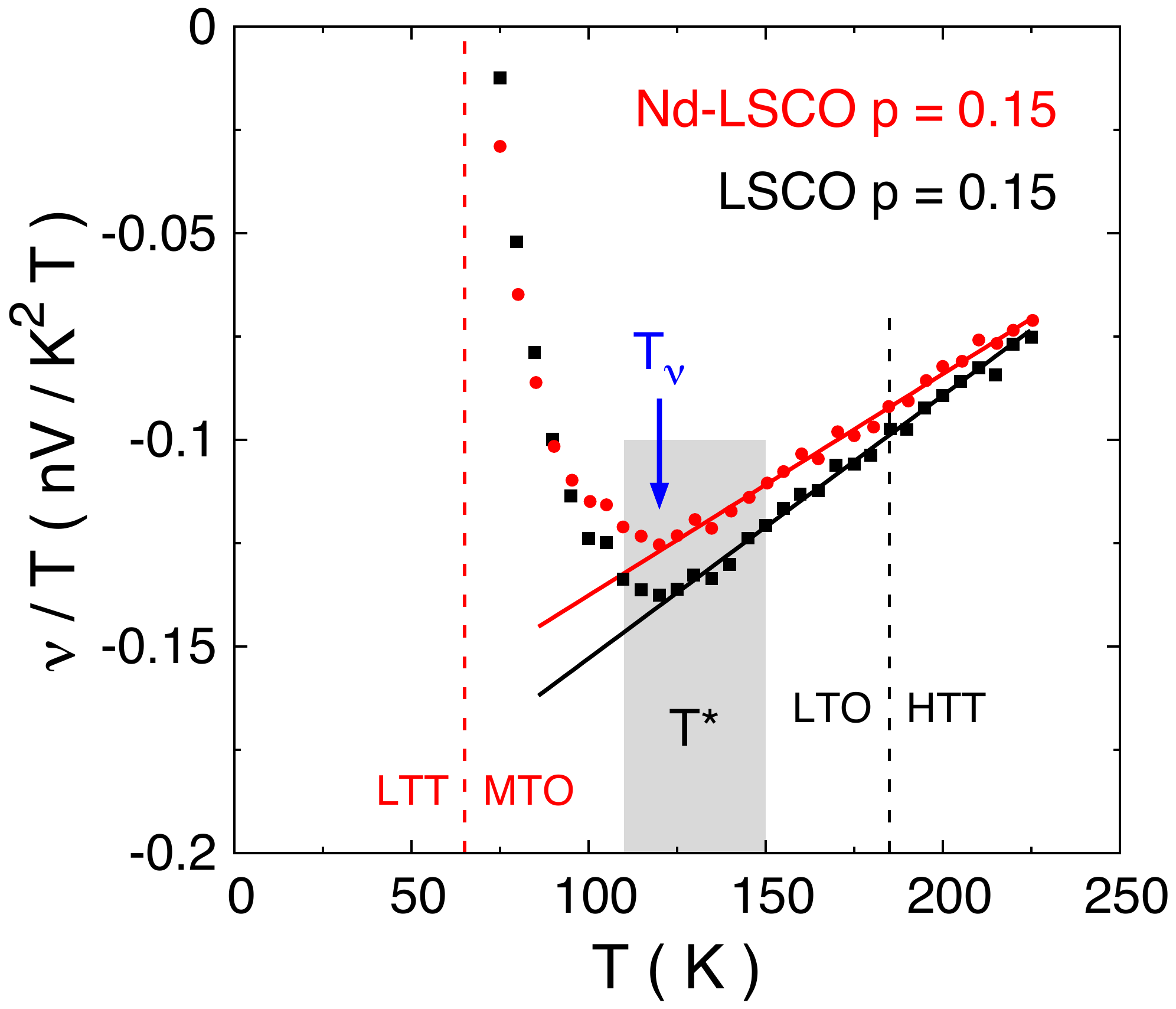}
\caption{(Color online) 
Nernst coefficient $\nu$ of Nd-LSCO (red circles) and LSCO (black squares) at $p$\,=\,$0.15$, plotted as $\nu$\,/\,$T$ versus $T$
(data from Ref.~[\onlinecite{Fujii2010}]).
Vertical dashed lines indicate structural transition temperatures: 
from middle-temperature orthorhombic (MTO) to low-temperature tetragonal (LTT) in Nd-LSCO ($70$\,K~[\onlinecite{Axe1994}], Fig.~\ref{LSCOp15}), 
and from high-temperature tetragonal (HTT) to low-temperature orthorhombic (LTO) in LSCO ($185$\,K~[\onlinecite{Keimer1992}]). 
One can see that the simultaneous rise in $\nu$\,/\,$T$~below \Tnu\,=\,$120$\,$\pm$\,$10$\,K (blue arrow)
in the two materials cannot be caused by their structural transitions, which take place well below and above, respectively.
The grey band marks the location of the pseudogap temperature measured by ARPES in LSCO at $p$\,=\,$0.15$~[\onlinecite{Yoshida2009}], 
at \Tstar\,=\,$130$\,$\pm$\,$20$\,K.
}
\label{Comparison-0p15}
\end{figure}
%
%%%%%%%%%%%%%%%%%%%%%%%%%%%%%%%%%%%%%%%%%%%%%%%%%%%%%%%%%%%%%%%%%%%%%%%

%%%%%%%%%%%%%%%%%%%%%%%%%%     FIGURE 14  %%%%%%%%%%%%%%%%%%%%%%%%%%%%%%%%%%%%%%
% Comparison-0p125
\begin{figure}
\centering
\includegraphics[width=0.48\textwidth]{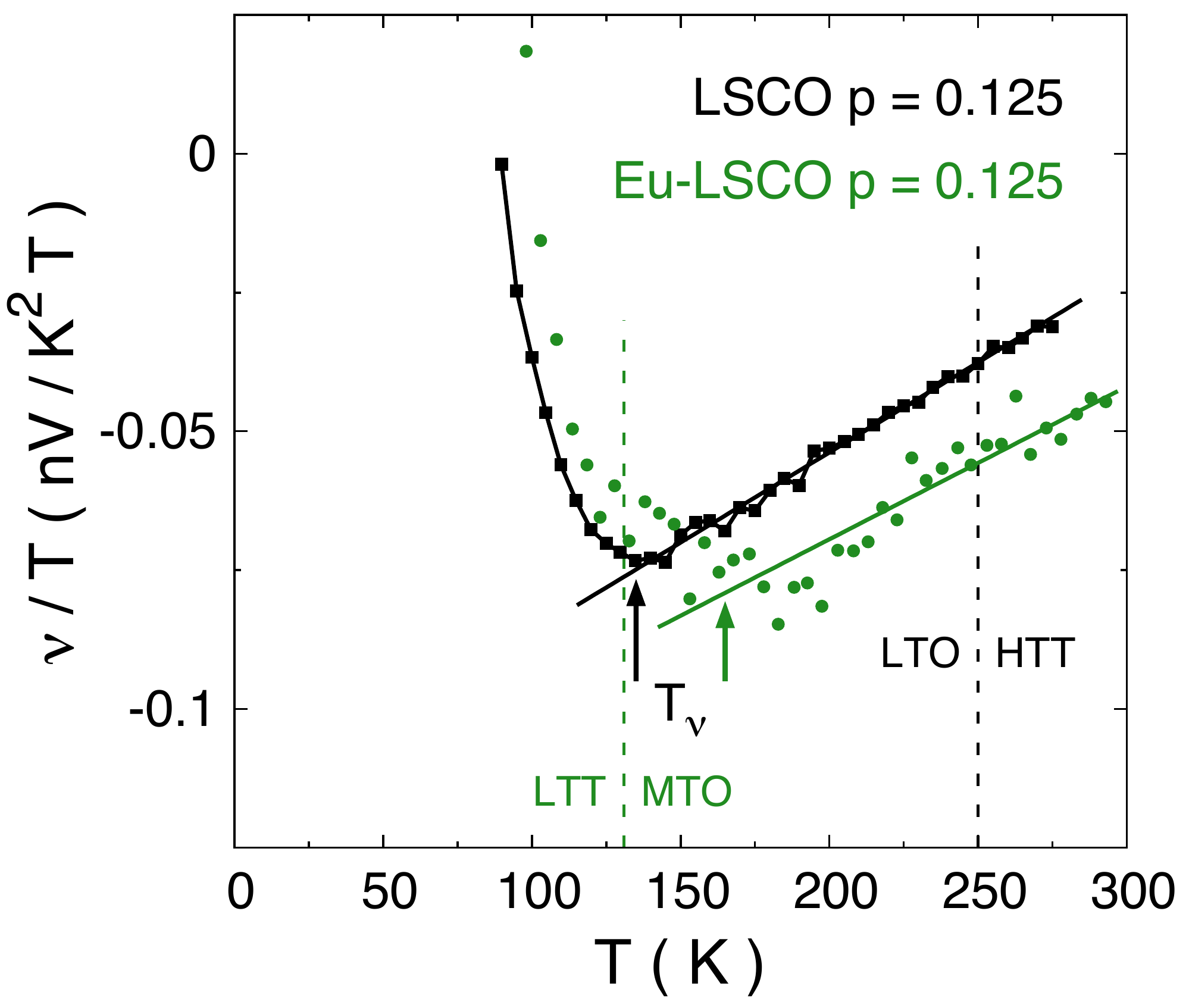}
\caption{(Color online) 
Nernst coefficient $\nu$ of Eu-LSCO (green circles; from Ref.~[\onlinecite{Hess2010}]) 
and LSCO (black squares; from Ref.~[\onlinecite{Fujii2010}]) at $p$\,=\,$0.125$, plotted as $\nu$\,/\,$T$ versus $T$.
Vertical dashed lines indicate structural transition temperatures: 
from middle-temperature orthorhombic (MTO) to low-temperature tetragonal (LTT) in Eu-LSCO ($131$\,K~[\onlinecite{Fink2011}]), 
and from high-temperature tetragonal (HTT) to low-temperature orthorhombic (LTO) in LSCO ($250$\,K~[\onlinecite{Keimer1992}]). 
As in Figs.~\ref{Comparison-0p21}~and~\ref{Comparison-0p15}, the simultaneous rise in $\nu$\,/\,$T$ below 
\Tnu\,=\,$165$\,$\pm$\,$20$\,K for Eu-LSCO and \Tnu\,=\,$135$\,$\pm$\,$20$\,K for LSCO 
is unrelated to their structural transitions.
}
\label{Comparison-0p125}
\end{figure}
%
%%%%%%%%%%%%%%%%%%%%%%%%%%%%%%%%%%%%%%%%%%%%%%%%%%%%%%%%%%%%%%%%%%%%%%%

Upon close inspection of the Nernst data on Nd-LSCO $p$\,=\,$0.20$ (Fig.~\ref{Nernst-0p20}), 
we see a small kink at $T $\,=\,$ 82$\,K, due to the structural transition into the low-temperature tetragonal (LTT) phase. 
To ascertain that this transition has only a small effect on the large upturn in $\nu$\,/\,$T$, we compare
Nernst data in the three LSCO-based cuprates, at three different dopings. 
In Fig.~\ref{Comparison-0p21}, we compare our own data at $p$\,=\,$0.21$ on Nd-LSCO and Eu-LSCO.
In our Nd-LSCO sample, there is a clear kink in $\rho(T)$ at 
$T_{\rm LTT}$\,=\,$84$\,K (red dotted line). 
In Eu-LSCO, the LTT transition at $p $\,=\,$ 0.21$ is expected at $T_{\rm LTT}$\,$\simeq$\,$140$\,K~[\onlinecite{Fink2011}] (green dotted line).
However, it has no detectable signature in our sample; even the $c$-axis resistivity
shows no feature whatsoever.
Be that as it may, any structural transition in Eu-LSCO at $p$\,=\,$0.21$ occurs well above $80$\,K. 
Yet, in both samples the Nernst data show very similar upturns.
We define \Tnu~as the temperature where the upturn in $\nu$\,/\,$T$ vs $T$ begins. 
At $p $\,=\,$ 0.21$, we find 
\Tnu\,=\,$ 75$\,$\pm$\,$10$\,K in Nd-LSCO and 
\Tnu\,=\,$ 80$\,$\pm$\,$10$\,K in Eu-LSCO; those values are added to the phase diagram (squares; Fig.~\ref{LSCO-phasediag}).

%%%%%%%%%%%%%%%%%%%%%%%%%%     FIGURE 15   %%%%%%%%%%%%%%%%%%%%%%%%%%%%%%%%%%%%%%
%
\begin{figure}
\centering
\includegraphics[width=0.48\textwidth]{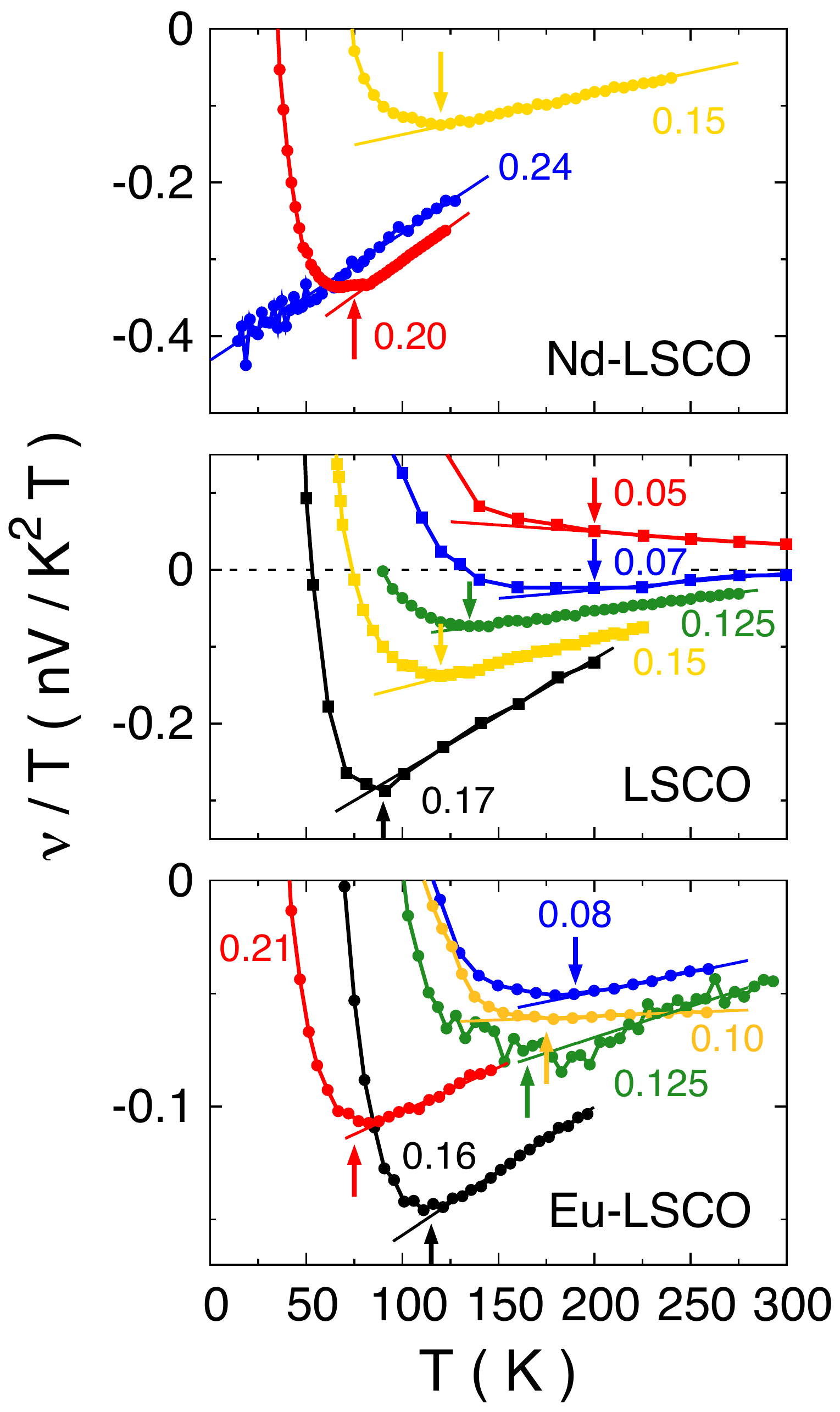}
\caption{(Color online) 
Nernst coefficient $\nu$ of 
Nd-LSCO (top), LSCO (middle) and Eu-LSCO (bottom),
at various dopings as indicated, plotted as $\nu$\,/\,$T$ versus temperature.
Lines are linear fits of the data at high temperature.
Arrows mark the temperature \Tnu{} below which the data start to deviate upward from linearity 
(see Figs.~\ref{Nernst-0p20},\ref{Comparison-0p21},\ref{Comparison-0p15} and \ref{Comparison-0p125} 
for a zoomed view of the data from which we can more easily identify \Tnu{}).
The values of \Tnu{} are (from low to high $p$):
$T_\nu$\,=\,$120$\,$\pm$\,$10$, $75$\,$\pm$\,$10$ and $0$\,K in Nd-LSCO,
$T_\nu $\,=\,$200$\,$\pm$\,$25$, $200$\,$\pm$\,$25$, $135$\,$\pm$\,$10$, $120$\,$\pm$\,$10$ and $90$\,$\pm$\,$10$\,K in LSCO,
and 
$T_\nu$\,=\,$190$\,$\pm$\,$10$, $175$\,$\pm$\,$10$, $165$\,$\pm$\,$20$, $115$\,$\pm$\,$10$ and $75$\,$\pm$\,$10$\,K in Eu-LSCO. 
%
%Note the decrease of $T_\nu$ with doping and the agreement in $T_\nu$ of all three materials for similar dopings. 
All values of $T_\nu$ are plotted on the phase diagram of Fig.~\ref{LSCO-phasediag}.
Nd-LSCO with $p$\,=\,$ 0.15$, LSCO with $p$\,=\,$ 0.15$ and $p$\,=\,$ 0.125$ were measured at $9$\,T (from Ref.~[\onlinecite{Fujii2010}]); 
Nd-LSCO with $p$\,=\,$ 0.20$ and Eu-LSCO with $p$\,=\,$ 0.21$ at $16$\,T (present work); 
Nd-LSCO with $p$\,=\,$0.24$, Eu-LSCO with $p$\,=\,$0.16$ (from Ref.~[\onlinecite{Cyr-Choiniere2009}]), 
Eu-LSCO with $p$\,=\,$0.08$ and $0.10$ (present work)
and Eu-LSCO with $p$\,=\,$ 0.125$~(from Ref.~[\onlinecite{Hess2010}]) at $10$\,T; 
LSCO with $p$\,=\,$ 0.17$ at $8$\,T (from Ref.~[\onlinecite{Xu2000}]) 
and LSCO with $p$\,=\,$ 0.05, 0.07$ (from Ref.~[\onlinecite{Wang2001}]) at $H \to 0$.
}
\label{nu-Nd-Eu-LSCO}	
\end{figure}
%
%%%%%%%%%%%%%%%%%%%%%%%%%%%%%%%%%%%%%%%%%%%%%%%%%%%%%%%%%%%%%%%%%%%%%%%

%%%%%%%%%%%%%%%%%%%%%%%%%%     FIGURE 16   %%%%%%%%%%%%%%%%%%%%%%%%%%%%%%%%%%%%%%
%
\begin{figure}
\centering
\includegraphics[width=0.47\textwidth]{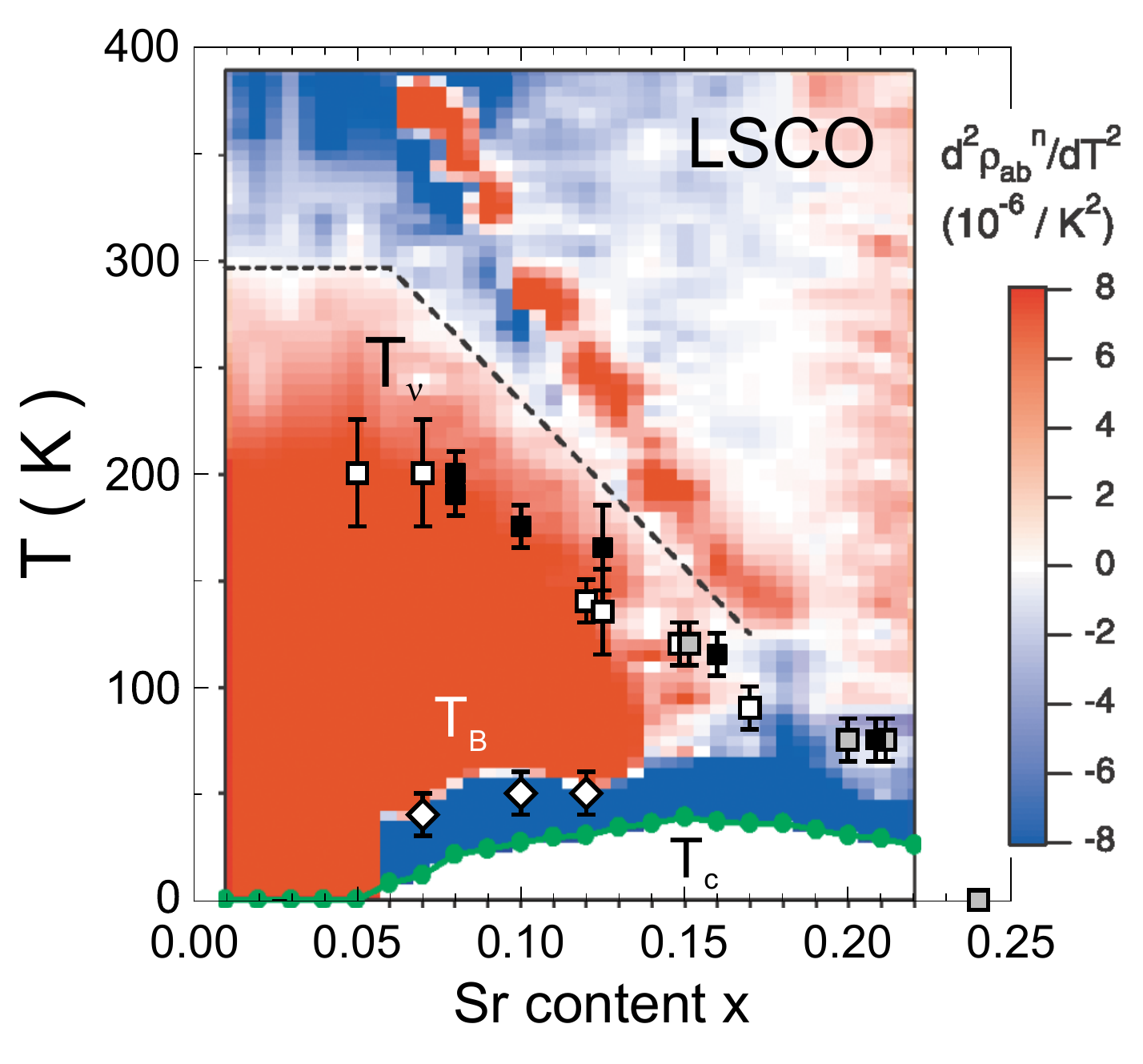}
\caption{(Color online) 
Resistivity curvature map from Ando {\it et al.}~[\onlinecite{Ando2004}] 
showing the second temperature derivative of their resistivity data on LSCO, with $T_{\rm c}$ as solid green circles. 
As in Fig.~\ref{Ando-map-YBCO}, regions in white correspond to linear, 
in blue to sub-linear (downward; $d^2\rho/dT^2$\,$<$\,$0$) 
and in red to super-linear (upward; $d^2\rho/dT^2$\,$>$\,$0$) 
behavior of resistivity with temperature. 
The red ridge inside the white region is due to the HTT-LTO structural transition in LSCO. 
The boundary of the pseudogap phase ($T^\star$) is the lower border of the white region (the dashed line is a guide to the eye). 
Our data points for $T_\nu$ from Fig.~\ref{LSCO-phasediag} are added,
for Nd-LSCO (grey squares), Eu-LSCO (black squares) and LSCO (open squares).
The $T_\nu$ data points agree reasonably well with the start of the upturn in the resistivity (as in Fig.~\ref{LSCO-phasediag}). 
The narrow blue region that tracks $T_{\rm c}$ is due to paraconductivity. 
The values of $T_{\rm B}$ for LSCO are added as open diamonds (from Fig.~\ref{TB-LSCO}) .
They agree well with the onset of paraconductivity. 
Together they delineate the regime of significant superconducting fluctuations in LSCO,
limited to $30$\,K above \Tc.
%This small boundary above $T_{\rm c}$ imposes an upper limit in temperature 
%regarding the importance of relevant superconducting fluctuations in Nernst signal at high temperatures. 
}
\label{Ando-map-LSCO}
\end{figure}
%
%%%%%%%%%%%%%%%%%%%%%%%%%%%%%%%%%%%%%%%%%%%%%%%%%%%%%%%%%%%%%%%%%%%%%%%

In Fig.~\ref{Comparison-0p15}, we compare data at $p$\,=\,$0.15$ on Nd-LSCO and LSCO (from Ref.~[\onlinecite{Fujii2010}]).
We see that the upturn in $\nu$\,/\,$T$ starts at a higher temperature than it did at $p$\,=\,$0.21$, with 
\Tnu\,=\,$120$\,$\pm$\,$10$\,K not only in Nd-LSCO but also in LSCO. 
The two samples exhibit essentially identical behavior, even though their respective crystal structures and structural transitions
are quite different: 
the LTT transition in Nd-LSCO is at 
$T_{\rm LTT}$\,=\,$70$\,K~[\onlinecite{Axe1994}] (red dashed line), 
$50$\,K below \Tnu,
while the LTO transition in LSCO is at $T_{\rm LTO}$\,$\simeq$\,$185$\,K~[\onlinecite{Keimer1992}] (black dashed line), 
$65$\,K above \Tnu.
This shows that the large upturns in $\nu$\,/\,$T$ are not caused by structural transitions.
Instead, they are caused by the opening of the pseudogap, as confirmed also in LSCO
by ARPES measurements at $p$\,=\,$0.15$, which yield 
\Tstar\,=\,$130$\,$\pm$\,$20$\,K (grey band in Fig.~\ref{Comparison-0p15})~[\onlinecite{Yoshida2009}]. 
As we did at $p$\,=\,$0.20$, we again find that 
\Tnu\,=\,\Trho\,=\,\Tstar{} at $p$\,=\,$0.15$, within error bars (Fig.~\ref{LSCO-phasediag}).

This conclusion is reinforced by yet another comparison, at $p$\,=\,$0.125$, between
Eu-LSCO (from Ref.~[\onlinecite{Hess2010}]) and LSCO (from Ref.~[\onlinecite{Fujii2010}]), as displayed in Fig.~\ref{Comparison-0p125}.
We see that in Eu-LSCO the upturn in $\nu$\,/\,$T$ now 
starts above the LTT transition 
at $T_{\rm LTT}$\,=\,$131$\,K (green dotted line),
whereas it started well below it at $p$\,=\,$0.21$ (Fig.~\ref{Comparison-0p21}).
In other words, the \Tstar~line in Eu-LSCO goes through the LTT transition unperturbed,
as in Nd-LSCO (Fig.~\ref{LSCO-phasediag}).
Similarly, the structural transition in LSCO has no effect on $\nu(T)$ and \Tstar~is well below.

In Fig.~\ref{nu-Nd-Eu-LSCO}, we collect data at several dopings for all three materials.
We see that the behavior is similar in all three: 
the upturn at low $T$ in $\nu$\,/\,$T$ onsets at a temperature \Tnu{} (arrows) that
increases monotonically with decreasing $p$.
In Fig.~\ref{LSCO-phasediag}, all values of \Tnu{} are plotted on a common phase diagram. 
The first thing to note is that \Tnu$(p)$ is the same in all three materials, within error bars, across the whole phase diagram.

In Fig.~\ref{LSCO-phasediag},
we also plot \Trho{} in Nd-LSCO~[\onlinecite{Daou2009},\onlinecite{Ichikawa2000}] (red circles), 
the temperature below which $\rho(T)$ 
deviates from its linear dependence at high temperature,
as illustrated in Fig.~\ref{rho-YBCO-LNSCO}(d).
(This is the same definition used for YBCO, except that here the deviation 
is upward instead of downward.)
We see that \Tnu\,=\,\Trho, within error bars, as also found in YBCO (Fig.~\ref{YBCO-phasediag}).

In Fig.~\ref{Ando-map-LSCO}, the $T_\nu$ values for LSCO, Nd-LSCO and Eu-LSCO 
are plotted on the curvature map of Ando and co-workers for LSCO~[\onlinecite{Ando2004}]. 
They are seen to coincide reasonably well with the upper boundary of the red region, 
where the upward deviation in $\rho(T)$ begins. 
Note that in LSCO the (white) region of linear-$T$ behaviour 
is contaminated near its lower bound by the structural transition, seen clearly as the red ridge inside the white region.
This anomaly in $\rho(T)$ can be mistaken for the pseudogap phase boundary in a resistive determination of $T^\star$. 
By contrast, a determination based on the Nernst coefficient is clear (Fig.~\ref{Comparison-0p15}), 
and it shows that the \Tstar$(p)$ line in LSCO lies well below its structural transition (Fig.~\ref{Ando-map-LSCO}). 

In Fig.~\ref{Ando-map-LSCO}, the region of paraconductivity, in which superconducting fluctuations cause a decrease
in $\rho(T)$ above \Tc, shows up very clearly as a blue band tracking the \Tc~dome, of width 30\,K or so.
We also plot $T_{\rm B}$ in LSCO (white diamonds), the temperature above which $\nu$ is independent of field (see Fig.~\ref{TB-LSCO}).
It agrees well with the upper limit of paraconductivity, both saying that superconducting fluctuations
have a negligible impact on either resistivity or Nernst above $\sim$\,\Tc\,$+$\,$30$\,K or so.
The long-held notion that superconducting fluctuations are detected in LSCO up to $\sim$\,\Tc\,$+$\,$100$\,K
is incorrect (see \hyperref[sec:Appendix]{Appendix} for further discussion).

%%%%%%%%%%%%%%%%%%%%%%%%%%     FIGURE 17   %%%%%%%%%%%%%%%%%%%%%%%%%%%%%%%%%%%%%%
%
\begin{figure*}[!t]
\centering
\includegraphics[width=1.0\textwidth]{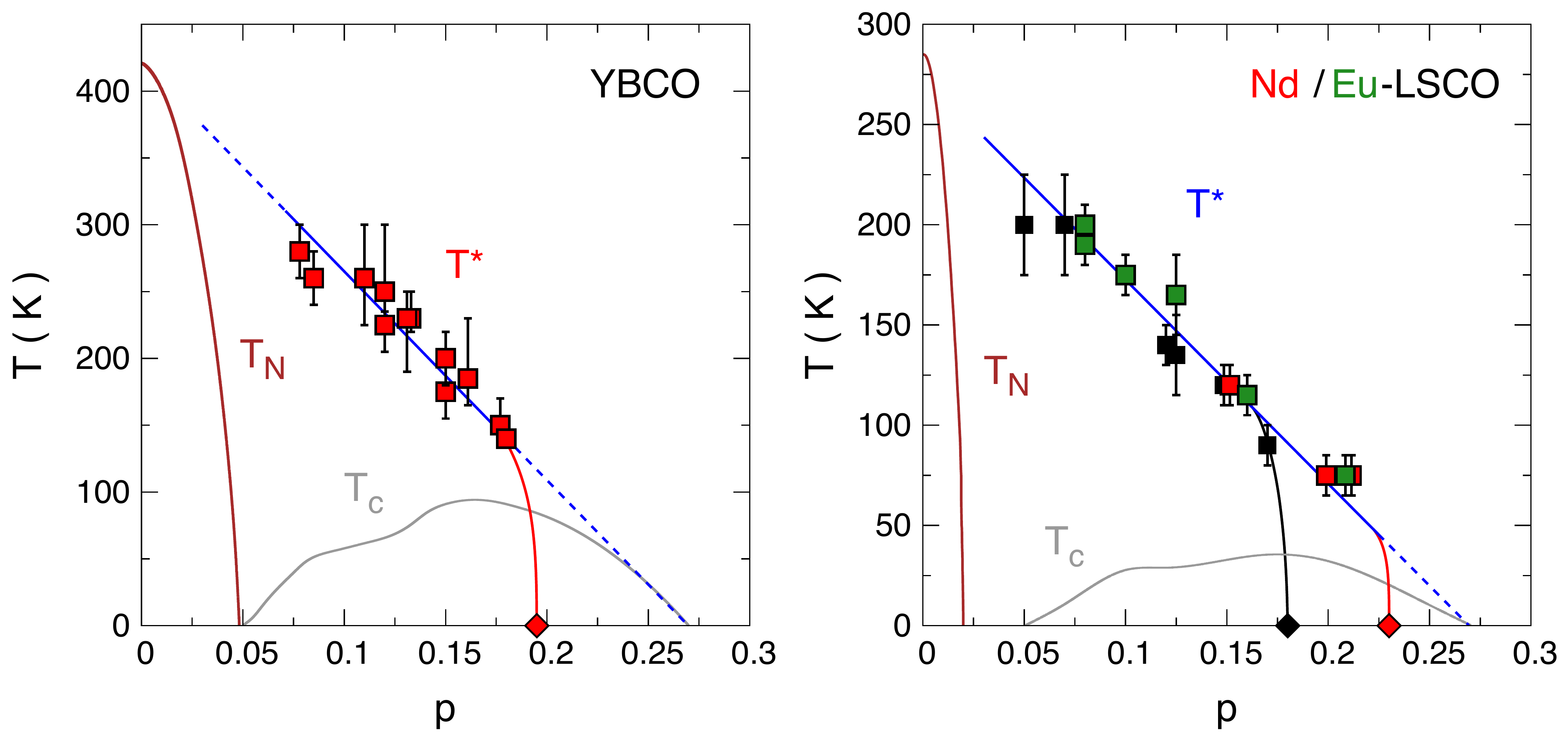}
\caption{(Color online)  
Temperature-doping phase diagrams of YBCO (a) and Nd/Eu-LSCO (b) 
showing the pseudogap temperature $T^\star$ ($T_\nu$, squares), 
the N\'{e}el temperature $T_{\rm N}$ (brown line) and the superconducting transition temperature $T_{\rm c}$ (grey line). 
The blue line is a linear guide to the eye showing that $T^\star$ extrapolates to $T_{\rm N}$ 
at half filling on the underdoped side ($p$\,=\,$0$) while it merges with $T_{\rm c}$ on the overdoped side where superconductivity disappears. 
Note that the $T^\star$ line of YBCO is proportional to but higher than that of LSCO: $T^\star_{\rm YBCO}$\,$\simeq$\,$1.5$\,$T^\star_{\rm LSCO}$. 
Roughly the same scaling applies to \TN{} at $p$\,=\,$0$: $T^{\rm YBCO}_{\rm N}(0)$\,$\simeq$\,$450$\,K~[\onlinecite{Brewer1988}] and 
$T^{\rm LSCO}_{\rm N}(0)$\,$\simeq$\,$280$\,K~[\onlinecite{Keimer1992}].
% so that $T^{\rm YBCO}_{\rm N}(0)$\,$\simeq$\,$1.6$\,$T^{\rm LSCO}_{\rm N}(0)$.
Diamonds mark the pseudogap critical points for 
YBCO (red) at $p^\star $\,=\,$ 0.195$\,$\pm$\,$0.005$~[\onlinecite{Badoux2016}] , 
LSCO (black) at $p^\star $\,=\,$ 0.18$\,$\pm$\,$0.01$~[\onlinecite{Cooper2009},\onlinecite{Laliberte2016}] and 
Nd-LSCO (red) at $p^\star $\,=\,$ 0.23$\,$\pm$\,$0.01$~[\onlinecite{Collignon2017}]. 
$T_\nu$ are taken from Fig.~\ref{YBCO-phasediag} for YBCO and from Fig.~\ref{LSCO-phasediag} for LSCO; 
$T_{\rm N}$ is taken from Ref.~[\onlinecite{Brewer1988}] for YBCO and from Ref.~[\onlinecite{Hucker1999}] for LSCO. 
}
\label{YBCO-LSCO-PhaseDiag}
\end{figure*}
%
%%%%%%%%%%%%%%%%%%%%%%%%%%%%%%%%%%%%%%%%%%%%%%%%%%%%%%%%%%%%%%%%%%%%%%%

In order to complete our determination of the pseudogap phase boundary in LSCO, we need to know the location of \pstar,
its end point at $T$\,=\,$0$. 
High-field measurements of the resistivity of LSCO reveal that $\rho(T)$ is perfectly linear below 70\,K or so,
down to the lowest $T$,
at $p$\,=\,$0.23$, $p$\,=\,$0.21$ and even $p$\,=\,$0.18$~[\onlinecite{Cooper2009}]. 
At $p$\,=\,$0.17$ and lower dopings, however, an upward deviation from linearity is observed at low $T$~[\onlinecite{Boebinger1996}]. 
Just as the appearance of an upturn was used to locate 
\pstar\,=\,$0.23$\,$\pm$\,$0.01$ in Nd-LSCO, we find that 
\pstar\,=\,$0.18$\,$\pm$\,$0.01$ in LSCO (black diamond, Fig.~\ref{LSCO-phasediag}).

In summary, the onset of the pseudogap phase at \Tstar$(p)$ 
causes an upturn in $\nu$\,/\,$T$ in the three La$_2$CuO$_4$-based cuprates,
which coincides with the upturn in $\rho(T)$, it has nothing to do with structural transitions,
and it is distinct from the upturn due to superconducting fluctuations close to \Tc.
In the $T-p$ phase diagram (Fig.~\ref{LSCO-phasediag}), 
the three materials are found to have the same \Tstar$(p)$ line, 
decreasing monotonically with $p$. 
However, the pseudogap phase ends sooner in LSCO, at \pstar\,=\,$0.18$, than in Nd-LSCO (or Eu-LSCO), 
where it extends up to \pstar\,=\,$0.23$.

%%%%%%%%%%%%%%%%%%%%%%%%%%%%%%%%%%%%%%%%%%%%%%%%%%%%%%%%%%%%%%%%%%%%%%%
%%%%%% Discussion
%%%%%%%%%%%%%%%%%%%%%%%%%%%%%%%%%%%%%%%%%%%%%%%%%%%%%%%%%%%%%%%%%%%%%%%

\section{Discussion} 
\label{sec:Discussion}

We have shown that it is possible to disentangle the superconducting and quasiparticle contributions to the Nernst coefficient $\nu(T)$~in cuprates.
The key difference is that the former depends strongly on magnetic field and not the latter.
In YBCO, they are also of opposite sign.
We then showed that the quasiparticle Nernst signal in Nd-LSCO and LSCO undergoes a pronounced change 
when temperature is reduced below \Tstar, the onset temperature of the pseudogap phase established by ARPES measurements.
A similar, albeit smaller, change in the resistivity $\rho(T)$ occurs simultaneously.
The onset of these changes, at \Tnu{} and \Trho{} respectively, can therefore be used to define \Tstar.
Using new and published Nernst data in four cuprates -- YBCO, LSCO, Nd-LSCO, and Eu-LSCO -- 
we identify \Tnu{} at various dopings and then map \Tstar{} across the temperature-doping phase diagram, 
in Fig.~\ref{YBCO-phasediag} for YBCO and in Fig.~\ref{LSCO-phasediag} for the other three.
We find that the latter three materials all have the same \Tstar$(p)$ line
(up to $p$\,$\simeq$\,$0.17$), irrespective of their different structural transitions.

%%%%%%%%%%%%%%%%%%%%%%%%%%%%%%%%%%%%%%%%%%%%%%%%%%%%%%%%%%%%%%%%%%%%%%%%%%%%

\subsection{Boundary of the pseudogap phase}

Having delineated the boundary \Tstar$(p)$ of the pseudogap phase, 
the question arises:
is it a transition or a crossover?
Detailed studies of the pseudogap opening via ARPES show a rather sharp onset with decreasing temperature,
as in optimally-doped Bi-2201~[\onlinecite{Kondo2011}] and Nd-LSCO at $p$\,=\,$0.20$~[\onlinecite{Matt2015}], pointing to a transition. 
By contrast, the change in $\rho(T)$ across \Tstar{} is always very gradual (Fig.~\ref{rho-YBCO-LNSCO}), suggestive of a crossover.
The change in $\nu(T)$ is also rather gradual when \Tstar{} is high, but it does get sharper when \Tstar{} is lower (Fig.~\ref{nu-Nd-Eu-LSCO}).
In the normal state at $T$\,$\to$\,$0$, the drop in Hall number \nH{} across \pstar{} (in either YBCO or Nd-LSCO) is as sharp as expected
theoretically for a quantum phase transition into a phase of long-range antiferromagnetic order~[\onlinecite{Collignon2017}]. 
In Nd-LSCO, the upturn in $\rho(T)$ appears very rapidly upon crossing below \pstar, going from no upturn to full upturn
over a doping interval of relative width
$\delta p$\,/\,\pstar\,$\simeq$\,$0.06$~[\onlinecite{Collignon2017}].

To better compare the phase diagrams of YBCO and LSCO, we display them side by side in Fig.~\ref{YBCO-LSCO-PhaseDiag}.
Some general features are immediately apparent.

%%%%%%%%%%%%%%%%%%%%%%%%%%%%%%%%%%%%%%%%%%%%%%%%%%%%%%%%%%%%%%%%%%%%%%%

\subsubsection{Pseudogap temperature \texorpdfstring{$T^\star$}{T*}}

\Tstar{} decreases monotonically with $p$, in both cases.
We see that the pseudogap temperature is 1.5 times larger in YBCO (and Bi-2212) 
than in LSCO (and Nd-LSCO and Eu-LSCO):
\Tstar$_{\rm YBCO}$\,$\simeq$\,$1.5$\,\Tstar$_{\rm LSCO}$ (up to $p$\,$\simeq$\,$0.17$). 
This is an important quantitative fact, which may reflect the strength of interactions and possibly the pairing strength.
The weaker maximal \Tc{} of LSCO (40\,K) compared to YBCO (93\,K)
may be related to its smaller \Tstar.

A linear fit to \Tstar{} vs $p$ gives a line that connects \TN$(0)$, the N\'{e}el temperature for the onset of 
commensurate antiferromagnetic order at $p$\,=\,$0$, to \pctwo, 
the upper end of the superconducting dome at $T$\,=\,$0$ (straight dashed lines in Fig.~\ref{YBCO-LSCO-PhaseDiag}).
The slope of that line is 1.5 times larger in YBCO
and so is \TN$(0)$:
$T^{\rm YBCO}_{\rm N}$\,$\simeq$\,$450$\,K~[\onlinecite{Brewer1988}] and 
$T^{\rm LSCO}_{\rm N}$\,$\simeq$\,$280$\,K~[\onlinecite{Keimer1992}], at $p$\,=\,$0$. 

These connections suggest a link between the pseudogap phase and the antiferromagnetism of the undoped Mott insulator.
They also suggest that the same interactions favour pseudogap formation and pairing.

%%%%%%%%%%%%%%%%%%%%%%%%%%%%%%%%%%%%%%%%%%%%%%%%%%%%%%%%%%%%%%%%%%%%%%%%%%%%

\subsubsection{Pseudogap critical doping \texorpdfstring{$p^\star$}{p*}}

If the linear decrease of \Tstar$(p)$ with doping continued all the way, 
\Tstar$(p)$ would go to zero at $p$\,$\simeq$\,\pctwo, 
the critical doping where \Tc{} goes to zero at high doping.
In Fig.~\ref{YBCO-LSCO-PhaseDiag}, we see that this is not the case, and the pseudogap phase instead comes to a rather abrupt end,
with \Tstar$(p)$ dropping precipitously to zero at \pstar, well below \pctwo. 
In Nd-LSCO, \Tstar$(p)$ extends up to $p \simeq 0.23$ (Fig.~\ref{LSCO-phasediag}), 
and only then does it drop suddenly to zero at \pstar\,=\,$0.23$~[\onlinecite{Collignon2017},\onlinecite{Cyr-Choiniere2010}],
slightly (but distinctly) below \pctwo\,$\simeq$\,$0.27$. 
In LSCO, \Tstar$(p)$  follows the very same line as in Nd-LSCO, up to $p$\,$\simeq$\,$0.16$, but then, in striking contrast,
it starts to drop at $p$\,=\,$0.17$ and goes to zero at \pstar\,$\simeq$\,$0.18$ (Fig.~\ref{YBCO-LSCO-PhaseDiag}).
The difference between those two materials is seen most clearly in their normal-state resistivity (measured to low $T$ in high fields):
in Nd-LSCO, $\rho(T)$ shows a huge upturn at $p$\,=\,$0.20$ and 0.22, for example~[\onlinecite{Collignon2017}],
while in LSCO $\rho(T)$ remains linear down to $T$\,$\to$\,$0$ at $p$\,=\,$0.18$ and 0.21~[\onlinecite{Cooper2009}] 
(see Fig.~\ref{rho-YBCO-LNSCO}).

This raises a crucial, and largely unexplored question: what controls the location of \pstar?
And specifically: why is \pstar{} so much higher in Nd-LSCO than in LSCO, 
when \Tstar$(p)$ is otherwise the same (below $p$\,$\simeq$\,$0.17$)?
An answer to these new questions could elucidate the fundamental nature of the pseudogap phase.
A potential ingredient in the answer is the interesting observation~[\onlinecite{Benhabib2015}] made in Bi-2212 that the end of the pseudogap phase
in the normal state (above \Tc) coincides with the (Lifshitz) transition that changes the topology of the Fermi surface 
(in one of the two CuO$_2$ planes of the bi-layer~[\onlinecite{Kaminski2006}]), from hole-like below to electron-like above
the critical doping \pFS\,=\,$0.225$~at which the van Hove singularity crosses the Fermi level~[\onlinecite{Kaminski2006}].
The idea would be that the pseudogap cannot form on an electron-like Fermi surface. 
This is consistent with data on LSCO~[\onlinecite{Yoshida2006}] and Nd-LSCO~[\onlinecite{Matt2015}] 
and, to our knowledge, no data on any cuprate contradicts this idea. 
This scenario requires further investigation.​
%
%
%Of course, this could simply be a coincidence.
%The idea would be that the pseudogap cannot form on an electron-like Fermi surface.
%This constraint would be presumably be strongest
%for single-layer materials, such as LSCO and Nd-LSCO (and Eu-LSCO and Bi-2201).
%
%To our knowledge, no data on any cuprate  contradicts this idea.
%
%It is verified  in both LSCO and Nd-LSCO
%(and in Bi2201 as well~[\onlinecite{Kawasaki2010},\onlinecite{Kondo2004}]). 
%In LSCO, for example, ARPES measurements show that
%\pFS\,$<$\,$0.22$ and there is no pseudogap at $p = 0.22$~[\onlinecite{Yoshida2006}].
%
%Similarly, in Nd-LSCO 
%\pFS\,$<$\,$0.24$ and no pseudogap is present at $p = 0.24$~[\onlinecite{Matt2015}].
%
%It is therefore possible that the significantly lower \pstar~in LSCO is imposed by the constraint of a lower \pFS.}
%
%This scenario requires further investigation.

%%%%%%%%%%%%%%%%%%%%%%%%%%%%%%%%%%%%%%%%%%%%%%%%%%%%%%%%%%%%%%%%%%%%%%%%%%%%

\subsection{Orders inside the pseudogap phase}

In hole-doped cuprates, a number of phases, sometimes with only short-range order, exist in the underdoped region of the phase diagram.
%All of these phases are confined within the pseudogap phase: they onset either at or below \Tstar, and either at or below \pstar.
%
%The only exception is the superconducting phase, which persists beyond \pstar.
%
Here we discuss four of the main phases that have been detected experimentally.

%%%%%%%%%%%%%%%%%%%%%%%%%%%%%%%%%%%%%%%%%%%%%%%%%%%%%%%%%%%%%%%%%%%%%%%%%%%%

\subsubsection{Spin density wave}

\begin{figure}
\centering
\includegraphics[width=0.46\textwidth]{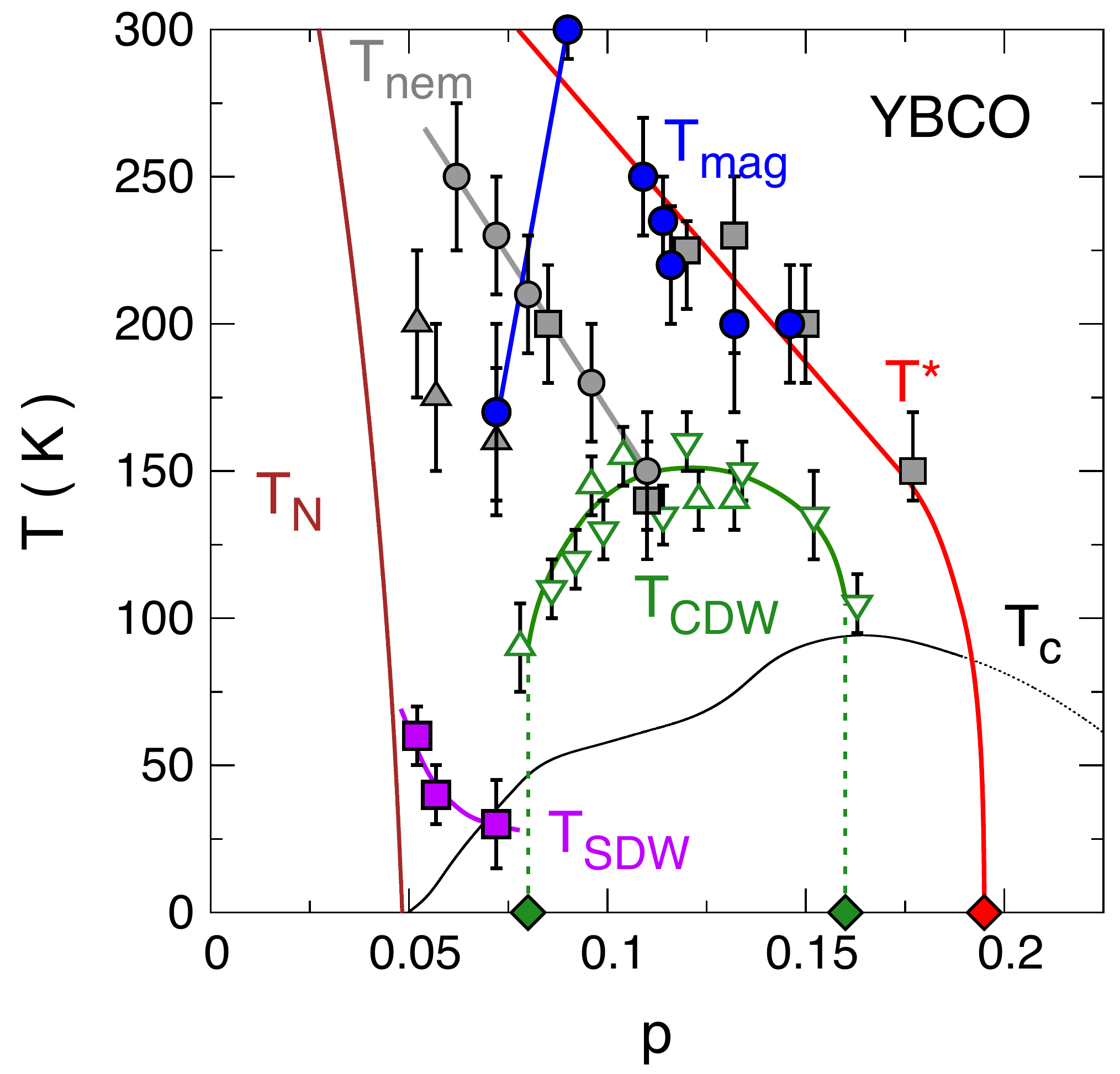}
\caption{(Color online)  
Temperature-doping phase diagram of YBCO
showing the N\'{e}el temperature \TN{} (brown line),
the superconducting transition temperature \Tc{} (black line),
and the pseudogap temperature \Tstar{} (red line) and critical point \pstar{} (red diamond), 
all from Figs.~\ref{YBCO-phasediag}~and~\ref{YBCO-LSCO-PhaseDiag}.
In addition, we show the charge-density-wave phase (CDW; green), delineated by the temperature \TCDW~below which 
short-range CDW correlations are detected by X-ray diffraction (up triangles~[\onlinecite{Hucker2014}]; down triangles~[\onlinecite{Blanco-Canosa2014}]). 
The two green diamonds mark the critical dopings at which the CDW phase begins (\pcdwone\,=\,$0.08$~[\onlinecite{LeBoeuf2011}]) 
and ends (\pcdwtwo\,=\,$0.16$~[\onlinecite{Badoux2016}])
at $T$\,=\,$0$ in the absence of superconductivity, as detected by high-field Hall effect measurements.
$T_{\rm SDW}$ (purple squares) marks the temperature below which incommensurate short-range spin-density-wave (SDW) correlations
are detected by neutron diffraction (in zero field)~[\onlinecite{Haug2010}].
Grey symbols mark \Tnem, the onset temperature of nematicity, an electronic in-plane anisotropy
detected in the resistivity~(circles~[\onlinecite{Ando2002},\onlinecite{Cyr-Choiniere2015}]), 
the Nernst coefficient (squares~[\onlinecite{Daou2010},\onlinecite{Cyr-Choiniere2015}]) and 
the spin fluctuation spectrum measured by inelastic neutron scattering (triangles~[\onlinecite{Haug2010}]). 
\Tmag~(blue circles) is the onset temperature of intra-unit-cell magnetic order
detected by polarized neutron diffraction~[\onlinecite{Fauque2006,Sidis2013,Mangin-Thro2017}]. 
The blue line highlights the drop in \Tmag~below $p$\,=\,$0.09$.
}
\label{YBCO-PhaseDiag-CDW}
\end{figure}

\begin{figure}
\centering
\includegraphics[width=0.48\textwidth]{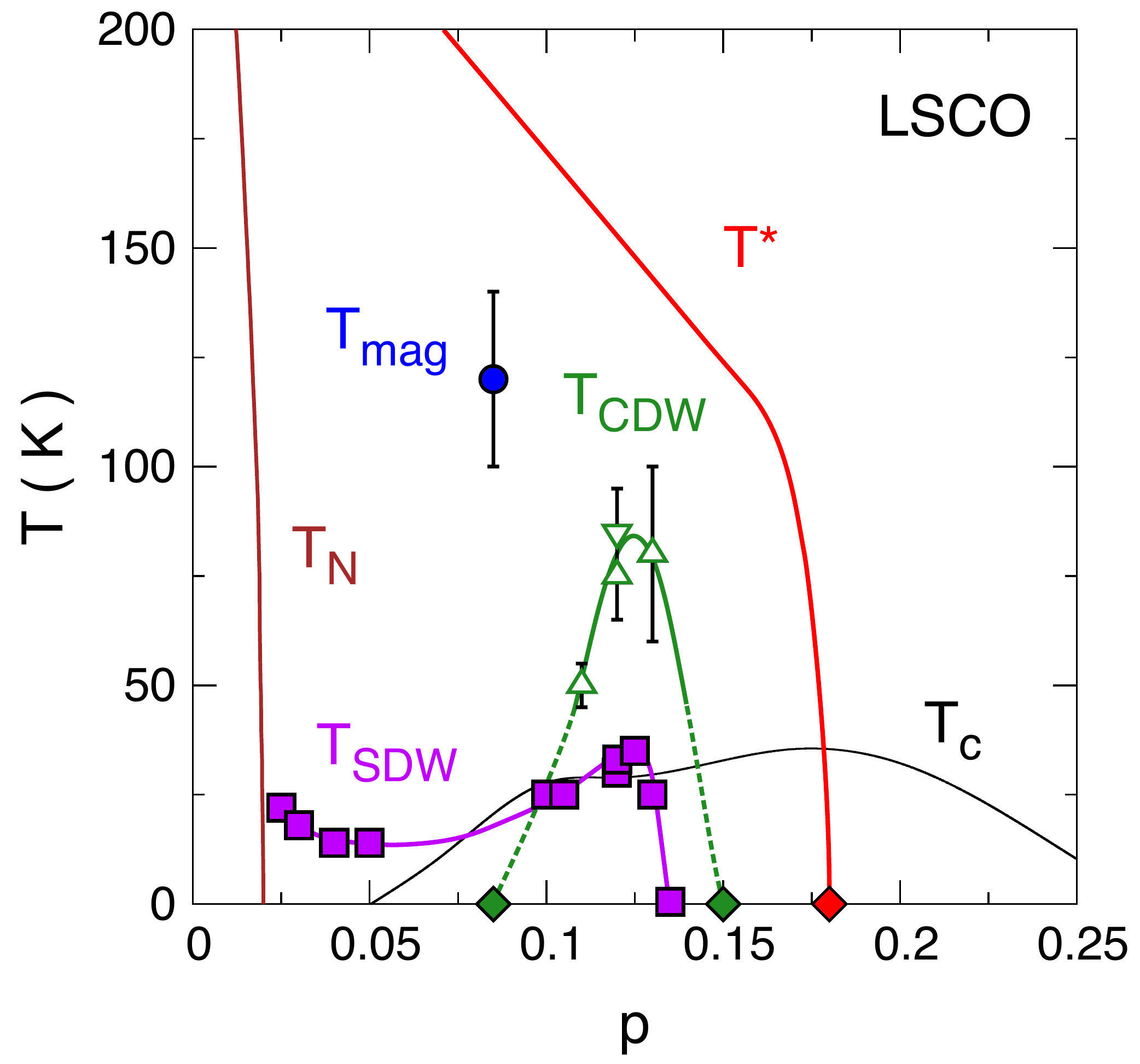}
\caption{(Color online) 
Temperature-doping phase diagram of LSCO 
showing the N\'{e}el temperature \TN{} (brown line),
the superconducting transition temperature \Tc{} (black line),
and the pseudogap temperature \Tstar{} (red line) and critical point \pstar{} (red diamond), 
all from Figs.~\ref{LSCO-phasediag}~and~\ref{YBCO-LSCO-PhaseDiag}.
In addition, we show the charge-density-wave phase (CDW; green), delineated by the temperature \TCDW{} below which 
short-range CDW correlations are detected by X-ray diffraction (up triangles~[\onlinecite{Croft2014}]; down triangles~[\onlinecite{Christensen2014}]). 
The two green diamonds mark the critical dopings at which the CDW phase begins (\pcdwone\,=\,$0.085$) and ends (\pcdwtwo\,=\,$0.15$) 
at $T$\,=\,$0$ in the absence of superconductivity, as detected by high-field thermopower measurements~[\onlinecite{Badoux2016a}]. 
$T_{\rm SDW}$ (purple squares) marks the temperature below which incommensurate short-range SDW order
is detected by neutron diffraction~[\onlinecite{Chang2008,Kofu2009,Wakimoto1999,Lake2002,Kimura1999}]. 
The blue circle at $p$\,=\,$0.085$ marks \Tmag, the onset temperature of intra-unit-cell magnetic order, 
detected by polarized neutron diffraction~[\onlinecite{Baledent2010}]. 
}
\label{LSCO-PhaseDiag-CDW}
\end{figure}

Long-range commensurate antiferromagnetic (AF) order dies out quickly with increasing $p$:
 \TN~goes to zero at the critical doping \pN\,=\,$0.05$ in YBCO and \pN\,$\simeq$\,$0.02$ in LSCO (Fig.~\ref{YBCO-LSCO-PhaseDiag}).
 Beyond \pN, incommensurate spin-density-wave (SDW) order is observed at low $T$, with correlation lengths that vary from rather short to fairly long amongst the various
 cuprates.
 In YBCO, short-range SDW order is observed up to \psdw\,$\simeq$\,$0.07$ in zero field (purple squares, Fig.~\ref{YBCO-PhaseDiag-CDW}).
 It stops when charge-density-wave (CDW) order starts, at \pcdwone\,$\simeq$\,$0.08$, evidence that the two orders compete
 (arguably because their periods do not match~[\onlinecite{Nie2017}]).
 In LSCO, SDW order extends up to \psdw\,$\simeq$\,$0.13$ in zero field (purple squares, Fig.~\ref{LSCO-PhaseDiag-CDW}),
 and it coexists with CDW order, evidence that the two orders do not compete (arguably because their periods match~[\onlinecite{Nie2017}]).
 A magnetic field which suppresses superconductivity enhances SDW order in both YBCO and LSCO~[\onlinecite{Khaykovich2005}].
 In LSCO, a field of 15\,T pushes the SDW critical point up to \psdw\,$\simeq$\,$0.15$~[\onlinecite{Chang2008}]. 
 Extrapolating to higher fields, it is conceivable that 
 \psdw\,=\,\pstar\,$\simeq$\,$0.18$ at $H$\,=\,\Hc\,$\simeq$\,$60$\,T.
 In other words, when the competing superconductivity is fully suppressed by a field, SDW order in LSCO
 could extend up to \pstar, \ie{} the non-superconducting ground state of the pseudogap phase could
 host SDW order.
This is confirmed by $\mu$SR studies on LSCO with Zn impurities used to suppress superconductivity, where magnetism is detected
up to $p = 0.19 \pm 0.01$~[\onlinecite{Kimura2003,Panagopoulos2003,Panagopoulos2004}].

%%%%%%%%%%%%%%%%%%%%%%%%%%     FIGURE 18   %%%%%%%%%%%%%%%%%%%%%%%%%%%%%%%%%%%%%%
%
%
%%%%%%%%%%%%%%%%%%%%%%%%%%%%%%%%%%%%%%%%%%%%%%%%%%%%%%%%%%%%%%%%%%%%%%%

%%%%%%%%%%%%%%%%%%%%%%%%%%     FIGURE 19   %%%%%%%%%%%%%%%%%%%%%%%%%%%%%%%%%%%%%%
%
%
%%%%%%%%%%%%%%%%%%%%%%%%%%%%%%%%%%%%%%%%%%%%%%%%%%%%%%%%%%%%%%%%%%%%%%%

This is also established in the case of Nd-LSCO, where magnetic Bragg peaks are detected by neutron diffraction~[\onlinecite{Tranquada1997}] 
up to $p$\,=\,$0.20$ and their onset temperature \TSDW{} and intensity both go to zero at 
$p$\,$\to$\,\pstar\,=\,$0.23$\,$\pm$\,$0.01$.
In Nd-LSCO, superconductivity is much weaker than in LSCO and a magnetic field is not needed
to help SDW order win the competition. 
Hence the magnetic Bragg peaks do not depend on field~[\onlinecite{Chang2008}]. 
Note, however, that the magnetism in Nd-LSCO at $p$\,=\,$0.20$ may not be fully static,
as it is not detected by $\mu$SR~[\onlinecite{Nachumi1998}]. 
 
In YBCO, suppressing superconductivity with a large field does not induce SDW order
in the range where there is CDW order, \ie{} between \pcdwone\,=\,$0.08$ and \pcdwtwo\,=\,$0.16$~[\onlinecite{Wu2013}]. 
However, adding Zn impurities to suppress superconductivity, \eg{} at $p$\,$\simeq$\,$0.12$, also suppresses CDW order, 
and this nucleates SDW order~[\onlinecite{Blanco-Canosa2013}]. 
In other words, there is a three-way phase competition. 
It is then conceivable that between \pcdwtwo{} and \pstar, SDW order 
could emerge if superconductivity is fully suppressed,
as we have proposed above for LSCO.
In YBCO, this would require fields of order $150$\,T, the maximal value of \Hc~[\onlinecite{Grissonnanche2014}]. 

In summary, magnetic order (AF or SDW) at low $T$ is ubiquitous in hole-doped cuprates and it may well 
exist at all dopings from $p$\,=\,$0$ up to \pstar{} when it is not suppressed by competition from 
superconductivity or CDW order.
It is therefore an important property of the pseudogap phase at $T$\,$\to$\,$0$ -- a second link between pseudogap and 
antiferromagnetism (the first being \Tstar\,$\simeq$\,\TN{} at $p$\,$\to$\,$0$).
Having said this, the pseudogap phase is not simply a phase of SDW order, since \TSDW\,$\ll$\,\Tstar~(Figs.~\ref{YBCO-PhaseDiag-CDW}~and~\ref{LSCO-PhaseDiag-CDW}).

%%%%%%%%%%%%%%%%%%%%%%%%%%%%%%%%%%%%%%%%%%%%%%%%%%%%%%%%%%%%%%%%%%%%%%%%%%%%
%%%%%    TABLE 1    %%%%%%%%%%%%%%%%%%%%%%%%%%%%%%%%%%%%%%%%%%%%%%%%%%%%%%%%%%%%%%%%
%%%%%%%%%%%%%%%%%%%%%%%%%%%%%%%%%%%%%%%%%%%%%%%%%%%%%%%%%%%%%%%%%%%%%%%%%%%%

\setlength{\tabcolsep}{15pt}

\begin{table*}
%[t]
\caption{
Critical dopings for the four cuprate materials discussed in this article,
measured at low temperature ($T$\,$\to$\,$0$).
The pseudogap critical point \pstar{} and the beginning and end of the CDW region, at \pcdwone{} and \pcdwtwo{} respectively, 
were measured in the normal state, reached by suppressing superconductivity with a large magnetic field.
The end of the SDW phase, at \psdw, is given here for zero field.
The doping \pFS{} at which the van Hove singularity occurs is determined by ARPES.
It is the doping where the large hole-like Fermi surface of overdoped cuprates undergoes a (Lifshitz) transition
to a large electron-like Fermi surface upon increasing $p$.
All single numbers with two (three) significant digits have an error bar $\pm$\,$0.01$ ($\pm$\,$0.005$).
When a doping interval is given, the critical doping is located inside that interval.
Information on how the critical dopings were defined can be found in the associated references.
%(and references therein).
}
\centering
\begin{tabular}{ l c  c  c  c  c }
\\  
\hline
%   Material        p*              p_FS          p_SDW    p_1^CDW      p_2^CDW
\\       
      Material   & 	\pstar    &	  \pFS       &    \psdw  &  \pcdwone  &  \pcdwtwo 	\\
\\ 
% & & (K) & (T) & $(\mu\Omega$~cm) & \multicolumn{2}{c|}{($\mu$W/K$^2$~cm)} &	 \\
		\hline\hline
%  Material           p*                       p_vHs                            p_SDW               p_1^CDW       p_2^CDW      
\\ 
YBCO	&	0.195~[\onlinecite{Badoux2016}]		&	?  & 0.07~[\onlinecite{Haug2010}]    &  0.08~[\onlinecite{LeBoeuf2011}]   &  0.16~[\onlinecite{Badoux2016}] \\ 
\\ 
LSCO	& 0.18~[\onlinecite{Laliberte2016}]		& 0.17\,-\,0.22~[\onlinecite{Yoshida2006},\onlinecite{Chang2008a}] 	& 0.13~[\onlinecite{Chang2008},\onlinecite{Kofu2009}]   & 0.085~[\onlinecite{Badoux2016a}]     &  0.15~[\onlinecite{Badoux2016a}]     \\
\\ 
Nd-LSCO	& 0.23~[\onlinecite{Collignon2017}]		&  0.20\,-\,0.24~[\onlinecite{Matt2015}]	&	0.24~[\onlinecite{Tranquada1997}]    &  ?	&  0.15\,-\,0.20~[\onlinecite{Daou2009a}] \\
\\ 
Eu-LSCO	& 0.24~[\onlinecite{Laliberte2011}]   &   ?      &          ?   & 0.09~[\onlinecite{Laliberte2011}]  &   0.16\,-\,0.21~[\onlinecite{Laliberte2011}]    \\      
\\ 
%      
%\hline
%0.127  &A  & 8.4 	&15		&1500	 &16.3	&5.6   &0.34			\\			%9270
%0.127  &B  & 9.3 	&15		&1130	 &21.7	&6.8   &0.31			\\		 	%9271
\hline
\end{tabular}
	
\label{table1}
\end{table*}

%%%%%%%%%%%%%%%%%%%%%%%%%%%%%%%%%%%%%%%%%%%%%%%%%%%%%%%%%%%%%%%%%%%%%%%%%%

%%%%%%%%%%%%%%%%%%%%%%%%%%%%%%%%%%%%%%%%%%%%%%%%%%%%%%%%%%%%%%%%%%%%%%%%%%%%

%%%%%%%%%%%%%%%%%%%%%%%%%%%%%%%%%%%%%%%%%%%%%%%%%%%%%%%%%%%%%%%%%%%%%%%%%%%%

 \subsubsection{Charge density wave}

Twenty years ago, CDW order was first detected in cuprates by neutron diffraction, in Nd-LSCO and LBCO at $p$\,$\simeq$\,$0.12$~[\onlinecite{Tranquada1995}]. 
Five years later, it was seen via STM in Bi-2212~[\onlinecite{Hoffman2002},\onlinecite{Howald2003}]. 
Another five years later, CDW order was first sighted in YBCO via its effect on the Fermi surface, reconstructed into 
small electron pockets~[\onlinecite{Doiron-Leyraud2007,LeBoeuf2007,Taillefer2009,Chang2010,Laliberte2011,LeBoeuf2011}], 
and then observed directly via NMR~[\onlinecite{Wu2013},\onlinecite{Wu2011}] 
and X-ray diffraction (XRD)~[\onlinecite{Ghiringhelli2012},\onlinecite{Chang2012a}].
In addition to YBCO, CDW order has been observed by XRD in
Nd-LSCO~[\onlinecite{Zimmermann1998},\onlinecite{Niemoller1999}],
Eu-LSCO~[\onlinecite{Fink2009},\onlinecite{Fink2011}],
LSCO~[\onlinecite{Croft2014},\onlinecite{Christensen2014}],
Hg-1201~[\onlinecite{Tabis2014}],
Bi-2212~[\onlinecite{daSilvaNeto2014}],
and Bi-2201~[\onlinecite{Comin2014}].
It is typically strongest at $p$\,$\simeq$\,$0.12$
and confined to a region entirely inside the pseudogap phase, between two critical dopings:
\pcdwone{} at low doping and \pcdwtwo{} at high doping.
For the four materials of particular focus here, all evidence to date indicates that \pcdwtwo{} is well below \pstar,
(see Table~\ref{table1} and Figs.~\ref{YBCO-PhaseDiag-CDW}~and~\ref{LSCO-PhaseDiag-CDW}).
This immediately implies that the pseudogap phase is not a phase of CDW order,
nor is it a high-temperature precursor of that order.
This is confirmed by the fact that the onset temperature of CDW order in these same materials is a dome peaked at $p$\,$\simeq$\,$0.12$,
while \Tstar{} rises monotonically with decreasing $p$ (Figs.~\ref{YBCO-PhaseDiag-CDW}~and~\ref{LSCO-PhaseDiag-CDW}).

In other cuprates, the location of \pcdwtwo{} and \pstar{} is still not fully established.
In Bi-2212, STM studies at $T$\,$\simeq$\,$10$\,K (below \Tc) detect CDW modulations up to $p$\,=\,$0.17$ 
and a transition from Fermi arcs (with pseudogap) at $p$\,=\,$0.17$ to a complete large Fermi surface (without pseudogap)
at $p$\,=\,$0.20$~[\onlinecite{Fujita2014}]. 
In other words, \pcdwtwo\,$\simeq$\,\pstar\,=\,$0.185$\,$\pm$\,$0.015$. 
However, normal-state measurements of the pseudogap (above \Tc),
such as ARPES and Raman, find \pstar\,=\,$0.22$\,$\pm$\,$0.01$~[\onlinecite{Vishik2012},\onlinecite{Benhabib2015}]. 
Given this uncertainty, it seems possible that 
\pstar\,$\simeq$\,\pcdwtwo\,$+$\,$0.03$,
much as in YBCO and LSCO (Table~\ref{table1} ).

We infer that CDW ordering is a secondary instability of the pseudogap phase.
Two open questions are 
why it tends to peak at $p$\,$\simeq$\,$0.12$ and 
why its onset at $T$\,=\,$0$ is delayed relative to \pstar.

\subsubsection{Nematicity}

In orthorhombic YBCO, the in-plane resistivity is anisotropic because the CuO chains that run along the $b$ axis 
conduct. But in addition to this chain-related anisotropy, another anisotropy emerges upon cooling at low doping~[\onlinecite{Ando2002}].
The onset of this additional anisotropy, which we will call nematicity, is at a temperature \Tnem~that runs 
parallel to \Tstar, some 100\,K below (Fig.~\ref{YBCO-PhaseDiag-CDW}).
\Tnem{} coincides with the inflexion point in $\rho_a(T)$~[\onlinecite{Cyr-Choiniere2015}], 
 \ie{} the white line that separates the red and blue regions
in the curvature map of Fig.~\ref{Ando-map-YBCO}.
Not surprisingly, this anisotropy is also detected in the Nernst coefficient~[\onlinecite{Cyr-Choiniere2015}].

Close to the \Tnem{} line in the phase diagram at low doping, an anisotropy develops in the spin fluctuation spectrum,
detected by inelastic neutron scattering as a splitting in the peak at $Q$\,=\,$(\pi,\pi)$ that appears for one direction 
and not the other~[\onlinecite{Haug2010}].
This ``spin nematicity'' may be responsible for the transport anisotropy below \Tnem.

Similarly, a ``charge nematicity'' is observed in the region of CDW order, at higher doping~[\onlinecite{Cyr-Choiniere2015}]. 
Here, the onset of nematicity occurs at $T$\,$\simeq$\,\Tstar~[\onlinecite{Daou2010}].
In other words, at temperatures above the SDW and CDW orders, there is a region of enhanced nematic susceptibility,
possibly associated with the precursor fluctuations of these two orders~[\onlinecite{Schutt2015}].

There are three problems with equating this nematic phase with the pseudogap phase.
The first is that \Tnem\,$<$\,\Tstar~at $p < 0.11$. 
The second is that nematic order does not open a gap (or a pseudogap).
The third is that nematic order does not cause a change in carrier density, and so cannot explain the 
main signature of \pstar.
But again, nematicity may well be a secondary instability of the pseudogap phase.
Or the pseudogap may cause an enhanced nematic susceptibility~[\onlinecite{Okamoto2010}].

%%%%%%%%%%%%%%%%%%%%%%%%%%%%%%%%%%%%%%%%%%%%%%%%%%%%%%%%%%%%%%%%%%%%%%%%%%%%

\subsubsection{Intra-unit cell magnetic order}

In the cuprates YBCO, Hg-1201 and Bi-2212, magnetic order has been detected by polarized neutron
diffraction, with an onset temperature \Tmag{} that coincides roughly with \Tstar.
This intra-unit-cell (IUC) order has a wavevector $Q$\,=\,$0$.
In Fig.~\ref{YBCO-PhaseDiag-CDW}, we reproduce the reported values of \Tmag{} for YBCO~[\onlinecite{Fauque2006,Sidis2013,Mangin-Thro2017}]. 
We see that in the range $0.09$\,$\leq$\,$p$\,$\leq$\,$0.15$, \Tmag\,=\,\Tstar, within error bars.
However, at lower doping ($p$\,$\simeq$\,$0.08$), the IUC signal weakens and it onsets
at a significantly lower temperature: \Tmag\,=\,$170$\,$\pm$\,$20$\,K~[\onlinecite{Sidis2013}], 
while \Tstar\,=\,$280$\,$\pm$\,$20$\,K (Figs.~\ref{YBCO-phasediag}~and~\ref{nu-YBCO-6p45}). 
It has been suggested that the weakening of the IUC magnetic order in YBCO at low $p$
may be due to a competition with SDW order (or correlations) that develops below the CDW phase,
\ie{} at $p$\,$<$\,\pcdwone\,=\,$0.08$. 
However, the pseudogap does not weaken at $p$\,$<$\,\pcdwone. 
Indeed, \Tstar{} is higher in our sample with $p$\,=\,$0.078$, clearly below the CDW region
(\ie{} with a positive Hall coefficient at low $T$)~[\onlinecite{LeBoeuf2011}], 
than it is in our sample with  $p$\,=\,$0.085$,
a doping above \pcdwone{} (Fig.~\ref{nu-YBCO-6p45}).

A similar discrepancy is observed in LSCO at $p$\,=\,$0.085$,
where \Tmag\,=\,$120$\,$\pm$\,$20$\,K,
while \Tstar\,=\,$185$\,$\pm$\,$20$\,K (Fig.~\ref{LSCO-PhaseDiag-CDW}). 
This weakening at low $p$ suggests that the IUC magnetic order
is more likely to be a secondary instability of the pseudogap phase, rather than its primary cause.
Note that as in the case of nematic order, another $Q$\,=\,$0$ order, it is difficult to see
how the IUC order can open a gap (or a pseudogap) and cause a change in carrier density across \pstar.

%%%%%%%%%%%%%%%%%%%%%%%%%%     FIGURE 20   %%%%%%%%%%%%%%%%%%%%%%%%%%%%%%%%%%%%%%
%
\begin{figure}[!t]
\centering
\includegraphics[width=0.4\textwidth]{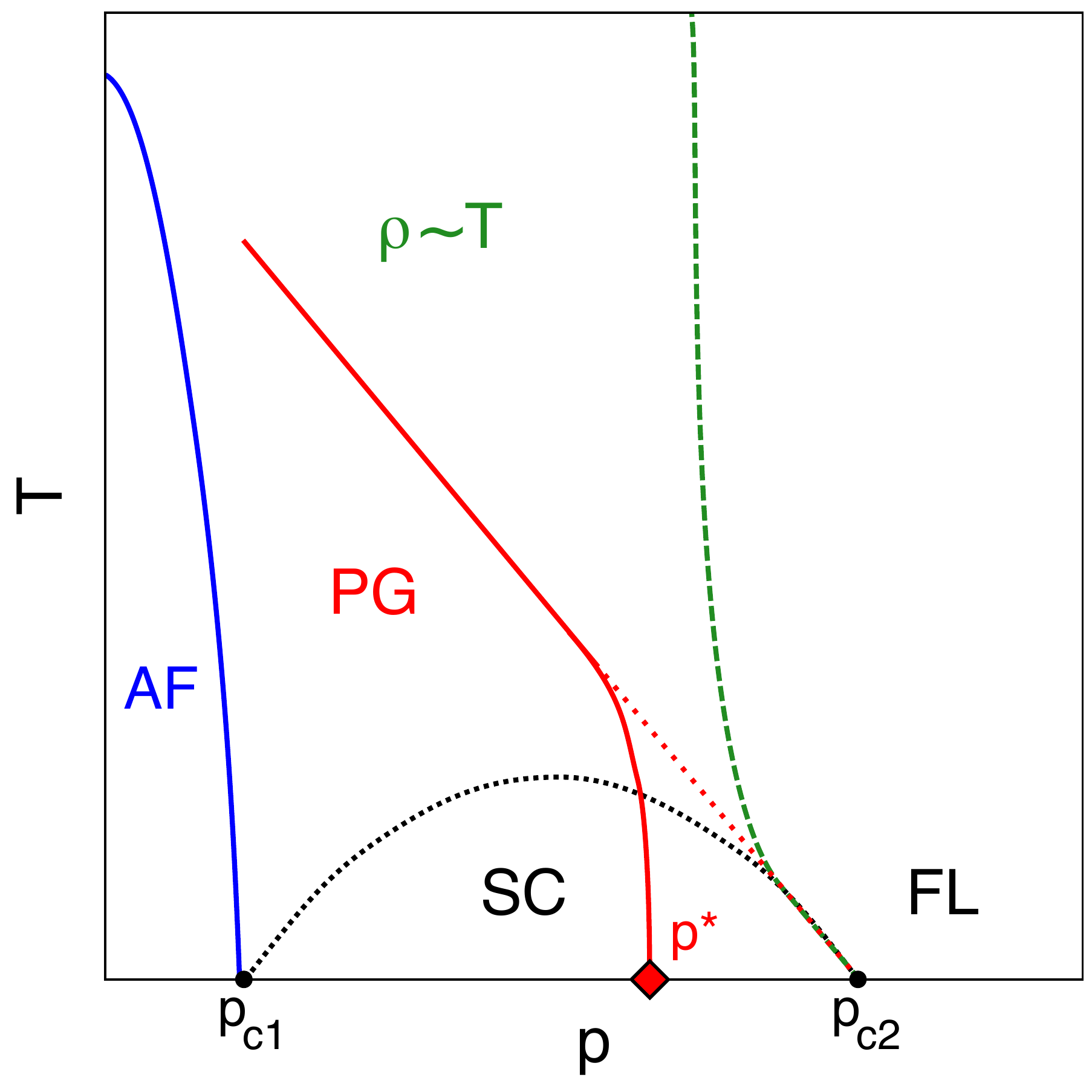}
\caption{(Color online)  
Schematic phase diagram of cuprates, showing 
the antiferromagnetic phase (AF, below the blue line),
the pseudogap phase (PG, below the red line),
and the superconducting dome between \pcone{} and \pctwo{} (SC, below the black dotted line).
As in Fig.~\ref{YBCO-LSCO-PhaseDiag}, a linear extension of the \Tstar{} line (red) extrapolates to zero at \pctwo.
In the region between the solid red line and the dashed green line, 
the normal-state resistivity $\rho(T)$ of cuprates is predominantly linear in temperature.
This linearity appears below \pctwo, along with the superconductivity (see text).
At $T \to 0$, it persists down to \pstar.
Above \pctwo, $\rho$\,$\propto$\,$T^2$ at low $T$, the signature of a Fermi-liquid-like metallic state (FL).
The green dashed line is drawn to capture the behavior of the linearity of resistivity as observed 
in LSCO in Fig.\,3 of Ref.~\onlinecite{Cooper2009} (the so-called ``red foot'' contour plot).
}
\label{PhaseDiag-sketch}
\end{figure}
%
%%%%%%%%%%%%%%%%%%%%%%%%%%%%%%%%%%%%%%%%%%%%%%%%%%%%%%%%%%%%%%%%%%%%%%%

%%%%%%%%%%%%%%%%%%%%%%%%%%%%%%%%%%%%%%%%%%%%%%%%%%%%%%%%%%%%%%%%%%%%%%%%%%%%

\subsection{Superconductivity}

Unlike the four phases discussed previously, which are all confined to the left of \pstar{} (and below \Tstar),
the superconducting phase extends beyond the pseudogap critical point.
The region of superconductivity in the phase diagram of cuprates is always a dome,
which starts at \pcone{} and ends at \pctwo, at low and high doping, respectively.
And this dome straddles \pstar, \ie{} \pcone\,$<$\,\pstar{} and \pctwo\,$>$\,\pstar, 
as we saw for YBCO, LSCO and Nd-LSCO (Fig.~\ref{YBCO-LSCO-PhaseDiag}).
The precise value of \pctwo{} may depend on the material, as does the precise value of \pstar; 
in LSCO and Nd-LSCO, \pctwo\,$\simeq$\,$0.27$, 
while \pctwo\,$\simeq$\,$0.31$ in Tl-2201~[\onlinecite{Bangura2010}]
and 
\pctwo\,$\simeq$\,$0.43$ in Bi-2201~[\onlinecite{Kondo2004}].
%The difference may be due to the pair-breaking effect of impurities or disorder,
%more severe in LSCO than in Tl2201.

Coming from high $p$, superconductivity with an order parameter of $d_{x^2-y^2}$ symmetry
emerges out of a Fermi-liquid-like metallic state,
characterized by a single large coherent hole-like Fermi surface~[\onlinecite{Vignolle2008}], 
with no pseudogap and no broken symmetry of any kind.
The big question is:
what electron-electron interaction in this simple-looking state causes the electrons to pair?
The phase diagrams in Fig.~\ref{YBCO-LSCO-PhaseDiag} may provide some clues.
We already pointed out that a linear extrapolation of the 
\Tstar$(p)$ line reaches $T$\,=\,$0$
at $p$\,$\simeq$\,\pctwo, suggesting that the same interactions which favour pairing
may also be responsible for the pseudogap. 

It turns out that \pctwo{} is also the onset of a third manifestation of electron-electron interactions:
the appearance of a linear term in the temperature dependence of the resistivity $\rho(T)$,
as sketched in Fig.~\ref{PhaseDiag-sketch}.
Detailed studies in overdoped Tl-2201~[\onlinecite{Manako1992,Mackenzie1996,Proust2002}] 
and LSCO~[\onlinecite{Cooper2009},\onlinecite{Nakamae2003}] reveal that 
a linear-$T$ term appears in $\rho(T)$ as soon as $p$\,$<$\,\pctwo,
while $\rho$\,$\propto$\,$T^2$ at $p$\,$>$\,\pctwo.
This empirical link between linear-$T$ resistivity and \Tc{}~[\onlinecite{Abdel-Jawad2007}] 
suggests that the interactions that cause the anomalous inelastic scattering also cause pairing~[\onlinecite{Taillefer2010}]. 
A similar link has been observed in iron-based and organic superconductors~[\onlinecite{Doiron-Leyraud2009}], 
materials whose phase diagrams consist of an antiferromagnetic quantum critical point (QCP)
surrounded by a dome of superconductivity. 
In both cases, the scattering and the pairing are attributed to antiferromagnetic spin fluctuations.

%%%%%%%%%%%%%%%%%%%%%%%%%%%%%%%%%%%%%%%%%%%%%%%%%%%%%%%%%%%%%%%%%%%%%%%%%%%%

In summary, three fundamental phenomena of cuprates emerge together below \pctwo:
superconductivity, pseudogap and anomalous scattering.
(Strictly speaking, the pseudogap opens slightly below \pctwo, at \pstar, but in some cases, such
as Nd-LSCO and Bi-2201, the separation is small:
\pstar\,=\,$0.23$ vs \pctwo\,$\simeq$\,$0.27$~[\onlinecite{Daou2009}], and 
\pstar\,=\,$0.38$ vs \pctwo\,$\simeq$\,$0.43$~[\onlinecite{Kawasaki2010},\onlinecite{Kondo2004}],
respectively.) 
Fig.~\ref{PhaseDiag-sketch} suggests another way to summarize the situation.
The two fundamental phases of cuprates -- superconductivity and pseudogap -- are both instabilities 
of a normal state that is characterized by a linear-$T$ resistivity.
Given that a linear-$T$ resistivity is generally observed on the border of antiferromagnetic order
and attributed to scattering by antiferromagnetic spin fluctuations, it is tempting to associate both the pseudogap 
and the $d$-wave superconductivity in cuprates to antiferromagnetic correlations (perhaps short-ranged). 
In this scenario, the fact that \Tc{} falls at low $p$ while \Tstar{} continues to rise (Fig.~\ref{YBCO-LSCO-PhaseDiag}) is attributed to
the competition suffered by the superconducting phase from the full sequence of other phases (Figs.~\ref{YBCO-PhaseDiag-CDW}~and~\ref{LSCO-PhaseDiag-CDW}):
first, the pseudogap phase below \pstar, then the CDW, SDW and AF orders below \pcdwtwo,
\psdw{} and \pN, respectively.

%%%%%%%%%%%%%%%%%%%%%%%%%%%%%%%%%%%%%%%%%%%%%%%%%%%%%%%%%%%%%%%%%%%%%%%
%%%%%%%%%%%%%%%%%%%%%%%%%%%%%%%%%%%%%%%%%%%%%%%%%%%%%%%%%%%%%%%%%%%%%%%

%%%%%%%%%%%%%%%%%%%%%%%%%%%%%%%%%%%%%%%%%%%%%%%%%%%%%%%%%%%%%%%%%%%%%%%
%%%%%%%%%%%%%%%%%%%                   SUMMARY           %%%%%%%%%%%%%%%%%%%%%%%%%%%%%%%%%%%%%
%%%%%%%%%%%%%%%%%%%%%%%%%%%%%%%%%%%%%%%%%%%%%%%%%%%%%%%%%%%%%%%%%%%%%%%

\section{Summary}
\label{sec:Summary}

We have shown how the quasiparticle and superconducting contributions to the Nernst effect in cuprates
can be disentangled.
We observe that the latter contribution is only significant in a narrow region of temperature above \Tc,
which extends up to roughly $1.5$~\Tc, much as the region of paraconductivity observed in the resistivity.
We showed how the quasiparticle Nernst signal can be used to detect the onset of the pseudogap phase,
at a temperature \Tnu.
In YBCO, LSCO and Nd-LSCO, we find that \Tnu\,=\,\Trho, the temperature below which the resistivity
deviates from its linear-$T$ dependence at high temperature, a standard signature of the 
pseudogap temperature \Tstar, consistent with ARPES measurements of the pseudogap.
The advantage of using Nernst over resistivity is its much greater sensitivity to \Tstar. 
By comparing Nernst data in three La$_2$CuO$_4$-based cuprates (LSCO, Nd-LSCO and Eu-LSCO),
we find that they have the same \Tstar$(p)$ line (up to $p$\,$\simeq$\,$0.17$), 
independent of their different structures and structural transitions.

We arrive at the temperature-doping phase diagram of two major families of cuprates,
YBCO and LSCO, which reveal some qualitative similarities and quantitative differences.
Qualitatively, 
\Tstar($p$) decreases monotonically with $p$ in both families,
along a line that stretches between \TN{} at $p$\,=\,$0$, 
where \TN{} is the N\'eel temperature for the onset of long-range commensurate antiferromagnetic order in the Mott insulator,
and \pctwo{} at $T$\,=\,$0$,
where 
\pctwo{} is the end point of the superconducting dome at high doping.
These empirical links suggest that the pseudogap phase is related to antiferromagnetism
and that pseudogap and pairing arise from the same interactions.

Quantitatively,
we find that \Tstar$(p)$ is 1.5 times larger in YBCO than in LSCO, as is \TN(0). 
We also find that although \Tstar{} is the same in LSCO and Nd-LSCO, 
the critical doping at which the pseudogap phase ends abruptly is much lower
in LSCO, where \pstar\,$\simeq$\,$0.18$, than in Nd-LSCO, where \pstar\,=\,$ 0.23$.
A possible explanation for this significant difference is the constraint
that the pseudogap can only open once the Fermi surface has undergone
its Lifshitz transition through the van Hove singularity, from a large electron-like
surface above \pFS{} to a large hole-like surface below \pFS,
\ie{} the constraint that \pstar\,$\leq$\,\pFS.

We briefly discussed four phases that occur inside the pseudogap phase, namely
spin density wave (SDW), charge density wave (CDW), nematicity, and intra-unit-cell magnetic order.
We conclude that all four are likely to be secondary instabilities of the pseudogap phase,
as opposed to its driving mechanism or origin.

Finally, we show that the three primary phenomena of cuprates -- 
the pseudogap, 
the $d$-wave superconductivity
and the anomalous metallic behaviour (linear-$T$ resistivity) --
are found to all emerge together, below \pctwo.
In analogy with other families of materials -- such as iron-based, heavy-fermion and organic superconductors --
where linear-$T$ resistivity and superconductivity are observed on the border of antiferromagnetism,
we suggest that antiferromagnetic spin fluctuations\,/\,correlations may play a common underlying 
role in these three phenomena.

%%%%%%%%%%%%%%%%%%%%%%%%%%%%%%%%%%%%%%%%%%%%%%%%%%%%%%%%%%%%%%%%%%%%%%%
%%%%%%%%%%%%%%%%%%%%%%%%%%%%%%%%%%%%%%%%%%%%%%%%%%%%%%%%%%%%%%%%%%%%%%%%%%

%%%%%%%%%%%%%%%%%%%%%%%%%%%%%%%%%%%%%%%%%%%%%%%%%%%%%%%%%%%%%%%%%%%%%%%

\section{ACKNOWLEDGEMENTS}

We thank 
D.~S\'{e}n\'{e}chal and A.-M.~Tremblay for useful discussions and
J. Corbin for his assistance with the experiments.
L.T. thanks ESPCI-ParisTech, Universit\'{e} Paris-Sud, CEA-Saclay and the Coll\`{e}ge de France for their hospitality and support, 
%and the \'{E}cole Polytechnique (ERC-319286 QMAC) 
and the European Research Council (Grant ERC-319286 QMAC)% to Antoine Georges) 
and LABEX PALM (ANR-10-LABX-0039-PALM) for their support, while this article was written. 
O.C.C. was supported by a fellowship from the Natural Sciences and Engineering Research Council of Canada (NSERC). 
J.C. was supported by a fellowship from the Swiss National Science foundation.
J.-S.Z. and J.B.G. were supported by a US National Science Foundation grant.
H.T. acknowledges MEXT Japan for a Grant-in-Aid for Scientific Research.
R.L., D.A.B. and W.N.H. acknowledge funding from the Natural Sciences and Engineering Research Council of Canada (NSERC). 
L.T. acknowledges support from the Canadian Institute for Advanced Research (CIFAR) and funding from 
the Natural Sciences and Engineering Research Council of Canada (NSERC; PIN:123817), 
the Fonds de recherche du Qu\'{e}bec - Nature et Technologies (FRQNT), 
the Canada Foundation for Innovation (CFI),
and a Canada Research Chair. 
Part of this work was funded by the Gordon and Betty Moore Foundation's EPiQS Initiative (Grant GBMF5306 to L.T.).

%%%%%%%%%%%%%%%%%%%%%%%%%%%%%%%%%%%%%%%%%%%%%%%%%%%%%%%%%%%%%%%%%%%%%%%
%%%%%%%%%%%%%%%%%%%%%%%%%%%%%%%%%%%%%%%%%%%%%%%%%%%%%%%%%%%%%%%%%%%%%%%

%%%%%%%%%%%%%%%%%%%%%%%%%%%%%%%%%%%%%%%%%%%%%%%%%%%%%%%%%%%%%%%%%%%%%%%%%%
%%%%%%%%%%%%%%%%%%%%%%%             APPENDIX          %%%%%%%%%%%%%%%%%%%%%%%%%%%%%%%%%%%%%%
%%%%%%%%%%%%%%%%%%%%%%%%%%%%%%%%%%%%%%%%%%%%%%%%%%%%%%%%%%%%%%%%%%%%%%%%%%

\section{APPENDIX}
\label{sec:Appendix}

\subsection*{Nernst signal from superconducting fluctuations}

%%%%%%%%%%%%%%%%%%%%%%%%%%%%%%%%%%%%%%%%%%%%%%%%%%%%%%%%%%%%%%%%%%%%%%%

In this Article, our main focus is on the quasiparticle Nernst signal and how it can be used to detect the onset of
 the pseudogap phase.
We only discussed briefly how that signal can be disentangled from the superconducting Nernst signal.
In this Appendix, we provide further information on the superconducting Nernst signal in cuprates.
We focus on the field dependence of $\nu$ as a way to isolate
$\nu_{\rm qp}$ in YBCO and LSCO.
We end by analyzing how prior interpretations of the Nernst effect in cuprates 
led to the mistaken notion that essentially all the Nernst signal above \Tc{} is due to
superconducting fluctuations.

%%%%%%%%%%%%%%%%%%%%%%%%%%%%%%%%%%%%%%%%%%%%%%%%%%%%%%%%%%%%%%%%%%%%%%%
%%%%%%      Gaussian fluctuations
%%%%%%%%%%%%%%%%%%%%%%%%%%%%%%%%%%%%%%%%%%%%%%%%%%%%%%%%%%%%%%%%%%%%%%%

\subsection{Gaussian fluctuations}
\label{subsec:Appendix-Gaussian}

Recent Nernst measurements in the electron-doped cuprate  PCCO have been used to show
that a Gaussian theory of superconducting fluctuations can account qualitatively and quantitatively for the observed superconducting signal $N_{\rm sc}$~[\onlinecite{Tafti2014}]. 
Because \Hc{} is very small in PCCO (at most 10\,T), one can fully suppress superconducting fluctuations by applying a field $H$\,$\simeq$\,$15$\,T.
This enables one to directly obtain $N_{\rm qp}$, which can then be subtracted from $N$ to get $N_{\rm sc}$,
and compare this $N_{\rm sc}$ to theory~[\onlinecite{Ussishkin2002,Serbyn2009,Michaeli2009}]. 

The authors find no difference in the nature of the superconducting fluctuations on the underdoped side of the \Tc{} dome
relative to the overdoped side~[\onlinecite{Tafti2014}]. 
This shows that the decrease of \Tc{} at low doping is not due to a growth of phase fluctuations, 
as originally proposed~[\onlinecite{Emery1995}]. 
Rather, the drop in \Tc{} below optimal doping is associated with the critical point where the Fermi surface of PCCO undergoes
a reconstruction~[\onlinecite{Dagan2004}].

A similar study was performed in the hole-doped cuprate Eu-LSCO, in the underdoped regime~[\onlinecite{Chang2012}]. 
The Nernst signal $N_{\rm sc}$ is here also found to agree with Gaussian theory,
as in more conventional superconductors, such as NbSi~[\onlinecite{Pourret2006}].

We note, however, that spectroscopic studies of ARPES~[\onlinecite{Reber2013},\onlinecite{Kondo2015}] 
and STM~[\onlinecite{Lee2009},\onlinecite{Gomes2007}] (see section~\ref{subsubsec:ARPES-STM}) 
show a superconducting gap persisting well above \Tc{} -- a fact that is hard to reconcile with Gaussian (amplitude) fluctuations.

The quantitative question of how far in temperature (or in magnetic field) superconducting fluctuations
extend above \Tc{} (or above \Hc) is in some sense meaningless, for it clearly depends on the 
sensitivity of the probe.
In NbSi, for example, a superconducting Nernst signal was detected up to 30\,\Tc{} and 5\,\Hc~[\onlinecite{Pourret2006}]. 
Nevertheless, because the extent of the fluctuation regime in cuprate superconductors has been the subject of much debate, 
we further explore that question in the following sections.
We emphasize that in this article no assumption is made about the nature of the SC fluctuations above \Tc{} nor is any use made of Gaussian theory. 
Readers interested in learning whether Gaussian theory can describe the SC fluctuations measured in cuprates 
are referred to Refs. \onlinecite{Tafti2014} and \onlinecite{Chang2012}.

%%%%%%%%%%%%%%%%%%%%%%%%%%%%%%%%%%%%%%%%%%%%%%%%%%%%%%%%%%%%%%%%%%%%%%%
%%%%%%      YBCO - SC fluctuations
%%%%%%%%%%%%%%%%%%%%%%%%%%%%%%%%%%%%%%%%%%%%%%%%%%%%%%%%%%%%%%%%%%%%%%%

\subsection{Field dependence and \texorpdfstring{\TB}{TB}}
\label{subsec:Appendix-TB}

In YBCO, the separation of quasiparticle and superconducting contributions is straightforward 
because the former is negative (below \Tstar) and the latter is positive.
In Fig.~\ref{nu-YBCO-one8},
the minimum in $\nu$\,/\,$T$ vs $T$ at \Tmin{} provides an immediate measure of the temperature
below which the superconducting signal becomes important.
A plot of \Tmin{} vs $p$ on the phase diagram reveals that the region of significant superconducting fluctuations
closely tracks \Tc, with $T_{\rm min}$\,$\simeq$\,$1.4$\,$T_{\rm c}$~(Fig.~\ref{YBCO-phasediag}).
The same conclusion is reached by looking at the paraconductivity in the resistivity,
as seen in the curvature map of Fig.~\ref{Ando-map-YBCO}.
This proves the essential point, that the pseudogap phase is not a phase of precursor superconductivity.
There is no evidence from Nernst data that short-lived Cooper pairs start to form at \Tstar.
% and superconducting fluctuations do not extend up to \Tstar.

The limitation is that \Tmin{} cannot be defined for a cuprate with $\nu_{\rm qp}$\,$>$\,$0$, like LSCO.
We therefore turn to another, more general criterion, based on the field dependence of $\nu$. 
Indeed, because 
$\nu_{\rm sc}$ always decreases with increasing $H$, we can say that when $\nu$
is independent of field, then $\nu_{\rm sc}$ is negligible compared to $\nu_{\rm qp}$.
We define \TB{} to be the temperature above which $\nu$ no longer decreases with $H$.

%%%%%%%%%%%%%%%%%%%%%%%%%%%%%%%%%%%%%%%%%%%%%%%%%%%%%%%%%%%%%%%%%%%%%%%%%%%%

\subsubsection{YBCO}

%%%%%%%%%%%%%%%%%%%%%%%%%%     FIGURE 21  %%%%%%%%%%%%%%%%%%%%%%%%%%%%%%%%%%%%%%

%
\begin{figure}[!t]
\centering
\includegraphics[width=0.45\textwidth]{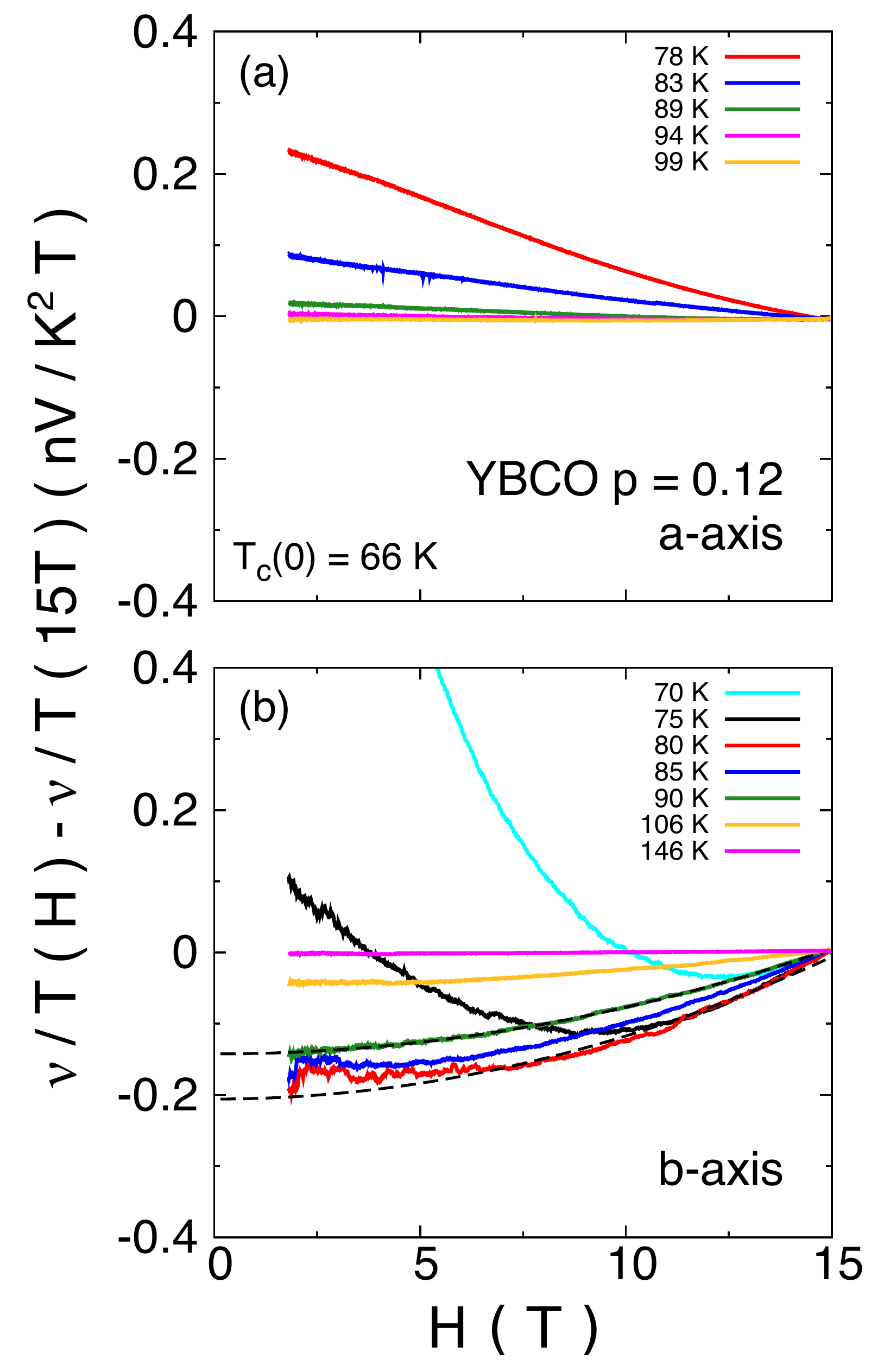}
\caption{(Color online) Nernst coefficient of YBCO at doping $p$\,=\,$0.12$ with thermal gradient applied along the $a$ axis (panel (a)) and $b$ axis (panel (b)),
		plotted as $\nu$\,/\,$T(H) - \nu$\,/\,$T$(15\,T), versus magnetic field, at various temperatures as indicated. 
		In (a), the isotherms above $T_{\rm c}(0)$\,=\,$66\,$K show a field-induced suppression, for 
		$T < T_{\rm B}$\,=\,$95$\,$\pm$\,$5$\,K. Above \TB, the field dependence of $\nu$ is negligible. 
		We use $T_{\rm B}$ as a second criterion to define the onset of superconducting fluctuations, in addition to $T_{\rm min}$. 
		In (b), isotherms immediately above $T_{\rm c}(0)$ (70~K and 75~K) also show a field-induced suppression of the superconducting signal.
		Isotherms far above \Tc~(90~K and 106~K) show a field-induced growth of $\nu(H)$, proportional to $H^2$ (dashed lines),
		due to a ``magneto-resistance'' in the quasiparticle contribution to the Nernst signal (see text).
		At low $H$, a superconducting signal is seen above the $H^2$ background (dashed line) at $T$\,=\,$80$\,K, for example.
		The temperature above which $\nu(H)$ is purely quadratic is $T_{\rm B}$\,=\,$90$\,$\pm$\,$5$\,K.
%
%Note that the $y$ range were chosen to express both panels (directions) on the same scale.
} 
\label{SC-fluc-nu-YBCO}
\end{figure}
%
%%%%%%%%%%%%%%%%%%%%%%%%%%%%%%%%%%%%%%%%%%%%%%%%%%%%%%%%%%%%%%%%%%%%%%

Fig.~\ref{SC-fluc-nu-YBCO} shows $\nu$\,/\,$T$ vs $H$ for YBCO at doping $p$\,=\,$0.12$ at different temperatures above $T_{\rm c}(0)$\,=\,$66$\,K. 
Note that the value of  $\nu$\,/\,$T$ at the maximum field ($15$\,T) is subtracted from the isotherms 
to remove most of the quasiparticle contribution. 
Let us first examine the $a$-axis data (panel (a)). 
For $T$\,$<$\,$90$\,K, the field is seen to suppress $\nu$, as expected. 
For $T$\,$>$\,$90$\,K, however, there is negligible field dependence. 
Using the lack of a detectable field dependence to define $T_{\rm B}$, 
we get $T_{\rm B}$\,=\,$95$\,$\pm$\,$5$\,K,
in agreement with \Tmin\,=\,$90$\,$\pm$\,$5$\,K in that sample (Fig.~\ref{YBCO-phasediag}).

In panel (b) of Fig.~\ref{SC-fluc-nu-YBCO}, we show the $b$-axis isotherms in YBCO at $p$\,=\,$0.12$.
At $T$\,=\,$70$\,K (pale blue curve) and $T$\,=\,$75$\,K (black curve), we see clearly that the field suppresses the superconducting signal.
But it also causes a positive rise in $\nu$, thereby producing a minimum in $\nu$ vs $H$.
We attribute this positive ``magneto-resistance'', which grows as $H^2$ (or as $H^3$ if plotted as $N$ vs $H$), 
to the quasiparticle component of the Nernst signal~[\onlinecite{Tafti2014}]. 
(All odd (even) powers of $H$ are allowed by symmetry in $N(\nu$).) 
The $H^2$ dependence is best seen at $T$\,=\,$90$\,K (green curve), where $\nu$\,/\,$T$ vs $H$ 
is perfectly described by a quadratic fit (dashed line in Fig.~\ref{SC-fluc-nu-YBCO}). 
(It is possible that the same $H^2$ contribution is in fact present in the $a$-axis data, but with a much reduced magnitude, 
perhaps in proportion with the ten-time smaller quasiparticle signal~[\onlinecite{Daou2010}].) 
At low $H$, a superconducting signal is seen above the $H^2$ background (dashed line) at $T$\,=\,$80$\,K, for example.
For the $b$-axis isotherms, we define \TB{} to be the temperature above which $\nu(H)$ is purely quadratic, 
giving \TB\,=\,$90$\,$\pm$\,$5$\,K for this doping, in agreement, within error bars, with the value obtained from the $a$-axis isotherms. 

In summary, we find that \Tmin\,$\simeq$\,\TB\,=\,$(1.4 \pm 0.05)$~\Tc~at $p$\,=\,$0.12$,
as also found at other nearby dopings (Fig.~\ref{YBCO-phasediag}).
Note that this is consistent with the onset of paraconductivity in the DC resistivity (Fig.~\ref{Ando-map-YBCO}) 
and microwave conductivity (see sec.~\ref{subsubsec:MW-THz}).

%diamagnetism detected via torque magnetometry, at $T_{\rm M}$. 
%Indeed, for an optimally-doped YBCO sample with $T_{\rm c} $\,=\,$ 90$\,K, $T_{\rm M} $\,=\,$125$\,$\pm$\,$5$\,K~[\onlinecite{Li2010}], 
%so that $T_{\rm M}$\,$\simeq$\,$1.4$\,$\,T_{\rm c}$. 

%%%%%%%%%%%%%%%%%%%%%%%%%%%%%%%%%%%%%%%%%%%%%%%%%%%%%%%%%%%%%%%%%%%%%%%
%%%%%%%%%%%%%%%%%%%%%%%%%%%%%%%%%%%%%%%%%%%%%%%%%%%%%%%%%%%%%%%%%%%%%%%

%%%%%%%%%%%%%%%%%%%%%%%%%%%%%%%%%%%%%%%%%%%%%%%%%%%%%%%%%%%%%%%%%%%%%%%
%%%%%%      LSCO - SC fluctuations
%%%%%%%%%%%%%%%%%%%%%%%%%%%%%%%%%%%%%%%%%%%%%%%%%%%%%%%%%%%%%%%%%%%%%%%

%%%%%%%%%%%%%%%%%%%%%%%%%%     FIGURE 22   %%%%%%%%%%%%%%%%%%%%%%%%%%%%%%%%%%%%%%

%
\begin{figure}
\centering
\includegraphics[width=0.45\textwidth]{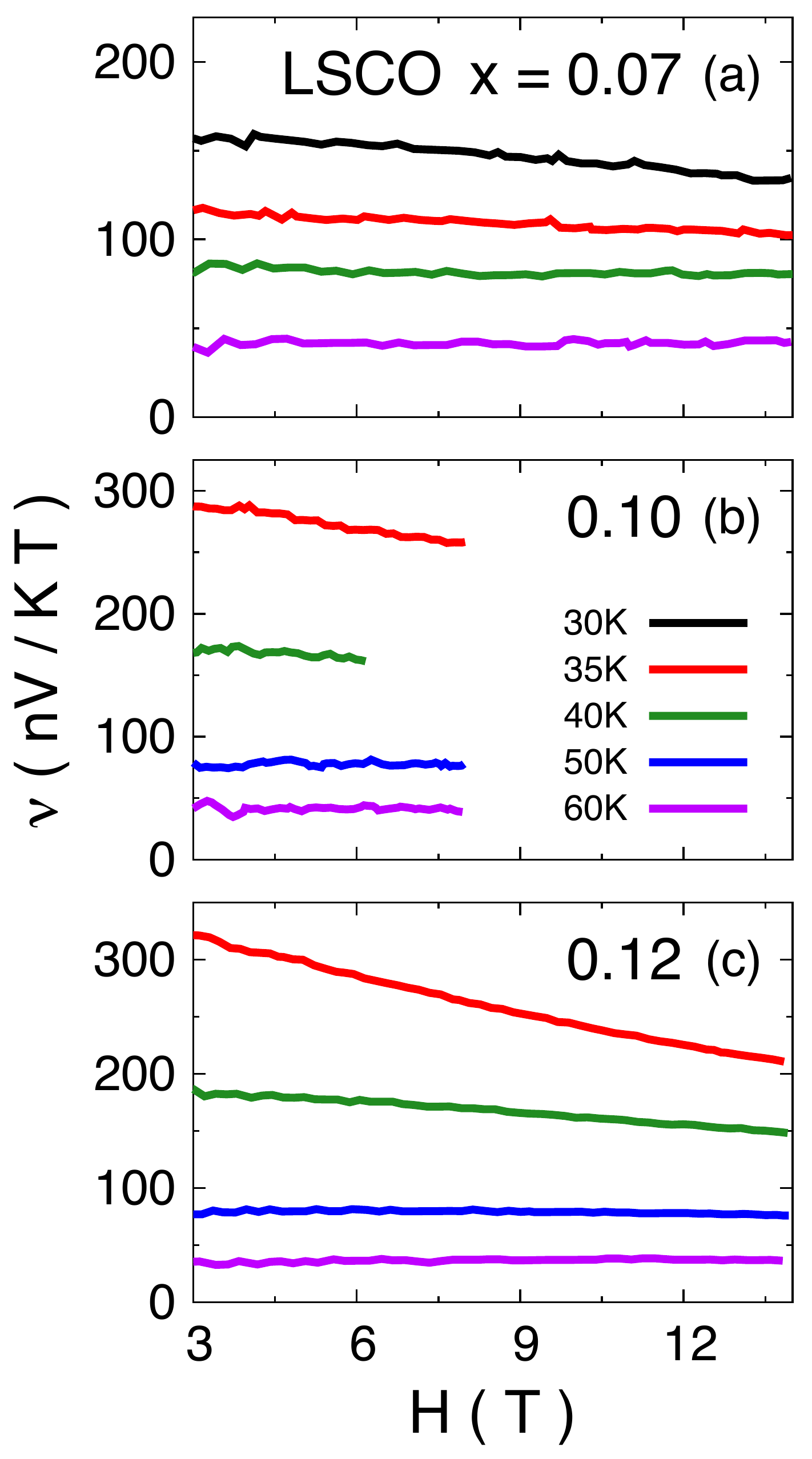}
\caption{(Color online) Field dependence of the Nernst coefficient in LSCO at doping $x$\,=\,$p$\,=\,$0.07 $
		((a); $T_{\rm c}$\,=\,$11$\,K), $0.10$ ((b); $T_{\rm c}$\,=\,$28$\,K) and $0.12$ ((c); $T_{\rm c}$\,=\,$29$\,K) 
		at various temperatures above \Tc~(color-coded legend). 
		These curves show that above a certain temperature, defined as $T_{\rm B}$, 
		the Nernst coefficient ($\nu$\,$\equiv$\,$N$\,/\,$H$) is essentially field-independent. 
		The strong field dependence associated with superconducting fluctuations gone, 
		the quasiparticle field-independent contribution dominates the signal. 
		We find $T_{\rm B}$\,=\,$40$\,$\pm$\,$10$, $50$\,$\pm$\,$10$ and $50$\,$\pm$\,$10$\,K for $x$\,=\,$0.07$, $0.10$ and $0.12$, respectively. 
		These $T_{\rm B}$ values are plotted on the curvature map of Fig.~\ref{Ando-map-LSCO} and are seen to fall on the boundary of the paraconductivity region. 
		Data at $p$\,=\,$0.07$ and 0.12 are taken from Ref.~[\onlinecite{Wang2006}], at $p$\,=\,$0.10$ from Ref.~[\onlinecite{Xu2000}]. 
		}
\label{TB-LSCO}
\end{figure}
%
%%%%%%%%%%%%%%%%%%%%%%%%%%%%%%%%%%%%%%%%%%%%%%%%%%%%%%%%%%%%%%%%%%%%%%%

\subsubsection{LSCO}
\label{subsubsec:Appendix-TB-LSCO}

In order to delineate the region of significant superconducting fluctuations in LSCO, we can use paraconductivity, as was done for YBCO. 
In Fig.~\ref{Ando-map-LSCO}, we see that the onset of paraconductivity in LSCO occurs 
at a temperature $T_{\rm para}$ between 50\,K and 65\,K, in the range $0.08$\,$<$\,$p$\,$<$\,$0.20$~[\onlinecite{Ando2004}]. 
%
%For example, $T_{\rm para} \simeq 65$\,K at $p $\,=\,$ 0.17$ and $T_{\rm para} \simeq 60$\,K at $p $\,=\,$ 0.09$.  
%
(Note that the weak $p$ dependence of $T_{\rm para}$ may come from some inhomogeneity in doping, 
whereby parts of all samples have some optimally-doped regions, where $T_{\rm c}$ is highest.)
At optimal doping, where \Tc\,=\,$40$\,K, $T_{\rm para}$\,$\simeq$\,$65$\,K, so that $T_{\rm para}$\,$\simeq$\,$1.6$\,\Tc.

%%%%%%%%%%%%%%%%%%%%%%%%%%     FIGURE 23   %%%%%%%%%%%%%%%%%%%%%%%%%%%%%%%%%%%%%%

%
\begin{figure*}
\centering
\includegraphics[width=0.85\textwidth]{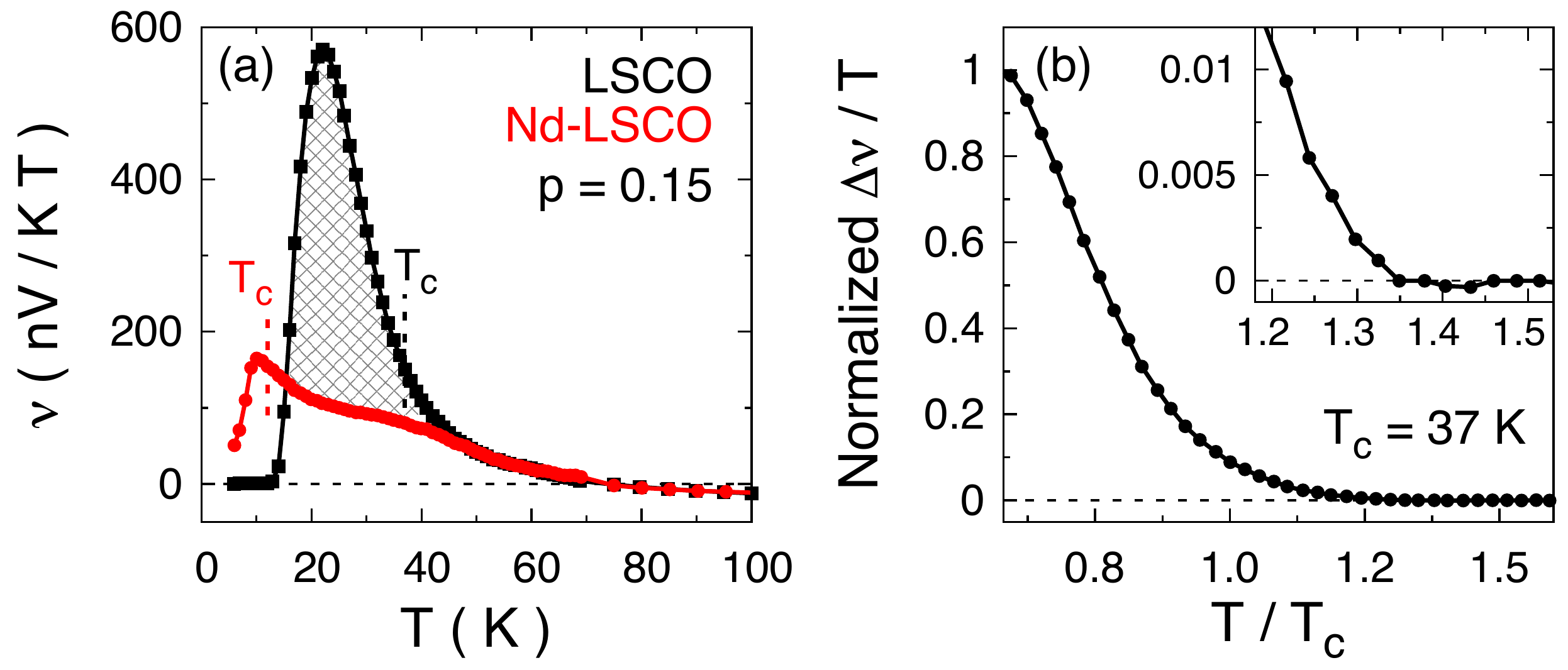}
\caption{(Color online) (a) Nernst coefficient $\nu$ of Nd-LSCO (red circles) and LSCO (black squares) 
		at  $p$\,=\,$0.15$ versus temperature, measured with $H$\,=\,$9$\,T. 
		At high $T$, there is good agreement between the two data sets (see also Fig.~\ref{LSCOp15}), until 
		 the superconducting fluctuations in LSCO start to grow, below $\sim 50$~K. 
		This difference (cross-hatched region) is attributable to their different 
		$T_{\rm c}$ ($37$\,K for LSCO and $12$\,K for Nd-LSCO) 
		and can be seen as the superconducting fluctuations contribution for LSCO. 
		(b) Difference between $\nu$ of LSCO and Nd-LSCO (cross-hatched region of panel (a)), 
		plotted as $\Delta\nu$\,/\,$T$, normalized at maximum value,
		versus normalized temperature 
		$T/$\Tc~(where \Tc~is the \Tc~of LSCO).
		Subtracting Nd-LSCO from LSCO has the effect of taking away the quasiparticle contribution and revealing 
		the superconducting contribution to the Nernst signal in LSCO. This superconducting contribution 
		is seen to decrease rapidly with increasing temperature, vanishing around 1.35\,\Tc~(see inset).
		The data are from Ref.~[\onlinecite{Fujii2010}].
		}
\label{SCfluc-LSCOp15}
\end{figure*}
%
%%%%%%%%%%%%%%%%%%%%%%%%%%%%%%%%%%%%%%%%%%%%%%%%%%%%%%%%%%%%%%%%%%%%%%%

It is harder to disentangle superconducting and quasiparticle contributions to the Nernst signal 
in LSCO-based materials because unlike YBCO the quasiparticle contribution also rises positively 
with decreasing temperature, so there is no equivalent of $T_{\rm min}$.
We therefore use the lack of a detectable field dependence to define $T_{\rm B}$, as our criterion for the onset of superconducting fluctuations. 
Fig.~\ref{TB-LSCO} shows $\nu$ as a function of magnetic field $H$ for LSCO at three dopings ($0.07, 0.10$ et $0.12$), at $T$\,$>$\,\Tc. 
We can extract $T_{\rm B}$ from these curves as the temperature above which the isotherms are flat: 
$T_{\rm B}$\,=\,$40$\,$\pm$\,$10, 50$\,$\pm$\,$10$ and  $50$\,$\pm$\,$10$\,K for $x$\,=\,$0.07, 0.10$ et $0.12$, respectively. 
We note that the available data is limited to $\approx$\,10\,T and it would be interesting to see if this linearity can be tracked at higher fields.
\TB{} is then plotted as a function of doping on the curvature map of LSCO (Fig.~\ref{Ando-map-LSCO}).
It is seen to fall more or less on the boundary of the paraconductivity region (blue band above \Tc),
\ie{} \TB\,$\simeq$\,$T_{\rm para}$.

%%%%%%%%%%%%%%%%%%%%%%%%%%%%%%%%%%%%%%%%%%%%%%%%%%%%%%%%%%%%%%%%%%%%%%%%%%%%

\subsubsection{Comparing LSCO to Nd-LSCO}

Another approach for disentangling $\nu_{\rm sc}$ and $\nu_{\rm qp}$ in LSCO is to compare with Nd-LSCO, 
its lower-\Tc{} counterpart, at the same doping. 
As seen in Fig.~\ref{SCfluc-LSCOp15}(a), at high temperature $\nu(T)$ is essentially identical in LSCO and Nd-LSCO, and it is not due to superconducting fluctuations. 
Therefore, comparing the two materials should reveal the onset of a detectable superconducting contribution 
in LSCO, since its \Tc{} is higher than in Nd-LSCO. 
Fig.~\ref{SCfluc-LSCOp15}(a) compares LSCO and Nd-LSCO at $p$\,=\,$0.15$, using data from Ref.~[\onlinecite{Fujii2010}], 
where $T_{\rm c}$\,=\,$37$\,K and $12$\,K, respectively. 
Down to 50\,K or so, the data are nearly identical, even through the LTT structural transition of Nd-LSCO (at 70\,K). 
Below 50\,K, the two curves deviate, with the LSCO curve showing a pronounced superconducting peak.
%while $\nu(T)$ of Nd-LSCO continues its smooth rise, associated with the quasiparticle signal. 
%
This difference between the two curves (shaded region in Fig.~\ref{SCfluc-LSCOp15}(a)) can be seen as the superconducting contribution of LSCO. 
Fig.~\ref{SCfluc-LSCOp15}(b) plots this difference $\Delta \nu$ between the two data sets (normalized at maximum value) 
vs $T$\,/\,$T_{\rm c}$, with \Tc\,=\,$37$\,K (in LSCO). 
In the inset, a zoom shows that the difference becomes non-zero below $\sim$\,$1.4$\,\Tc. 
This puts a reasonable upper bound on a detectable superconducting Nernst signal in LSCO. 

We conclude that the regime of significant superconducting fluctuations in LSCO extends up to $1.5$\,$\pm$\,$0.1$\,\Tc,
with the error bar covering the various criteria (paraconductivity in the resistivity, field independence in the Nernst signal,
comparison to Nd-LSCO).

%%%%%%%%%%%%%%%%%%%%%%%%%%%%%%%%%%%%%%%%%%%%%%%%%%%%%%%%%%%%%%%%%%%%%%%
%%%%%%%%%%%%%%%%%%%%%%%%%%%%%%%%%%%%%%%%%%%%%%%%%%%%%%%%%%%%%%%%%%%%%%%

\subsection{Other probes and materials}
\label{subsec:other}

\subsubsection{Nernst effect in Bi-2201 and Hg-1201} 
\label{subsubsec:Nernst-Bi-Hg}

\begin{figure}
\centering
\includegraphics[width=0.45\textwidth]{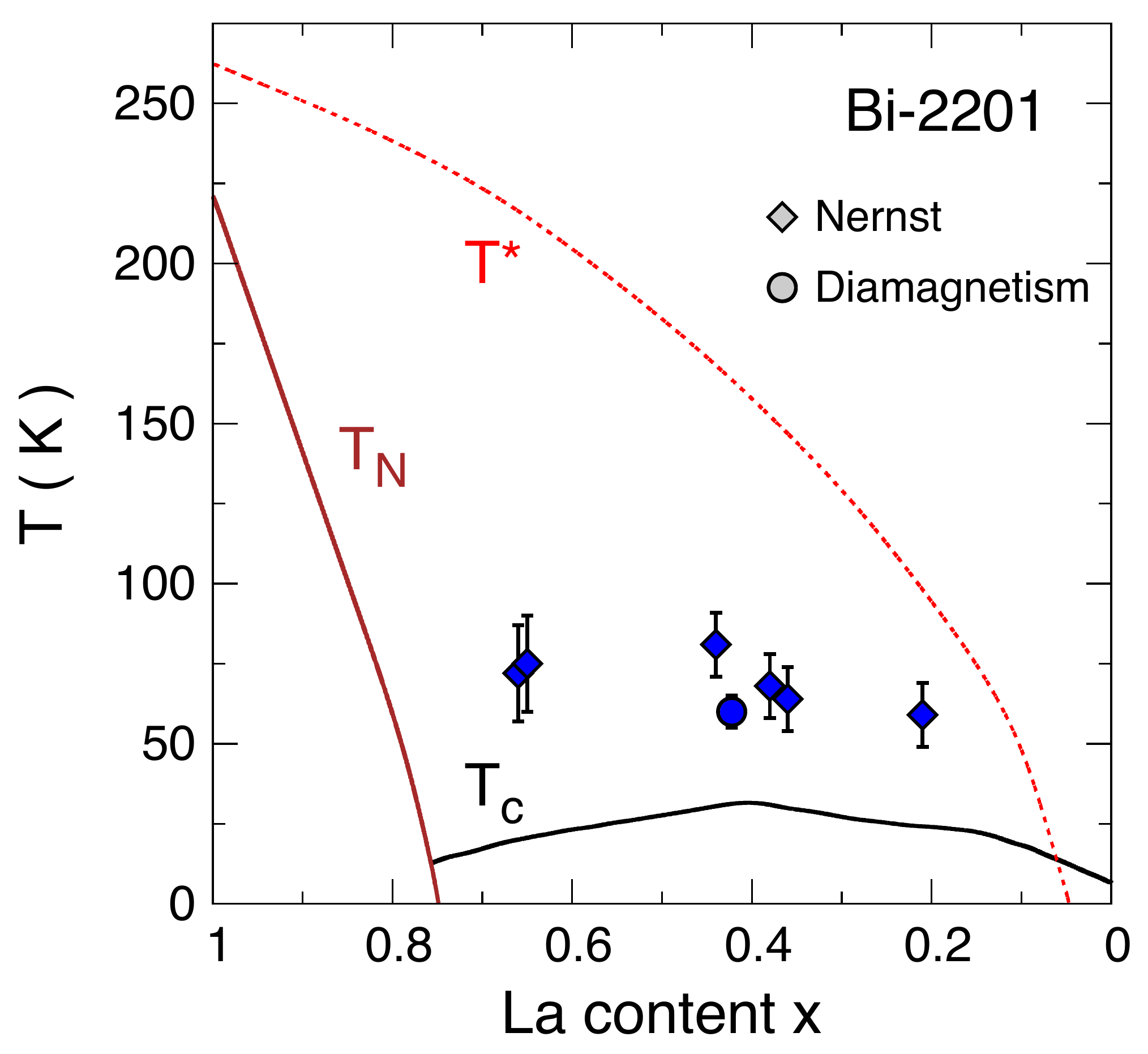}
\caption{Temperature-doping phase diagram of Bi-2201 as a function of La doping $x$, showing 
		\TN~(brown line), \Tc~(black line), and \Tstar~detected by NMR (dashed red line) (from [\onlinecite{Kawasaki2010}]).
		Blue diamonds mark the onset of a finite Nernst signal (\Tonset; [\onlinecite{Okada2010}]), attributed to SC fluctuations.
		\Tonset~of Bi-2201 doped with Eu is also plotted for comparison (\Tonset~open blue diamonds; [\onlinecite{Okada2010}]). 
		Also shown is the onset temperature for SC fluctuations in Bi-2201 detected by torque magnetometry 
		near optimal doping (blue circle; [\onlinecite{Yu2012},\onlinecite{Yu2017}]).
		}
\label{Bi2201-Tonset}
\end{figure}

In this Article on the Nernst effect in cuprates, we have focused on YBCO and LSCO (as well as Nd-LSCO and Eu-LSCO).
Now, studies of the Nernst effect have also been carried out on other cuprates, such as Bi-2212 and Bi-2201.
They lead to the same basic finding that the regime of SC fluctuations tracks \Tc~and ends well below the
pseudogap temperature \Tstar.
In Fig.~\ref{Bi2201-Tonset}, we plot the temperature \Tonset~below which the Nernst signal in Bi-2201 becomes detectable
in the data of ref.~\onlinecite{Okada2010}.
Note that \Tonset~is necessarily an upper bound on the regime of SC fluctuations.
Looking closely at the data across the doping range, one finds no trace of any signal above $70$\,K. 
As discussed below (sec.~\ref{subsubsec:Torque}), 
this value is consistent with the upper limit on detectable SC fluctuations in torque magnetometry data on Bi-2201.
What is clear from Fig.~\ref{Bi2201-Tonset} is that \Tonset~in Bi-2201 is flat vs doping, with \Tonset~$\simeq$\,$65$\,K across the phase diagram, 
whereas \Tstar~rises with underdoping, to values as high as \Tstar~$\simeq$\,$250$\,K at low doping.
This is therefore very similar to the phase diagram of LSCO shown in Fig.~\ref{LNSCO-ongplot-2002}. 
Both LSCO and Bi-2201 lead us to the same conclusion as reached for YBCO: 
the regime of SC fluctuations tracks \Tc, 
and it lies well below the pseudogap temperature \Tstar.

As for YBCO, Nernst measurements on Hg-1201 have the advantage of a negative quasiparticle signal, so 
that the onset of SC fluctuations can immediately be detected as a minimum occurring at \Tmin.
For a sample with \Tc\,=\,$65$\,K, \Tmin\,=\,$100$\,$\pm$\,5\,K~[\onlinecite{Doiron-Leyraud2013}].
In Fig.~\ref{YBCO-SCF}, we show how this compares to the \Tmin~values in YBCO,
where for the same \Tc~one gets \Tmin\,=\,$90$\,$\pm$\,$5$\,K~(Fig.~\ref{YBCO-phasediag}).

%%%%%%%%%%%%%%%%%%%%%%%%%%     FIGURE 24   %%%%%%%%%%%%%%%%%%%%%%%%%%%%%%%%%%%%%%

%
%
%%%%%%%%%%%%%%%%%%%%%%%%%%%%%%%%%%%%%%%%%%%%%%%%%%%%%%%%%%%%%%%%%%%%%%%

%%%%%%%%%%%%%%%%%%%%%%%%%%%%%%%%%%%%%%%%%%%%%%%%%%%%%%%%%%%%%%%%%%%%%%%

\subsubsection{Torque magnetometry} 
\label{subsubsec:Torque}

In this Article, we have seen that the resistivity and the Nernst coefficient can both be used to detect
the onset of SC fluctuations above \Tc.
Magnetization is another probe of such fluctuations, and torque magnetometry
measurements have been carried out on several cuprates.
Detailed high-sensitivity torque measurements of three different cuprates~[\onlinecite{Yu2012},\onlinecite{Yu2017}] reveal
that SC fluctuations can no longer be detected above 
$T$\,=\,$45$\,$\pm$\,$5$\,K in LSCO at $p$\,=\,$0.125$ 
(in good agreement with \TB\,=\,$50$\,$\pm$\,$10$\,K; see Fig.~\ref{LNSCO-ongplot-2002}),
$T$\,=\,$60$\,$\pm$\,$5$\,K in Bi-2201 at optimal doping 
(in good agreement with \Tonset\,$\simeq$\,{65}\,K; see Fig.~\ref{Bi2201-Tonset}),
and
$T$\,=\,$100$\,$\pm$\,$5$\,K in underdoped Hg-1201
(in good agreement with \Tmin\,=\,$100$\,K; see Fig.~\ref{YBCO-SCF}).

%%%%%%%%%%%%%%%%%%%%%%%%%%%%%%%%%%%%%%%%%%%%%%%%%%%%%%%%%%%%%%%%%%%%%%%

\subsubsection{Microwave and THz conductivity} 
\label{subsubsec:MW-THz}

\begin{figure}
\centering
\includegraphics[width=0.45\textwidth]{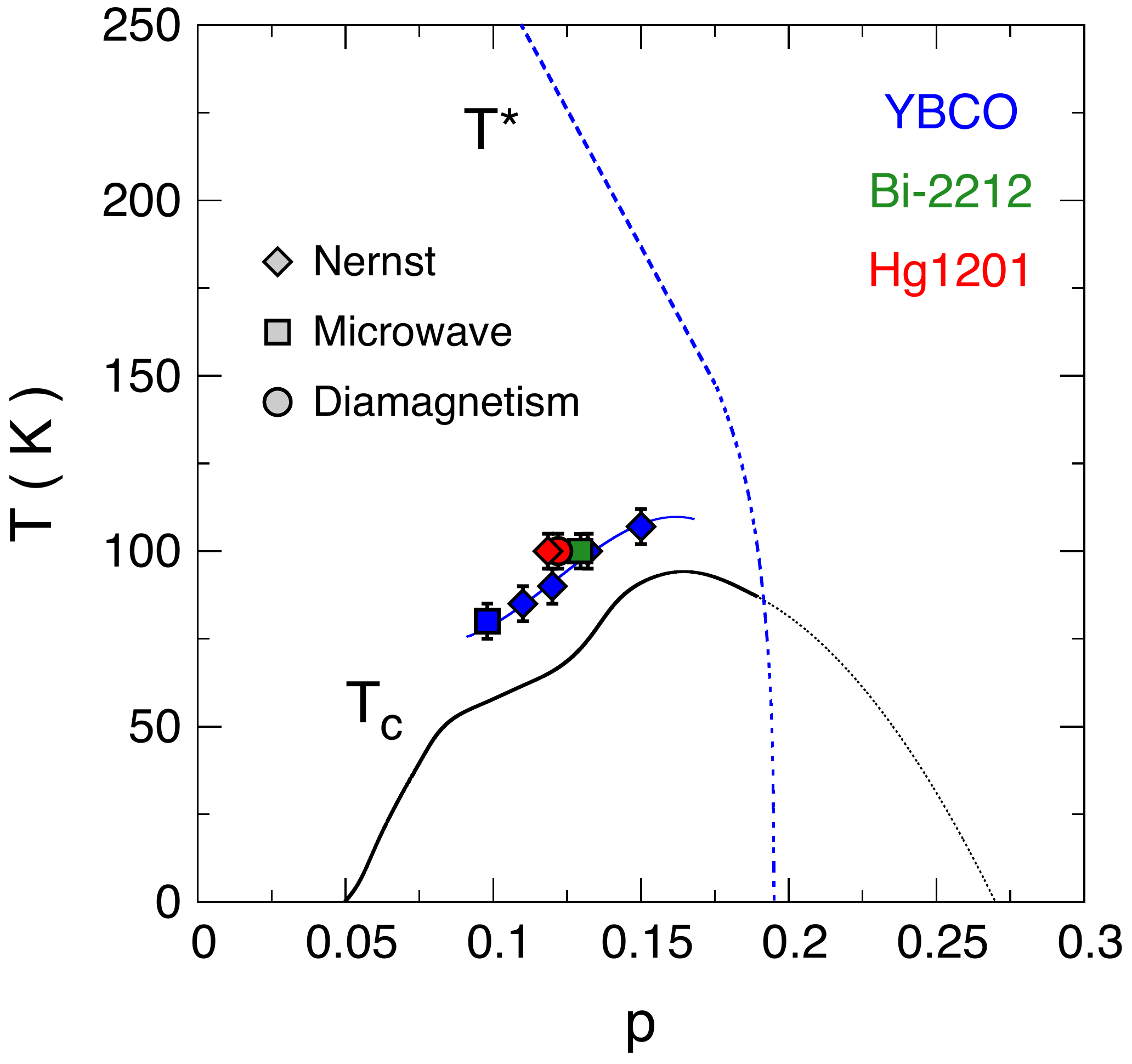}
\caption{Temperature-doping phase diagram of YBCO, showing the pseudogap temperature \Tstar~(dashed blue line) 
		and the onset of SC fluctuations detected in the Nernst signal (\Tmin, blue diamonds; from Fig.~\ref{YBCO-phasediag})
		and in the microwave conductivity (blue square; from [\onlinecite{Grbic2011}]).
		In addition, the onset temperature for SC fluctuations in Bi-2212 (green) and Hg-1201 (red) is also displayed,
		for samples with a \Tc~value given by the solid black line, from Nernst (red diamond, \Tmin; [\onlinecite{Doiron-Leyraud2013}]), 
		microwave (green square; [\onlinecite{Corson1999}]) and 
		magnetization (red circle; [\onlinecite{Yu2012},\onlinecite{Yu2017}]) data. 
		Three different measurements on three different cuprates are seen to yield a very similar
		regime of SC fluctuations, close to \Tc~and well below \Tstar.
%		The green band marks the range of \Tstar values measured in Bi-2212~[\onlinecite{Vishik2012}] (see Fig.~\ref{YBCO-phasediag}).
}
\label{YBCO-SCF}
\end{figure}

SC fluctuations can also be detected via measurements at microwave and THz frequencies.
In a seminal study using microwave measurements, Corson and co-workers detected SC fluctuations in an underdoped sample of Bi-2212
with \Tc\,=\,$74$\,K up to at most $T$\,=\,$100$\,K~[\onlinecite{Corson1999}]. 
As shown in Fig.~\ref{YBCO-SCF}, this upper limit for the fluctuation regime in Bi-2212 
agrees perfectly with \Tmin\,=\,$100$\,$\pm$\,$10$\,K measured in YBCO for the same value of \Tc.
More recent microwave measurements on YBCO itself~[\onlinecite{Grbic2011}] 
confirm the excellent agreement with \Tmin~(Fig.~\ref{YBCO-SCF}).

\begin{figure}
\centering
\includegraphics[width=0.45\textwidth]{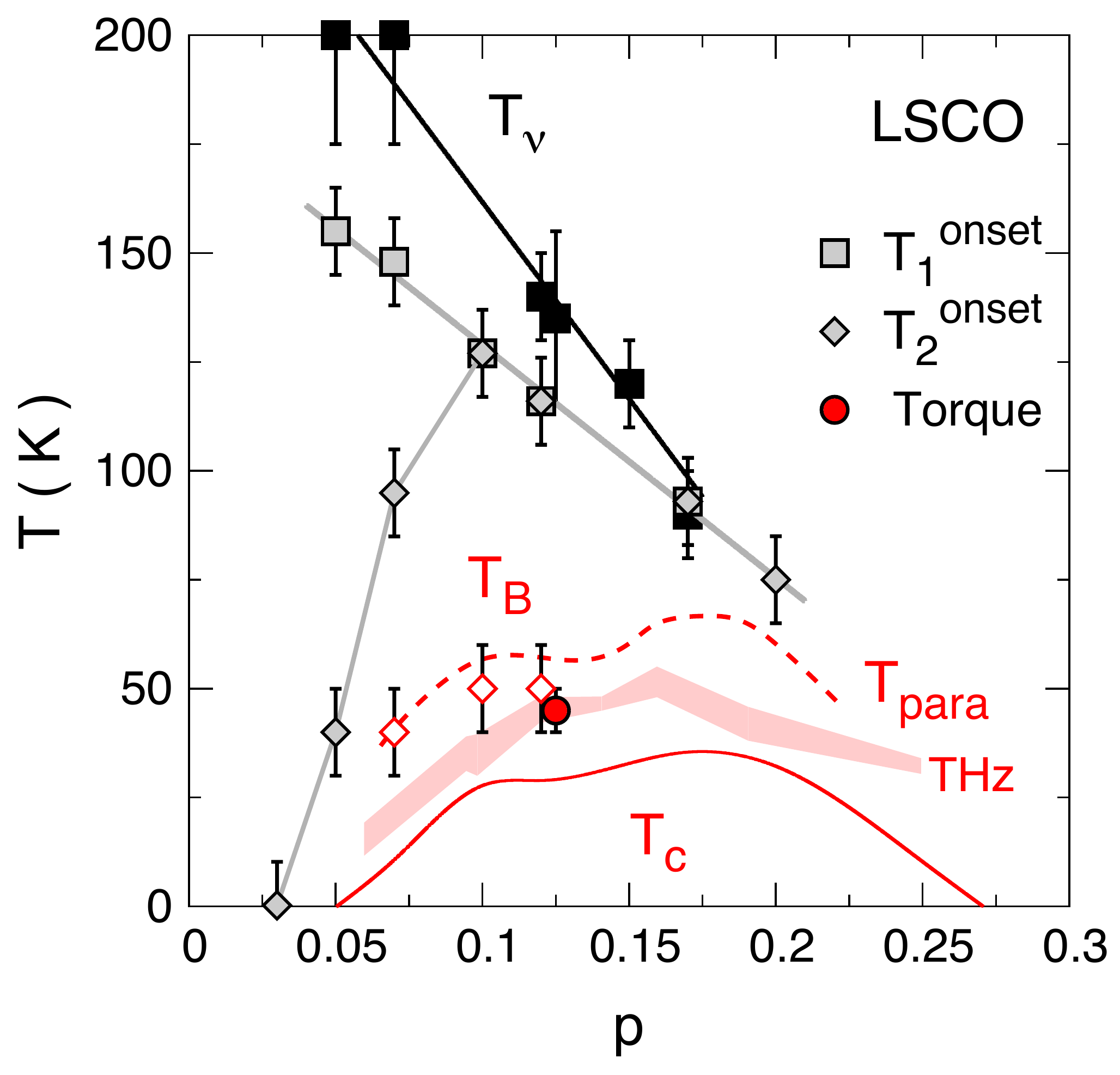}
\caption{Temperature-doping phase diagram of LSCO from the Princeton group (grey).
		The early version of their phase diagram~[\onlinecite{Xu2000}] plots an onset temperature, labelled $T_1^{\rm{onset}}$ (grey squares), 
		defined as the temperature below which $\nu(T)$ starts to rise upon cooling. 
		In a later version of their phase diagram~[\onlinecite{Wang2006},\onlinecite{Wang2001}], they plot a revised onset temperature, 
		which we label $T_2^{\rm{onset}}$ (grey diamonds).
For $p \geq 0.10$, $T_2^{\rm{onset}} \equiv T_1^{\rm{onset}}$;
for $p$\,$<$\,$0.10$, $T_2^{\rm{onset}}$ is
the temperature below which $\nu$\,+\,$\mu$\,$S$ starts to rise upon cooling,
where $\mu$ is the mobility and $S$ the Seebeck coefficient (see text).
%the inflexion point in $\alpha_{xy}$\,/\,$\sigma_{xx}$ vs $T$ (see text). 
		For comparison, we also plot the \Tnu~values reported here for LSCO (black squares, Fig.~\ref{LSCO-phasediag}), 
		\ie{} the pseudogap temperature \Tstar~defined as the deviation from linearity in $\nu$\,/\,$T$ vs $T$ (Fig.~\ref{nu-Nd-Eu-LSCO}). 
		In red, we show the various signatures of superconductivity: 
		\Tc{} (solid line); \TB{} (open diamonds, Fig.~\ref{Ando-map-LSCO}); 
		$T_{\rm para}$, the onset of paraconductivity (dashed line, Fig.~\ref{Ando-map-LSCO}); 
		the onset of superconducting fluctuations probed by terahertz spectroscopy (pink shading~[\onlinecite{Bilbro2011}]) 
		and by torque magnetization (red circle [\onlinecite{Yu2012},\onlinecite{Yu2017}]). 
}
\label{LNSCO-ongplot-2002}
\end{figure}

Measurements on LSCO at THz frequencies find that the regime of SC fluctuations tracks \Tc~closely,
as displayed in Fig.~\ref{LNSCO-ongplot-2002}, in excellent agreement with the torque magnetization 
and with \TB~from the Nernst effect.

%%%%%%%%%%%%%%%%%%%%%%%%%%     FIGURE 25   %%%%%%%%%%%%%%%%%%%%%%%%%%%%%%%%%%%%%%

%
%
%%%%%%%%%%%%%%%%%%%%%%%%%%%%%%%%%%%%%%%%%%%%%%%%%%%%%%%%%%%%%%%%%%%%%%%

%%%%%%%%%%%%%%%%%%%%%%%%%%%%%%%%%%%%%%%%%%%%%%%%%%%%%%%%%%%%%%%%%%%%%%%

\subsubsection{ARPES and STM} 
\label{subsubsec:ARPES-STM}

Although some ARPES studies (e.g. [\onlinecite{Lee2007}]) find that the superconducting gap closes at \Tc,
other studies find a superconducting gap persisting above \Tc. 
For example, Reber and co-workers argue that in underdoped Bi-2212
such a gap extrapolates to zero only at $T$\,=\,$1.4$~\Tc~[\onlinecite{Reber2013}], 
an observation confirmed by a recent high-resolution laser-ARPES study~[\onlinecite{Kondo2015}].
This is roughly consistent with the microwave data mentioned in the previous section.

STM studies on Bi-2212 also find superconductivity persisting above \Tc~[\onlinecite{Lee2009}],
in one case [\onlinecite{Gomes2007}] up to temperatures much higher than the limit imposed by the ARPES and microwave data.

%%%%%%%%%%%%%%%%%%%%%%%%%%     FIGURE 26   %%%%%%%%%%%%%%%%%%%%%%%%%%%%%%%%%%%%%%

%
%
%%%%%%%%%%%%%%%%%%%%%%%%%%%%%%%%%%%%%%%%%%%%%%%%%%%%%%%%%%%%%%%%%%%%%%%

%%%%%%%%%%%%%%%%%%%%%%%%%%%%%%%%%%%%%%%%%%%%%%%%%%%%%%%%%%%%%%%%%%%%%%%
%%%%%% Comparison with Princeton group
%%%%%%%%%%%%%%%%%%%%%%%%%%%%%%%%%%%%%%%%%%%%%%%%%%%%%%%%%%%%%%%%%%%%%%%

\subsection{The Princeton interpretation}
\label{subsec:Appendix-Princeton}

Following the seminal work of the Princeton group in the period 
2000\,-\,2006~[\onlinecite{Wang2006},\onlinecite{Ong2004},\onlinecite{Xu2000},\onlinecite{Wang2001},\onlinecite{Wang2002},\onlinecite{Wang2003}], 
the Nernst effect in cuprates has been widely attributed to superconducting fluctuations, 
and in the underdoped regime those have been mostly interpreted as {\it phase} fluctuations, detectable in some cases up to $\sim$\,$5$\,\Tc.
This has been viewed as evidence that short-lived Cooper pairs without phase coherence form at temperatures well above \Tc.
In this Article, we have argued that the superconducting Nernst signal does not, in fact, extend very far above \Tc, 
becoming negligible above $\sim$\,$1.5$\,\Tc.
Moreover, recent studies suggest that even in the underdoped regime these fluctuations are not {\it phase} fluctuations,
but rather Gaussian fluctuations of both the amplitude and the phase of the order parameter~[\onlinecite{Chang2012},\onlinecite{Tafti2014}]. 
(Phase fluctuations may appear at very low doping.)

In this section, we examine the analysis performed by the Princeton group to understand why their interpretation is different from our own.
We emphasize that the data themselves are perfectly consistent amongst the various groups,
so that the disagreement is on the analysis and interpretation only.
This discussion will focus on LSCO data.

%Our Nernst effect results in the pseudogap of cuprates may seems in contradiction with those of the Princeton group, but it is not the case. 
%In terms of actual raw data, both sets are correct and reliable; differences between groups lies in the analysis and interpretation. 
%Here are some comparison and discussion of these differences that reveal flaws in their approach and justify ours. 

A first difference in the analysis lies in the definition of the onset temperature.
The Princeton group defines the onset of the low-temperature rise in the Nernst signal of LSCO (and other cuprates)
as the temperature $T^{\rm{onset}}$ below which $\nu(T)$ (rather than $\nu(T)$\,/\,$T$) starts to rise upon cooling.
In general, this $T^{\rm{onset}}$ is not equal to our $T_\nu$ (defined as the temperature below which $\nu$\,/\,$T$ starts to rise).
For example, data on LSCO at $p$\,=\,$0.15$, plotted as $\nu$ vs $T$ in Fig.~\ref{SCfluc-LSCOp15}(a),
yield $T^{\rm{onset}}$\,$\simeq$\,$100$\,K,
while we get \Tnu\,=\,$120$\,$\pm$\,$10$\,K from the same data plotted as $\nu$\,/\,$T$ vs $T$ (Fig.~\ref{Comparison-0p15}).

As shown in Fig.~\ref{LNSCO-ongplot-2002}, a plot of $T^{\rm{onset}}$ vs $p$ 
($T_1^{\rm{onset}}$, open squares~[\onlinecite{Xu2000}]) yields a line that is qualitatively similar to 
the $T_\nu$ line in Fig.~\ref{LSCO-phasediag}full squares in Fig.~\ref{LNSCO-ongplot-2002}), but slightly lower. 
Although the difference is not huge, it is nevertheless significant,
and adopting the correct definition is important to arrive at a meaningful onset temperature.

For the same reason that one should plot $C$\,/\,$T$, $\kappa$\,/\,$T$ and $S$\,/\,$T$ when analyzing 
the specific heat $C$, thermal conductivity $\kappa$ and thermopower $S$ of a metal,
one should plot $\nu$\,/\,$T$ rather than $\nu$ when analyzing the Nernst coefficient (see Eq.~\ref{eq:nuKamran}).
Because the laws of thermodynamics require that all four quantities ($C$, $\kappa$, $S$ and $\nu$) go to zero as $T$\,$\to$\,$0$, 
the negative $\nu$ observed at high $T$ in LSCO (Fig.~\ref{SCfluc-LSCOp15}(a))
must inevitably rise upon cooling, but this rise may not reflect any change in the electronic behaviour. 
This point is illustrated by the data on Nd-LSCO at $p$\,=\,$0.24$ (Fig.~\ref{Nernst-0p20}), which show a monotonic decrease  
of $\nu$\,/\,$T$ vs $T$ as $T$\,$\to$\,$0$. There is no upturn and so $T_{\nu}$\,=\,$0$. 
The absence of a pseudogap temperature (or any other characteristic temperature) is confirmed by the fact that 
the resistivity is featureless and perfectly linear below 50\,K (Fig.~\ref{rho-YBCO-LNSCO}(d)).
By contrast, if we were to plot $\nu$ vs $T$ instead, we would necessarily obtain $T^{\rm{onset}}$\,$>$\,$0$, 
suggesting that there is a meaningful crossover, in contradiction with the featureless $\rho(T)$.
Furthermore, the good agreement between $T_\nu$ and $T_\rho$ for both YBCO (Fig.~\ref{YBCO-phasediag}) and Nd-LSCO (Fig.~\ref{LSCO-phasediag}) 
validates the use of $\nu$\,/\,$T$ to define the onset of the change in $\nu(T)$ at high temperature. 

Beyond the issue of the correct definition (whether \Tnu~or \Tonset), 
the real question is what causes $\nu$~to initially rise upon cooling below \Tonset ?
We attribute the initial rise in $\nu(T)$\,=\,$\nu_{\rm qp}(T)$\,+\,$\nu_{\rm sc}(T)$ (coming down from high temperature)
to a change in the quasiparticle component $\nu_{\rm qp}(T)$,
while the Princeton group attributes this rise to a growth in the superconducting component $\nu_{\rm sc}(T)$.
In 2000, this was their interpretation for all dopings~[\onlinecite{Xu2000}],
down to $x$\,=\,$0.05$, their lowest doping (Fig.~\ref{LNSCO-ongplot-2002}).

In 2001, they realized that this interpretation is incorrect at low doping [\onlinecite{Wang2001}],
by examining the behavior of $\nu$\,+\,$\mu$\,$S$,
where $\mu$\,=\,$\tan\theta_{\rm H}$\,/\,$H$ is the mobility and $S$ is the Seebeck coefficient.
At $x$\,=\,$0.05$, they recognized that the initial rise in $\nu(T)$ from high temperature, 
reaching $+ 40$~nV\,/\,KT at $T$\,=\,$60$~K is in fact due to an increase in the quasiparticle term $\nu_{\rm qp}(T)$. 
Only below 40~K is there an additional rise coming from superconducting fluctuations.
They therefore revised the estimated temperature for the onset of superconducting fluctuations
from $T_1^{\rm{onset}}$\,=\,$150$~K~[\onlinecite{Xu2000}] 
down to $T_2^{\rm{onset}}$\,=\,$40$~K~[\onlinecite{Wang2001}] (see Fig.~\ref{LNSCO-ongplot-2002}). 
However, the Princeton group adopted the view that such a revision was only needed for 
$x$\,$<$\,$0.10$, arguing that any rise in $\nu_{\rm qp}(T)$ is negligible for $x$\,$\geq$\,$0.10$. 
This is where we disagree. 
At $x$\,=\,$0.10$, $\nu(T)$ also rises up to $+40$~nV\,/\,KT at $T$\,=\,$60$~K~[\onlinecite{Xu2000}], 
a rise that is very similar to the above-mentioned rise seen at $x$\,=\,$0.05$. 
Why, then, would the rise in $\nu(T)$ at $x$\,=\,$0.10$ not also come from $\nu_{\rm qp}(T)$? 
A rough estimate of $\nu_{\rm qp}$ can be obtained by looking at $\mu$\,$S$~[\onlinecite{Behnia2009}].
The Princeton data shows that $\mu$\,$S$ at $T$\,=\,$60$\,K is actually larger at $x$\,=\,$0.10$, not smaller. 
Indeed, $\mu$\,$S$\,$\simeq$\,$90$\,nV\,/\,KT (Fig.~3a, ref.~\onlinecite{Xu2000}), 
while $\mu$\,$S$\,$\simeq$\,$60$\,nV\,/\,KT at $x$\,=\,$0.05$ (Fig.~3b, ref.~\onlinecite{Wang2001}). 
Moreover, the measured $\nu(60{\rm \,K})$ is comparable, namely $\nu$\,=\,$40$~nV\,/\,KT at both dopings $x$\,=\,$0.05$ and $x$\,=\,$0.10$.
There numbers show clearly that there is no reason to assume that $\nu_{\rm qp}$ can be neglected at $x$\,=\,$0.10$.

We see that in LSCO, just as in Eu-LSCO (see Fig.~\ref{nu-LESCO-low-x}c) and Nd-LSCO (see Fig.~\ref{LSCOp15}), 
the initial rise in $\nu(T)$, below \Tonset, is in fact due to $\nu_{\rm qp}$, 
and the rise in $\nu_{\rm sc}$ only starts at much lower temperature.

Not surprisingly, the fact of using different criteria for $T^{\rm{onset}}$ for dopings above and below 
$p$\,=\,$0.10$ causes a sharp change in $T^{\rm{onset}}$ at that doping, 
producing an artifical peak at $p$\,=\,$0.10$
(see $T_{2}^{\rm{onset}}$; grey diamonds in Fig.~\ref{LNSCO-ongplot-2002}). 
The resulting \Tonset~line has no clear relation to the real onset of superconducting fluctuations. 
For example, the peak value, at $p$\,=\,$0.10$, is $T^{\rm{onset}}$\,=\,$125$\,$\pm$\,$10$\,K, 
whereas at that doping the onset of 
paraconductivity occurs at $T_{\rm{para}}$\,$\simeq$\,$60$\,K (Fig.~\ref{Ando-map-LSCO}) 
and the onset of field dependence in $\nu(T)$ occurs at $T_{\rm B}$\,=\,$50$\,$\pm$\,$10$\,K (Fig.~\ref{TB-LSCO}).
Moreover, the onset of superconducting fluctuations detected in both THz conductivity (sec.~\ref{subsubsec:MW-THz}) 
and torque magnetization (sec.~\ref{subsubsec:Torque}) is $45 \pm 5$~K (Fig.~\ref{LNSCO-ongplot-2002}).

In summary, the widely used Nernst phase diagram of $T_2^{\rm{onset}}$ vs $p$ in LSCO (Fig.~\ref{LNSCO-ongplot-2002})
does not correspond to the region of superconducting fluctuations in LSCO.

%%%%%%%%%%%%%%%%%%%%%%%%%%%%%%%%%%%%%%%%%%%%%%%%%%%%%%%%%%%%%%%%%%%%%%%
%%%%%  Comparison of diamagnetism from Princeton group and others. 
%%%%%%%%%%%%%%%%%%%%%%%%%%%%%%%%%%%%%%%%%%%%%%%%%%%%%%%%%%%%%%%%%%%%%%%

The Princeton group has also used torque magnetometry as a separate way to detect superconducting fluctuations above \Tc~[\onlinecite{Li2010}]. 
They define an onset temperature of diamagnetism (from superconducting fluctuations), $T^{\rm{M}}_{\rm{onset}}$, as the temperature 
below which the magnetization (or susceptibility) deviates downwards, towards negative values, from a positive paramagnetic background presumed to have
a linear temperature dependence. 
The values of $T^{\rm{M}}_{\rm{onset}}$ they extract as a function of doping
agree with the $T_2^{\rm{onset}}$ vs $p$ in Fig.~\ref{LNSCO-ongplot-2002}. 
They argue that this reinforces their interpretation of $T_2^{\rm{onset}}$ as being the onset of superconducting fluctuations above $T_{\rm c}$ 
in the phase diagram [\onlinecite{Li2010}]. 

The assumption of a linear-in-temperature magnetization background has been questioned [\onlinecite{Yu2015},\onlinecite{Yu2012},\onlinecite{Yu2017}]. 
In particular, it neglects the effect of the pseudogap phase on the paramagnetic susceptibility [\onlinecite{Yu2015}].
(To attribute a downward drop in the susceptibility from its linear-$T$ dependence at high $T$ to diamagnetism
is a bit like attributing the downward drop in the resistivity of YBCO from its linear-$T$ dependence at \Trho~(Fig.~\ref{rho-YBCO-LNSCO}(a))
to paraconductivity.)
To properly identify the diamagnetism that comes from superconductivity, Yu and coworkers [\onlinecite{Yu2012},\onlinecite{Yu2017}] used
its non-linear field dependence (and the emergence of higher harmonics in its angular dependence).
This is similar to our definition of \TB{} from the Nernst signal.
With this criterion, Yu {\it et al.} find that superconducting fluctuations are significant 
(in the magnetization signal) only in a narrow temperature region above the superconducting dome,
up to at most $\sim$\,$1.5$\,$T_{\rm c}$, in LSCO, Bi-2201 and Hg-1201~[\onlinecite{Yu2012},\onlinecite{Yu2017}].
This narrow regime of SC fluctuations, much narrower than that reported by the Princeton group,
is consistent with several probes (\TB~from Nernst, paraconductivity from DC resistivity,
microwave and THz conductivity) applied to several cuprates (YBCO, LSCO, Bi-2201, Bi-2212, Hg-1201),
as shown in Figs.~\ref{Bi2201-Tonset},~\ref{YBCO-SCF}~and~\ref{LNSCO-ongplot-2002}. 

Note that the field and temperature dependence of the magnetization data by the Princeton group can
be explained in terms of a Gaussian Ginzburg-Landau approach~[\onlinecite{Rey2013}].
The theory of Gaussian superconducting fluctuations was also shown to provide a valid quantitative description
of diamagnetism data in YBCO~[\onlinecite{Kokanovic2013}]. 
We conclude that the scenario of strong phase fluctuations in underdoped cuprates is neither supported 
by Nernst data nor by magnetization data,
except perhaps close to $p$\,=\,\pcone\,$\simeq$\,$0.05$.

\subsection{Summary}

To summarize this Appendix, several different measurements and properties, including the Nernst effect, 
paraconductivity,
magnetization,
terahertz spectroscopy, 
and microwave conductivity -- applied to a variety of materials, including YBCO, LSCO, Bi-2201, Bi-2212, Hg-1201 --
%and specific heat~[\onlinecite{Wen2009}] 
point to the same conclusion: 
significant superconducting fluctuations are present in cuprates only in a temperature interval close to \Tc, and well below \Tstar.

%%%%%%%%%%%%%%%%%%%%%%%%%%%%%%%%%%%%%%%%%%%%%%%%%%%%%%%%%%%%%%%%%%%%%%%%%%

\bibliographystyle{apsrev4-1}

%

%%%%%%%%%%%%%%%%%%%%%%%%%%%%%%%%%%%%%%%%%%%%%%%%%%%%%%%%%%%%%%%%%%%%%%%%%%%

\end{document}